\documentclass[10pt,twocolumn,letterpaper]{article}

\usepackage[pagenumbers]{cvpr} 

\usepackage[dvipsnames]{xcolor}

\definecolor{cvprblue}{rgb}{0.21,0.49,0.74}
\usepackage[pagebackref,breaklinks,colorlinks,citecolor=cvprblue]{hyperref}
\usepackage{algorithm}
\usepackage{algorithmic}
\usepackage{multirow}
\usepackage{pifont}
\newcommand{\xmark}{\ding{55}}%
\newcommand{\cmark}{\ding{51}}%

\def\paperID{*****} 
\def\confName{CVPR}
\def\confYear{2024}

\title{Modality-agnostic Domain Generalizable Medical Image Segmentation by Multi-Frequency in Multi-Scale Attention}

\author{Ju-Hyeon Nam$^{1}$ \qquad Nur Suriza Syazwany$^{1}$ \qquad Su Jung Kim$^{1}$ \qquad Sang-Chul Lee$^{1, 2}$ \\
Department of Electrical and Computer Engineering, Inha University$^{1}$,  DeepCardio$^{2}$\\
{\tt\small \{jhnam0514, surizasyazwany, sk2266\}@inha.edu \qquad sclee@\{inha.ac.kr, deepcardio.com\}}
}

\begin{document}
\maketitle
\begin{abstract}
Generalizability in deep neural networks plays a pivotal role in medical image segmentation. However, deep learning-based medical image analyses tend to overlook the importance of frequency variance, which is critical element for achieving a model that is both modality-agnostic and domain-generalizable.  Additionally, various models fail to account for the potential information loss that can arise from multi-task learning under deep supervision, a factor that can impair the model’s representation ability. To address these challenges, we propose a Modality-agnostic Domain Generalizable Network (MADGNet) for medical image segmentation, which comprises two key components: a Multi-Frequency in Multi-Scale Attention (MFMSA) block and Ensemble Sub-Decoding Module (E-SDM). The MFMSA block refines the process of spatial feature extraction, particularly in capturing boundary features, by incorporating multi-frequency and multi-scale features, thereby offering informative cues for tissue outline and anatomical structures. Moreover, we propose E-SDM to mitigate information loss in multi-task learning with deep supervision, especially during substantial upsampling from low resolution. We evaluate the segmentation performance of MADGNet across six modalities and fifteen datasets. Through extensive experiments, we demonstrate that MADGNet consistently outperforms  state-of-the-art models across various modalities, showcasing superior segmentation performance.  This affirms MADGNet as a robust solution for medical image segmentation that excels in diverse imaging scenarios. Our MADGNet code is available in \href{https://github.com/Inha-CVAI/MADGNet}{GitHub Link}.
\end{abstract}    
\section{Introduction}
\label{sec:intro}

Various types of cancers continue to pose a substantial threat to human life, contributing significantly to global mortality rates. In this context, the importance of medical image analysis becomes evident, serving as a linchpin in the early detection of malignant tumors or abnormal cells, ultimately contributing to extending patients' lives \cite{coates2015tailoring}. However, the manual analysis of noisy and blurred medical images presents notable challenges, rendering it susceptible to human errors \cite{chen2019learning}. In response to these issues, computer-aided diagnosis using traditional segmentation algorithms \cite{otsu1979threshold, kass1988snakes, tizhoosh2005image, haralick1987image} has garnered attention from clinical experts. Nevertheless, these algorithms still lack generalizability to new patient cases due to issues such as uneven intensity distribution, unexpected artifacts, and severe noise in medical images \cite{riccio2018new}.

\begin{figure}[t]
    \centering
    \includegraphics[width=0.5\textwidth]{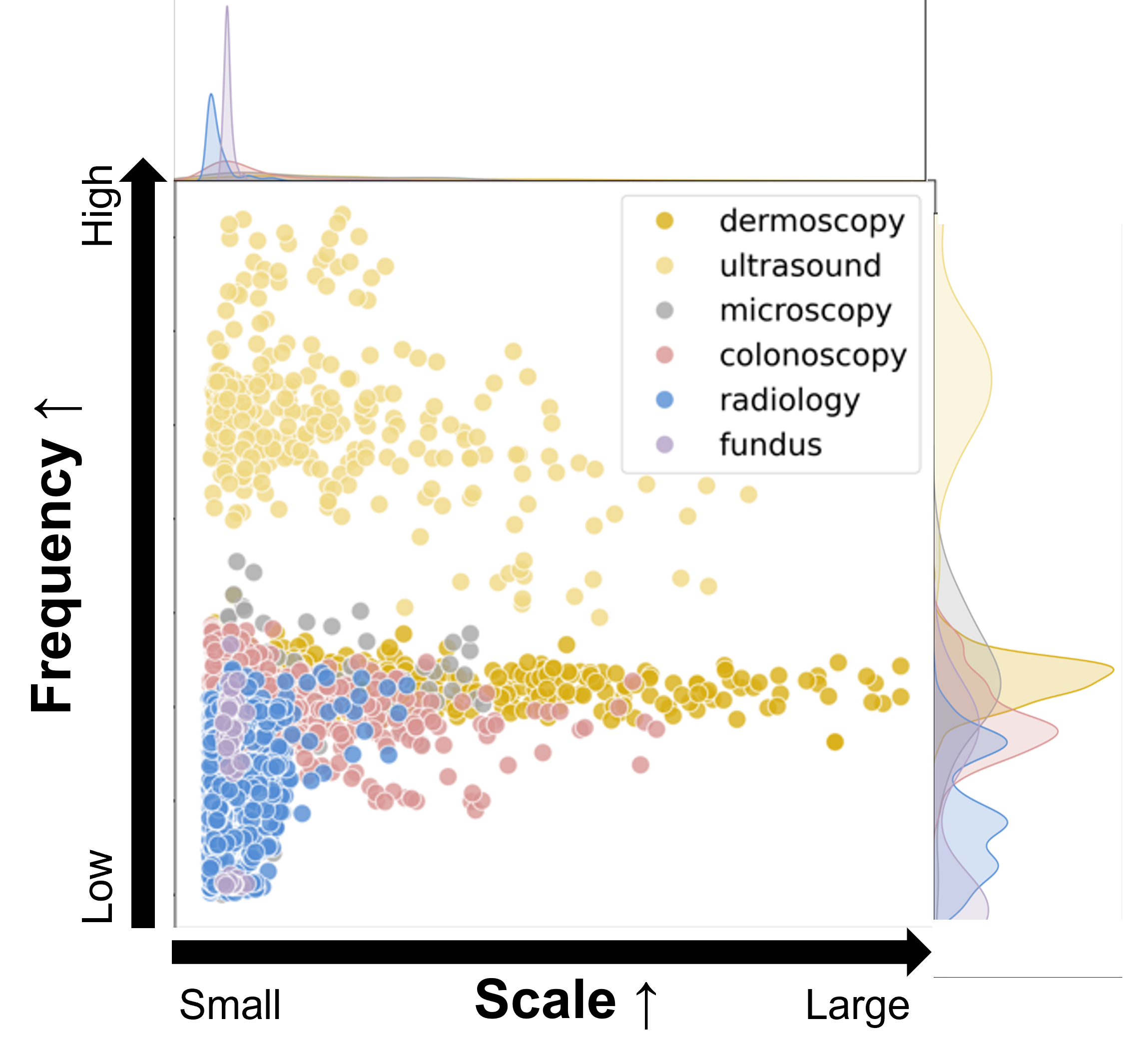}     \caption{Scale vs Frequency distribution per modality. The scale denotes the size of lesions, measured as the ratio of foreground pixels to the total number of pixels. Frequency is calculated by the power spectrum ratio of the high-frequency and full-frequency. We observe that the frequency variance is higher than the scale, which mainly focuses on solving the various sizes in other methods.}
    \label{fig:scale_frequency_distribution}
\end{figure}

Recent advancements in deep learning \cite{lecun1998gradient, vaswani2017attention} have paved the way for their application in medical image segmentation. In particular, UNet \cite{ronneberger2015u} has emerged as a fundamental approach for medical image segmentation, owing to its effective utilization of skip connections within its U-shaped architecture. An example of this evolution is UNet++ \cite{zhou2018unet++}, which employs nested UNet and deep supervision with skip connections to reduce the semantic gap between the encoder and decoder. With the recent surge in attention mechanisms \cite{wang2017residual, hu2018squeeze, park2018bam, woo2018cbam}, the evolution of UNet has witnessed the integration of attention, extracting discriminative features from noisy medical images, as seen in Attention UNet \cite{oktay1804attention} and Focus UNet \cite{yeung2021focus}.  Nevertheless, owing to disparities in medical image acquisition methods (modality) and the limited accessibility of datasets by patient privacy, these models often lack generalizability for unseen clinical settings, rendering practical implementation challenging. Therefore, exploring the question, \textit{“How a well-trained model in a modality can be generalized into other modalities (\textbf{modality-agnostic}) and be applicable in unseen clinical settings (\textbf{domain generalizability})?”} is a critical task to address these challenges.

Plotting the scale and frequency distribution graph (Fig. \ref{fig:scale_frequency_distribution}) unveiled a significant finding: medical images exhibit distinct distributions in both dimensions. While acknowledging the inherent interdependence between multi-frequency and multi-scale, Fig. \ref{fig:scale_frequency_distribution} emphasizes that they still reveal distinctive information. Multi-frequency, in particular, displays more variance, highlighting unique and valuable insights. This challenges the prevailing focus on either multi-scale \cite{feng2020cpfnet, gu2019net, srivastava2021msrf}  or multi-frequency information \cite{azad2021deep, yang2022fddl} and underscores the necessity of recognizing the independent and complementary nature of both dimensions in medical image segmentation. This discovery prompted us to efficiently utilize both dimensions, recognizing that even with interdependence, each dimension contributes distinct and valuable information.

To address the aforementioned problem, we propose a novel attention mechanism called Multi-Frequency in Multi-Scale Attention (MFMSA) block. This block employs multi-frequency channel attention (MFCA) with 2D Discrete Cosine Transform (2D DCT) \cite{ahmed1974discrete} to produce a channel attention map by extracting frequency statistics. Subsequently, multi-scale spatial attention (MSSA) is applied to extract discriminative boundary features and aggregate them from each scale. By effectively leveraging the sequential attention method to suppress the influence of noisy channels via MFCA and extracting discriminative feature maps for boundaries across various scales through MSSA, the MFMSA block's dual utilization enables the comprehensive capture of a broad spectrum of information from medical images. In more detail through multi-frequency analysis, it extract diverse features across different frequency bands while simultaneously performing multi-scale analysis to capture intricate details and broader structural information, ensuring a comprehensive understanding of the image content. Additionally, we introduce a Ensemble Sub-Decoding Module (E-SDM) to prevent information loss caused by drastic upsampling during multi-task learning with deep supervision. The resulting model, MADGNet, which mainly comprises the MFMSA block and E-SDM, achieved the highest segmentation performance in various modalities and clinical settings. The contributions of this paper can be summarized as follows:

\begin{itemize}
    \item We propose \textbf{MADGNet} for medical image segmentation in various modalities (\textbf{modality-agnostic}) and unseen clinical settings (\textbf{domain generalizability}) by efficiently fusing multi-frequency and multi-scale features. 

    \item Our novel MFMSA block is an efficient solution for integrating multi-scale and multi-frequency information, both of which are crucial for generalizing across various modalities and unseen clinical settings. E-SDM prevents information loss caused by drastic upsampling during multi-task learning with deep supervision in an ensemble manner.

    \item We evaluated MADGNet on fifteen different datasets with six different modalities and demonstrated that our model achieves the highest segmentation performance on various datasets. Furthermore, we showed that MADGNet can be utilized for multi-label medical image segmentation.
\end{itemize}
\section{Related Works}

\begin{figure*}[t]
    \centering
    \includegraphics[width=\textwidth]{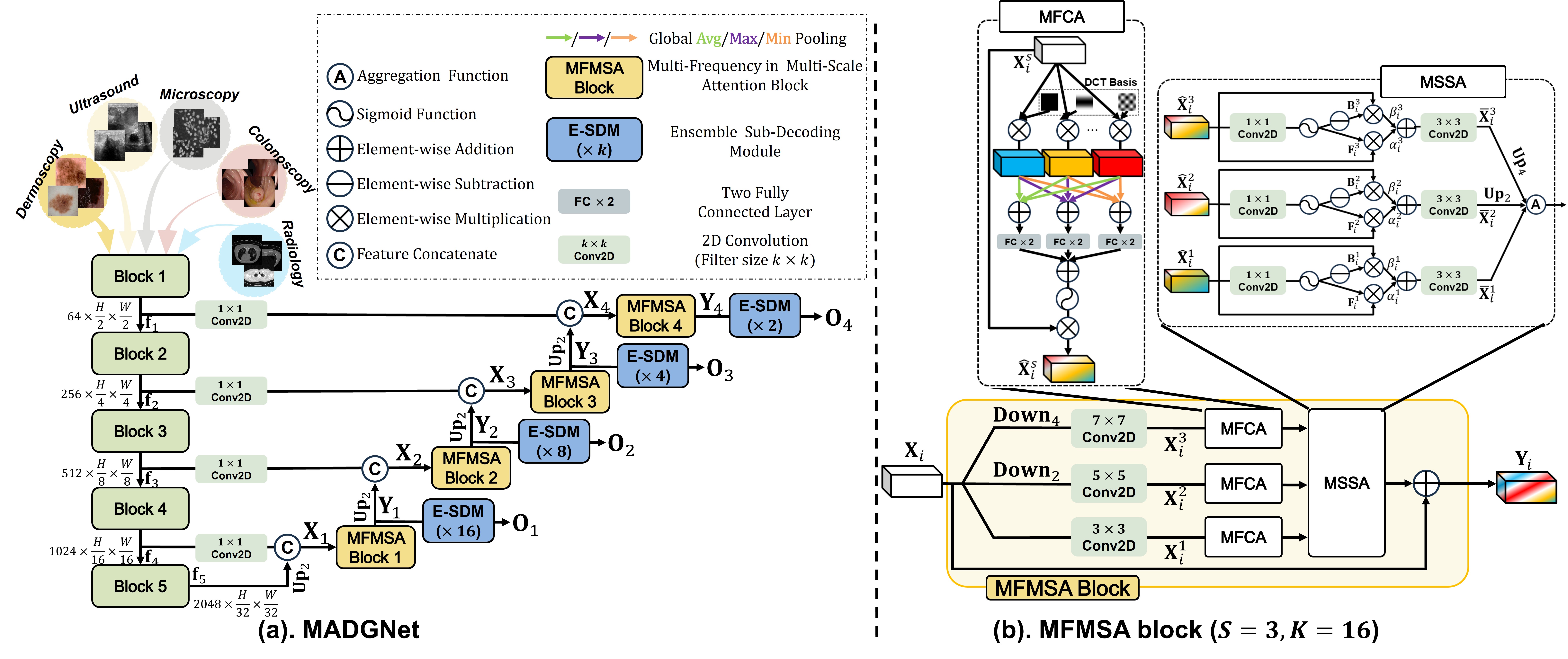}
    \caption{(a) The overall architecture of the proposed MADGNet mainly comprises MFMSA block and E-SDM (See Fig. \ref{fig:parallel_vs_cascade}). (b) MFMSA block contains $S$ scale branches ($S = 3$ in this figure) where the $s$-th branch input feature map are downsampled into $\eta^{s - 1}$ ($\eta = \frac{1}{2}$ in this figure). As our MFMSA block considers two dimensions (scale and frequency), MADGNet achieves the highest performance in various modalities and other clinical settings. Additionally, since E-SDM predicts a core task from sub-tasks, the final output is more accurate than when processed parallelly.}
    \label{fig:MFMSNet}
\end{figure*}

Recently, the significance of multi-scale information in the context of semantic segmentation has garnered increasing attention. For instance, DCSAUNet \cite{xu2023dcsau} retains the main features of the input image while combining feature maps with various receptive fields, resulting in enhanced segmentation quality. Simultaneously, several works \cite{chen2014semantic, yang2018denseaspp, liu2018receptive, lin2017feature, tan2020efficientdet} have demonstrated the effectiveness of enlarging the receptive field to capture multi-scale objects, often in conjunction with attention mechanisms. MS-CNN \cite{ji2018salient} improves context alignment through attention layers and multi-scale features, whereas DMSANet \cite{sagar2022dmsanet} proposes a lightweight module that integrates local and global features with spatial and channel attention.

In parallel, research efforts have surged to integrate multi-frequency techniques and attention mechanisms, aiming to enhance the extraction of local and global context for fine to coarse information by implementing various frequency transformation methods \cite{ahmed1974discrete, 1085978, ref1}. In particular, 2D DCT is widely used in computer vision to extract the frequency statistic feature for increasing the representation power due to its compression ability. For instance, FCANet \cite{qin2021fcanet} proposed a generalized SE block \cite{hu2018squeeze} based on 2D DCT basis functions. However, current medical image segmentation methods tend to focus on multi-scale or multi-frequency aspects, ignoring the potential benefits of combining both. To address this limitation, we integrate multi-frequency information with multi-scale features, thereby improving the model’s ability to detect subtle variations in lesion characteristics, which is crucial for accurate medical image segmentation.

Additionally, multi-task learning and deep supervision are recognized strategies for enhancing the representation power of models by training various tasks at low levels. These strategies are widely used in medical image segmentation, which demands dense predictions \cite{zhou2018unet++, fan2020pranet, fan2020inf, huang2020unet, cong2022boundary, nam2023m3fpolypsegnet}. However, when training multiple tasks with deep supervision, a challenge arises during upsampling of a low-resolution feature map to a high resolution, potentially leading to information loss. This effect is particularly pronounced when predicting detailed boundaries, negatively impacting model training. To address this challenge, we propose a multi-task learning scheme with deep supervision in a ensemble manner.
\section{Method}
\subsection{Multi-Frequency in Multi-Scale Attention Block}

\textit{Motivation:} The human visual system seamlessly integrates multi-scale and multi-frequency information to accurately interpret the environment \cite{nishida1997dual, mallat1989multifrequency}. Our approach mirrors a comprehensive analysis of visual information across diverse scales and frequencies designed to enhance tasks in the medical domain, motivated by Fig. \ref{fig:scale_frequency_distribution} and the operational algorithm of the human visual system. This process encompasses the wide variation in lesion  sizes in medical images, necessitating multi-scale features for precisely segmenting regions such as tumors, polyps, and cells. Moreover, as medical images exhibit higher frequency variance than scale due to modality characteristics, facilitating multi-frequency information is vital for crafting an effective  medical image segmentation model. Propelled by these insights, we propose the innovative \textit{Multi-Frequency in Multi-Scale Attention (MFMSA) block}, efficiently integrating multi-scale and multi-frequency information to effectively address a critical aspect often neglected in prior approaches. The overall architecture of the MADGNet and MFMSA block is illustrated on Fig. \ref{fig:MFMSNet}. And, the MFMSA block can be divided into three steps: \textit{1) Scale Decomposition}, \textit{2) MFCA}, and \textit{3) MSSA}.

\noindent \textbf{Feature Extraction.} We use a pre-trained ResNeSt \cite{zhang2022resnest} consisting of split-attention residual blocks to extract feature maps from input images due to enhanced feature representations. We also conducted experiments on various backbone (ResNet \cite{he2016deep}, Res2Net \cite{gao2019res2net}, Vision Transformer \cite{dosovitskiy2020image}), in Appendix \ref{appendix_ablation_on_backbone_model}. Let $\mathbf{f}_{i}$ be the feature maps from the  $i$-th stage encoder block for $i = 1, 2, 3, 4, 5$.  As the number of channels in each stage influences  a decoder's complexity, we reduce the number of channels into $C_{e}$ using 2D convolution. Furthermore, during decoding to restore the features to input image resolutions, we fuse two features from the encoder and previous decoder blocks as follows:

\begin{equation}
    \mathbf{X}_{i} = \textbf{Cat} (\textbf{Conv2D}_{1} (\mathbf{f}_{5 - i}), \textbf{Up}_{2} ( \mathbf{Y}_{i - 1} ) ),
\end{equation}

\noindent where $\textbf{Conv2D}_{k} ( \cdot ), \textbf{Cat} ( \cdot, \cdot )$ and $\textbf{Up}_{m} ( \cdot )$ denote the 2D convolution with a kernel size of $k$, concatenation between feature maps along the channel dimension and a upsampling with scale factor $2^{m - 1}$, respectively. And, $\mathbf{Y}_{0} = \mathbf{f}_{5}$. For convenience, we assume that $C_{i} = C_{e} + C_{i - 1} = C, H_{i} = H, W_{i} = W$. 

\noindent \textbf{Scale Decomposition.} To produce multi-scale features from input feature map $\mathbf{X}_{i}$, we assume a total of $S$ branches operating on $S$ different scales. For each $s$-th scale branch ($1 \le s \le S$), we reduce the resolution and channels of input feature map $\mathbf{X}_{i}$ with a channel and resolution reduction ratio $\gamma \in (0, 1)$, respectively, for computational efficiency. Thus, each input feature map at the $s$-th scale branch can be written as follows:

\begin{equation}
\mathbf{X}^{s}_{i} = \textbf{Conv2D}_{2s + 1} \left( \textbf{Down}_{s} \left( \mathbf{X}_{i} \right) \right) \in \mathbb{R}^{C_{s} \times H_{s} \times W_{s}},
\end{equation} 

\noindent where $\textbf{Down}_{s} (\cdot)$ is a downsampling with scale factor $2^{s - 1}$. Additionally , we denote the number of channels, height, and width at the $s$-th scale branch as $C_{s} = \text{max}(\frac{C}{\gamma^{s - 1}}, C_{\text{min}}), H_{s} = \text{max}(\frac{H}{2^{s - 1}}, H_{\text{min}})$, and $W_{s} = \text{max}(\frac{W}{2^{s - 1}}, W_{\text{min}})$, respectively. 

\noindent\textbf{Multi-Frequency Channel Attention (MFCA).} Recently, the compression ability of 2D DCT by expressing an image into a weighted sum of basis images produced by cosine functions oscillating at different frequencies has gained  attention for feature extraction in the frequency domain \cite{qin2021fcanet, gu2022fbi, sang2022multi}. Features at each scale branch can be characterized using 2D DCT with basis images $\mathbf{D}$ as follows:

\begin{equation}
    \mathbf{X}^{s, k}_{i} = \sum_{h = 0}^{H_{s} - 1} \sum_{w = 0}^{W_{s} - 1} \left( \mathbf{X}^{s}_{i} \right)_{:, h, w} \mathbf{D}^{u_{k}, v_{k}}_{h, w},
\end{equation}

\noindent where $(u_{k}, v_{k})$ is the frequency indices corresponding to $\mathbf{X}^{s, k}_{i}$. Moreover, 2D DCT basis images at the $s$-th scale branch are defined as $\mathbf{D}^{u_{k}, v_{k}}_{h, w} = \cos ( \frac{\pi h}{H_{s}} (u_{k} + \frac{1}{2}) ) \cos( \frac{\pi w}{W_{s}} (v_{k} + \frac{1}{2}) )$ with a top-$K$ selection strategy \cite{qin2021fcanet}. Subsequently, each $\mathbf{X}^{s, k}_{i}$ is compressed into $\mathbf{Z}_{\text{avg}}, \mathbf{Z}_{\text{max}}$, and $\mathbf{Z}_{\text{min}}$  using Global Average Pooling, Global Max Pooling, and Global Min Pooling  and aggregates each statistic of frequency to produce channel attention map at the $s$-th scale branch by using two fully-connected layers $\mathbf{W}_{1} \in \mathbb{R}^{C_{s} \times \frac{C_{s}}{r}}$ and $\mathbf{W}_{2}  \in \mathbb{R}^{\frac{C_{s}}{r} \times C_{s}}$ with reduction ratio $r$, as follows:

\begin{equation}
    \mathbf{M}^{s}_{i} = \sigma ( \sum_{d \in \{ \text{avg}, \text{max}, \text{min} \}} \mathbf{W}_{2} (\delta (\mathbf{W}_{1} \mathbf{Z}_{d})) ) \in \mathbb{R}^{C_{s}},
\end{equation}

\noindent where $\delta$ and $\sigma$ denote ReLU and Sigmoid activation functions, respectively. Finally, we recalibrate the feature map $\mathbf{X}^{s}_{i}$ using $\mathbf{M}^{s}_{i}$ at the $s$-th scale branch as $\hat{\mathbf{X}}^{s}_{i} = \mathbf{X}^{s}_{i} \times \mathbf{M}^{s}_{i}$.

\noindent\textbf{Multi-Scale Spatial Attention (MSSA).} The channel-recalibrated feature map $\hat{\mathbf{X}}^{s}_{i}$ is used to determine discriminative boundary cues with various scales in the spatial domain. At this stage, we introduce the two learnable parameters ($\alpha^{s}_{i}$ and $\beta^{s}_{i}$) for each scale branch to control the information flow between foreground and background, respectively, as follows:

\begin{equation}
    \overline{\mathbf{X}}^{s}_{i} = \textbf{Conv2D}_{3} \left( \alpha^{s}_{i} (\hat{\mathbf{X}}^{s}_{i} \times \mathbf{F}^{s}_{i}) + \beta^{s}_{i} (\hat{\mathbf{X}}^{s}_{i} \times \mathbf{B}^{s}_{i})  \right),
\end{equation}

\noindent where $\mathbf{F}^{s}_{i} = \sigma ( \textbf{Conv2D}_{1} (\hat{\mathbf{X}}^{s}_{i}) )$ is a foreground attention map. Accordingly, the background attention map can be derived via elementary-wise subtraction between $\mathbf{F}^{s}_{i}$ and a matrix filled with one, that is, $\mathbf{B}^{s}_{i} = 1 - \mathbf{F}^{s}_{i}$. At that point, we restore the number channel at the $s$-th scale branch $C_{s}$ in to $C$. Finally, to aggregate each refined feature from different scale branches, we apply upsampling to match the resolutions. For more stable training, we apply residual connection after the spatially refined feature map as follows: 

\begin{equation}
        \mathbf{Y}_{i} = \mathbf{X}_{i} + \mathbf{A}(\overline{\mathbf{X}}^{1}_{i}, \textbf{Up}_{2} ( \overline{\mathbf{X}}^{2}_{i} ) , \dots, \textbf{Up}_{S} ( \overline{\mathbf{X}}^{S}_{i} ) ),    
\end{equation}

\noindent where $\mathbf{A} ( \cdot )$ is the feature aggregation function. Subsequently, $\mathbf{Y}_{i}$ is concatenated with feature map from encoder and E-SDM to apply stable deep supervision. As the channel attention is initially applied in the MFCA, the channel information is enriched for representation, reducing the effect of noisy contained medical images. Subsequently, more discriminative feature maps for boundary cues robust to various scale and noise are produced using the MSSA. This dual process allows our model understand the subtle anatomical differences between various modalities and complex characteristics of irregular lesions in noisy medical images.

\noindent\textbf{Parameter Analysis.} To further analyze the MFMSA block, we compare its efficiency with the most simple variant of our block, a single-frequency in single-scale block. A theoretical proof of the number of parameters for each step is reported in Appendix \ref{appendix_parameter_efficiency_proof}. Consequently, the ratio of the number of parameters to the original scale branch at the $s$-th scale branch is $\frac{p_{s}}{p} = \gamma^{s - 1}$, where $p_{1} = p$. Based on this result, we conclude that the MFMSA block contains more $\sum_{s = 1}^{S} \frac{p_{s}}{p} = \frac{1 - \gamma^{S}}{1 - \gamma}$ than the single-frequency in single-scale block.

\begin{figure}[t]
    \centering
    \includegraphics[width=0.5\textwidth]{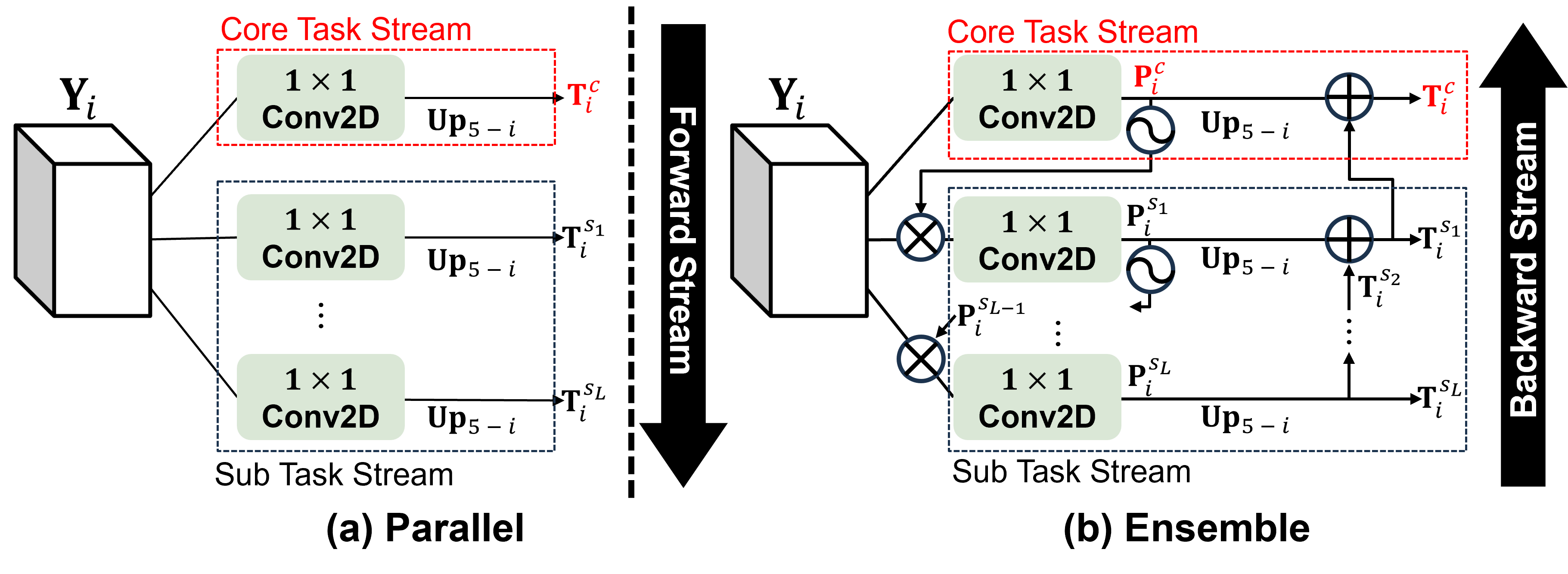}
    \caption{Comparison of multi-task learning method between (a) parallel and (b) ensemble manners.}
    \label{fig:parallel_vs_cascade}
\end{figure}

\subsection{Ensemble Multi-task Learning with Deep Supervision}
\textit{Motivation:} Multi-task learning with deep supervision is a well-known training strategy for enhancing the model representation power and preventing gradient vanishing problems; this is achieved through co-training with core task and other $L$ sub-tasks $\mathbf{O}_{i} = \{ \mathbf{T}^{c}_{i}, \mathbf{T}^{s_{1}}_{i}, \dots, \mathbf{T}^{s_{L}}_{i} \}$ at each $i$-th decoder stage \cite{fan2020pranet, zhao2023m, fan2020inf}. However, low-resolution feature maps must be upsampled into a high resolution to calculate the loss function between the ground truth and prediction. This drastic upsampling interferes with the model’s representation ability due to information loss for predicting detailed boundaries and structures.

To solve this problem, we propose a \textit{ensemble sub-decoding module (E-SDM)}, a novel training strategy for multi-task learning with deep supervision. Fig. \ref{fig:parallel_vs_cascade} illustrates the difference between (a) parallel and (b) ensemble manners. The main idea is to cascadingly supplement the information loss by incorporating sub-task predictions after upsampling, thereby improving the core task prediction.

\noindent\textbf{Forward Stream.} During the forward stream, core and sub-task pseudo predictions $\{ \mathbf{P}^{c}_{i}, \mathbf{P}^{s_{1}}_{i}, \dots, \mathbf{P}^{s_{L}}_{i} \}$ are produced at the $i$-th decoder stage as follows:

\begin{equation}
    \begin{cases}
        &\mathbf{P}^{c}_{i} = \textbf{Conv2D}_{1} (\mathbf{Y}_{i}) \\ 
        &\mathbf{P}^{s_{l}}_{i} = \textbf{Conv2D}_{1} (\mathbf{Y}_{i} \times \sigma ( \mathbf{P}_{i}^{s_{l - 1}} )) \text{ for } 1 \le l \le L
    \end{cases}
\end{equation}

\noindent where $\mathbf{P}_{i}^{s_{0}} = \mathbf{P}_{i}^{c}$. This stream ensures that the following sub-task prediction focuses on the region through a spatial attention mechanism, starting from the core pseudo prediction $\mathbf{P}^{c}_{i}$. 

\noindent\textbf{Backward Stream.} After producing the $L$-th sub-task pseudo prediction $\mathbf{P}^{s_{L}}_{i}$, to obtain the final core task prediction $\mathbf{T}^{c}_{i}$, we apply the backward stream as follows: 

\begin{equation}
\label{eq:backward_stream}
    \begin{cases}
        &\mathbf{T}^{s_{L}}_{i} = \textbf{Up}_{5 - i} \left( \mathbf{P}^{s_{L}}_{i} \right) \\ 
        &\mathbf{T}^{s_{l}}_{i} = \textbf{Up}_{5 - i} \left( \mathbf{P}^{s_{l}}_{i} \right) + \mathbf{T}^{s_{l+1}}_{i}  \text{ for } 1 \le l \le L - 1
    \end{cases},
\end{equation}

\noindent where $\mathbf{T}^{s_{0}}_{i} = \mathbf{T}^{c}_{i}$. To further analyze the Eq \ref{eq:backward_stream}, we can recursively rewrite from core task $\mathbf{T}^{c}_{i}$ as follows:

\begin{equation}
\label{eq:backward_stream_recursive}
    \begin{split}
        \mathbf{T}^{c}_{i} &= \mathbf{T}^{s_{0}}_{i} = \textbf{Up}_{5 - i} \left( \mathbf{P}^{s_{0}}_{i} \right) + \mathbf{T}^{s_{1}}_{i} \\ 
                           &= \left[ \textbf{Up}_{5 - i} \left( \mathbf{P}^{s_{0}}_{i} \right) + \textbf{Up}_{5 - i} \left( \mathbf{P}^{s_{1}}_{i} \right) \right] + \mathbf{T}^{s_{2}}_{i} = \cdots \\
                           &= \sum_{l = 0}^{L} \textbf{Up}_{5 - i} \left( \mathbf{P}^{s_{l}}_{i} \right)
    \end{split}
\end{equation}

\noindent Consequently, E-SDM can be interpreted as an ensemble of predictions between different tasks for describing the same legion. Algorithm \ref{alg_CSD} describes the detailed training algorithm for E-SDM.

\begin{algorithm}[t]
\caption{Ensemble Sub-Decoding Module for Multi-task Learning with Deep Supervision}
\label{alg_CSD}
\textbf{Input}: Refined feature map $\mathbf{Y}_{i}$ from $i$-th MFMSA block \\
\textbf{Output}: Core task prediction $\textbf{T}^{c}_{i}$ and sub-task predictions $\{ \mathbf{T}^{s_{1}}_{i}, \dots, \mathbf{T}^{s_{L}}_{i} \}$ at $i$-th decoder
\begin{algorithmic}[1] 
\STATE $\mathbf{P}_{i}^{c} = \textbf{Conv2D}_{1} (\mathbf{Y}_{i}) $ 
\FOR{$l = 1, 2, \dots, L$ }
    \STATE $\mathbf{P}_{i}^{s_{l}} = \textbf{Conv2D}_{1} (\mathbf{Y}_{i} \times \sigma \left( \mathbf{P}_{i}^{s_{l - 1}}\right) ) $.
\ENDFOR
\STATE $\mathbf{T}_{i}^{s_{L}} = \textbf{Up}_{5 - i} \left( \mathbf{P}_{i}^{s_{L}} \right) $ 
\FOR{$l = L - 1, \dots, 0$}
    \STATE $\mathbf{T}_{i}^{s_{l}} = \textbf{Up}_{5 - i} \left( \mathbf{P}_{i}^{s_{l}} \right) + \mathbf{T}_{i}^{s_{l + 1}}$
\ENDFOR
\STATE \textbf{return} $\textbf{O}_{i} = \{ \textbf{T}^{c}_{i}, \mathbf{T}^{s_{1}}_{i}, \dots, \mathbf{T}^{s_{L}}_{i} \}$
\end{algorithmic}
\end{algorithm} 

\begin{table*}
    \centering
    \scriptsize
    \setlength\tabcolsep{5pt} 
    \begin{tabular}{c|cc|cc|cc|cc|cc|cc|c}
    \hline
    \multicolumn{1}{c|}{\multirow{3}{*}{Method}} & \multicolumn{2}{c|}{Dermoscopy} & \multicolumn{2}{c|}{Radiology}  & \multicolumn{2}{c|}{Ultrasound} & \multicolumn{2}{c|}{Microscopy}  & \multicolumn{4}{c|}{Colonoscopy} & \multicolumn{1}{c}{\multirow{3}{*}{$P$-value}} \\ \cline{2-13}
     & \multicolumn{2}{c|}{ISIC2018 \cite{gutman2016skin}} & \multicolumn{2}{c|}{COVID19-1 \cite{ma_jun_2020_3757476}} & \multicolumn{2}{c|}{BUSI \cite{al2020dataset}} & \multicolumn{2}{c|}{DSB2018 \cite{caicedo2019nucleus}} & \multicolumn{2}{c}{CVC-ClinicDB \cite{bernal2015wm}} & \multicolumn{2}{c|}{Kvasir \cite{jha2020kvasir}} &  \\ \cline{2-13}
     & DSC & mIoU & DSC & mIoU & DSC & mIoU & DSC & mIoU & DSC & mIoU & DSC & mIoU & \\
     \hline
     UNet \cite{ronneberger2015u}               & 87.3 \tiny{(0.8)} & 80.2 \tiny{(0.7)} & 47.7 \tiny{(0.6)} & 38.6 \tiny{(0.6)} & 69.5 \tiny{(0.3)} & 60.2 \tiny{(0.2)} & 91.1 \tiny{(0.2)} & 84.3 \tiny{(0.3)} & 76.5 \tiny{(0.8)} & 69.1 \tiny{(0.9)} & 80.5 \tiny{(0.3)} & 72.6 \tiny{(0.4)} & 5.2E-06 \\
     AttUNet \cite{oktay1804attention}          & 87.8 \tiny{(0.1)} & 80.5 \tiny{(0.1)} & 57.5 \tiny{(0.2)} & 48.4 \tiny{(0.2)} & 71.3 \tiny{(0.4)} & 62.3 \tiny{(0.6)} & 91.6 \tiny{(0.1)} & 85.0 \tiny{(0.1)} & 80.1 \tiny{(0.6)} & 74.2 \tiny{(0.5)} & 83.9 \tiny{(0.1)} & 77.1 \tiny{(0.1)} & 4.1E-06 \\
     UNet++ \cite{zhou2018unet++}               & 87.3 \tiny{(0.2)} & 80.2 \tiny{(0.1)} & 65.6 \tiny{(0.7)} & 57.1 \tiny{(0.8)} & 72.4 \tiny{(0.1)} & 62.5 \tiny{(0.2)} & 91.6 \tiny{(0.1)} & 85.0 \tiny{(0.1)} & 79.7 \tiny{(0.2)} & 73.6 \tiny{(0.4)} & 84.3 \tiny{(0.3)} & 77.4 \tiny{(0.2)} & 7.5E-07 \\
     CENet \cite{gu2019net}                     & 89.1 \tiny{(0.2)} & 82.1 \tiny{(0.1)} & 76.3 \tiny{(0.4)} & 69.2 \tiny{(0.5)} & 79.7 \tiny{(0.6)} & 71.5 \tiny{(0.5)} & 91.3 \tiny{(0.1)} & 84.6 \tiny{(0.1)} & 89.3 \tiny{(0.3)} & 83.4 \tiny{(0.2)} & 89.5 \tiny{(0.7)} & 83.9 \tiny{(0.7)} & 1.0E-05 \\
     TransUNet \cite{chen2021transunet}         & 87.3 \tiny{(0.2)} & 81.2 \tiny{(0.8)} & 75.6 \tiny{(0.4)} & 68.8 \tiny{(0.2)} & 75.5 \tiny{(0.5)} & 68.4 \tiny{(0.1)} & 91.8 \tiny{(0.3)} & 85.2 \tiny{(0.2)} & 87.4 \tiny{(0.2)} & 82.9 \tiny{(0.1)} & 86.4 \tiny{(0.4)} & 81.3 \tiny{(0.4)} & 9.9E-08 \\
     FRCUNet \cite{azad2021deep}                & 88.9 \tiny{(0.1)} & 83.1 \tiny{(0.2)} & 77.3 \tiny{(0.3)} & 70.4 \tiny{(0.2)} & \textcolor{blue}{\textbf{\textit{81.2}}} \tiny{(0.2)} & \textcolor{blue}{\textbf{\textit{73.3}}} \tiny{(0.3)} & 90.8 \tiny{(0.3)} & 83.8 \tiny{(0.4)} & 91.8 \tiny{(0.2)} & 87.0 \tiny{(0.2)} & 88.8 \tiny{(0.4)} & 83.5 \tiny{(0.6)} & 6.6E-02 \\
     MSRFNet \cite{srivastava2021msrf}          & 88.2 \tiny{(0.2)} & 81.3 \tiny{(0.2)} & 75.2 \tiny{(0.4)} & 68.0 \tiny{(0.4)} & 76.6 \tiny{(0.7)} & 68.1 \tiny{(0.7)} & \textcolor{blue}{\textbf{\textit{91.9}}} \tiny{(0.1)} & \textcolor{blue}{\textbf{\textit{85.3}}} \tiny{(0.1)} & 83.2 \tiny{(0.9)} & 76.5 \tiny{(1.1)} & 86.1 \tiny{(0.5)} & 79.3 \tiny{(0.4)} & 8.8E-07 \\
     HiFormer \cite{heidari2023hiformer}        & 88.7 \tiny{(0.5)} & 81.9 \tiny{(0.5)} & 72.9 \tiny{(1.4)} & 63.3 \tiny{(1.5)} & 79.3 \tiny{(0.2)} & 70.8 \tiny{(0.1)} & 90.7 \tiny{(0.2)} & 83.8 \tiny{(0.4)} & 89.1 \tiny{(0.6)} & 83.7 \tiny{(0.6)} & 88.1 \tiny{(1.0)} & 82.3 \tiny{(1.2)} & 1.8E-05 \\
     DCSAUNet \cite{xu2023dcsau}                & 89.0 \tiny{(0.3)} & 82.0 \tiny{(0.3)} & 75.3 \tiny{(0.4)} & 68.2 \tiny{(0.4)} & 73.7 \tiny{(0.5)} & 65.0 \tiny{(0.5)} & 91.1 \tiny{(0.2)} & 84.4 \tiny{(0.2)} & 80.5 \tiny{(1.2)} & 73.7 \tiny{(1.1)} & 82.6 \tiny{(0.5)} & 75.2 \tiny{(0.5)} & 6.2E-07 \\
     M2SNet \cite{zhao2023m}                    & \textcolor{blue}{\textbf{\textit{89.2}}} \tiny{(0.2)} & \textcolor{blue}{\textbf{\textit{83.4}}} \tiny{(0.2)} & \textcolor{blue}{\textbf{\textit{81.7}}} \tiny{(0.4)} & \textcolor{blue}{\textbf{\textit{74.7}}} \tiny{(0.5)} & 80.4 \tiny{(0.8)} & 72.5 \tiny{(0.7)} & 91.6 \tiny{(0.2)} & 85.1 \tiny{(0.3)} & \textcolor{blue}{\textbf{\textit{92.8}}} \tiny{(0.8)} & \textcolor{blue}{\textbf{\textit{88.2}}} \tiny{(0.8)} & \textcolor{blue}{\textbf{\textit{90.2}}} \tiny{(0.5)} & \textcolor{blue}{\textbf{\textit{85.1}}} \tiny{(0.6)} & 2.0E-05 \\
    \hline
    SFSSNet                                    & 88.8 \tiny{(0.3)} & 81.9 \tiny{(0.2)} & 80.3 \tiny{(0.8)} & 73.0 \tiny{(0.7)} & 66.1 \tiny{(0.6)} & 59.3 \tiny{(0.8)} & 91.5 \tiny{(0.2)} & 84.0 \tiny{(0.2)} & 90.7 \tiny{(0.4)} & 83.0 \tiny{(0.7)} & 88.1 \tiny{(0.6)} & 82.2 \tiny{(0.7)} & 2.2E-06   \\
    MFSSNet                                    & 88.5 \tiny{(0.2)} & 81.8 \tiny{(0.2)} & 80.4 \tiny{(0.7)} & 73.1 \tiny{(0.4)} & 81.0 \tiny{(0.1)} & 73.2 \tiny{(0.2)} & 91.6 \tiny{(0.1)} & 85.1 \tiny{(0.2)} & 92.3 \tiny{(0.5)} & 87.7 \tiny{(0.5)} & 89.9 \tiny{(0.6)} & 84.7 \tiny{(0.7)} & 5.1E-07   \\
    SFMSNet                                    & 89.2 \tiny{(0.3)} & 82.5 \tiny{(0.3)} & 81.4 \tiny{(0.3)} & 74.5 \tiny{(0.3)} & 80.8 \tiny{(0.4)} & 73.0 \tiny{(0.3)} & 91.5 \tiny{(0.2)} & 84.9 \tiny{(0.4)} & 92.3 \tiny{(0.3)} & 88.0 \tiny{(0.3)} & 89.0 \tiny{(0.6)} & 84.1 \tiny{(0.5)} & 1.4E-04   \\
    \hline
    \textbf{MADGNet}                           & \textcolor{red}{\textbf{\underline{90.2}}} \tiny{(0.1)} & \textcolor{red}{\textbf{\underline{83.7}}} \tiny{(0.2)} & \textcolor{red}{\textbf{\underline{83.7}}} \tiny{(0.2)} & \textcolor{red}{\textbf{\underline{76.8}}} \tiny{(0.2)} & \textcolor{red}{\textbf{\underline{81.3}}} \tiny{(0.4)} & \textcolor{red}{\textbf{\underline{73.4}}} \tiny{(0.5)} & \textcolor{red}{\textbf{\underline{92.0}}} \tiny{(0.0)} & \textcolor{red}{\textbf{\underline{85.5}}} \tiny{(0.1)} & \textcolor{red}{\textbf{\underline{93.9}}} \tiny{(0.6)} & \textcolor{red}{\textbf{\underline{89.5}}} \tiny{(0.5)} & \textcolor{red}{\textbf{\underline{90.7}}} \tiny{(0.8)} & \textcolor{red}{\textbf{\underline{85.3}}} \tiny{(0.8)} & \multicolumn{1}{c}{-} \\
    \hline
    \end{tabular}
    \caption{Segmentation results on five different modalities with \textit{seen} clinical settings. We also provide one tailed \textit{t}-Test results (\textit{P}-value) compared to our method and other methods. $( \cdot )$ denotes the standard deviations of multiple experiment results.}
    \label{tab:comparison_sota_in_domain}
\end{table*}

\begin{table*}
    \centering
    \scriptsize
    \setlength\tabcolsep{2.5pt} 
    \begin{tabular}{c|cc|cc|cc|cc|cc|cc|cc|c}
    \hline
    \multicolumn{1}{c|}{\multirow{3}{*}{Method}} & \multicolumn{2}{c|}{Dermoscopy} & \multicolumn{2}{c|}{Radiology}  & \multicolumn{2}{c|}{Ultrasound} & \multicolumn{2}{c|}{Microscopy}  & \multicolumn{6}{c|}{Colonoscopy} & \multicolumn{1}{c}{\multirow{3}{*}{$P$-value}} \\ \cline{2-15} 
     & \multicolumn{2}{c|}{PH2 \cite{mendoncca2013ph}} & \multicolumn{2}{c|}{COVID19-2 \cite{COVID19_2}} & \multicolumn{2}{c|}{STU \cite{zhuang2019rdau}} & \multicolumn{2}{c|}{MonuSeg2018 \cite{dinh2021breast}} & \multicolumn{2}{c}{CVC-300 \cite{vazquez2017benchmark}} & \multicolumn{2}{c}{CVC-ColonDB \cite{tajbakhsh2015automated}}  & \multicolumn{2}{c|}{ETIS \cite{silva2014toward}} &  \\ \cline{2-15}
     & DSC & mIoU & DSC & mIoU & DSC & mIoU & DSC & mIoU & DSC & mIoU & DSC & mIoU & DSC & mIoU & \\
     \hline
     UNet \cite{ronneberger2015u}              & 90.3 \tiny{(0.1)} & \textcolor{blue}{\textbf{\textit{83.5}}} \tiny{(0.1)} & 47.1 \tiny{(0.7)} & 37.7 \tiny{(0.6)} & 71.6 \tiny{(1.0)} & 61.6 \tiny{(0.7)} & 29.2 \tiny{(5.1)} & 18.9 \tiny{(3.5)} & 66.1 \tiny{(2.3)} & 58.5 \tiny{(2.1)} & 56.8 \tiny{(1.3)} & 49.0 \tiny{(1.2)} & 41.6 \tiny{(1.1)} & 35.4 \tiny{(1.0)} & 1.1E-09 \\
     AttUNet \cite{oktay1804attention}         & 89.9 \tiny{(0.2)} & 82.6 \tiny{(0.3)} & 43.7 \tiny{(0.8)} & 35.2 \tiny{(0.8)} & 77.0 \tiny{(1.6)} & 68.0 \tiny{(1.7)} & \textcolor{blue}{\textbf{\textit{39.0}}} \tiny{(3.1)} & \textcolor{blue}{\textbf{\textit{26.5}}} \tiny{(2.4)} & 63.0 \tiny{(0.3)} & 57.2 \tiny{(0.4)} & 56.8 \tiny{(1.6)} & 50.0 \tiny{(1.5)} & 38.4 \tiny{(0.3)} & 33.5 \tiny{(0.1)} & 6.7E-09	\\
     UNet++ \cite{zhou2018unet++}              & 88.0 \tiny{(0.3)} & 80.1 \tiny{(0.3)} & 50.5 \tiny{(3.8)} & 40.9 \tiny{(3.7)} & 77.3 \tiny{(0.4)} & 67.8 \tiny{(0.3)} & 25.4 \tiny{(0.8)} & 15.3 \tiny{(0.5)} & 64.3 \tiny{(2.2)} & 58.4 \tiny{(2.0)} & 57.5 \tiny{(0.4)} & 50.2 \tiny{(0.4)} & 39.1 \tiny{(2.4)} & 34.0 \tiny{(2.1)} & 1.0E-05	\\
     CENet \cite{gu2019net}                    & 90.5 \tiny{(0.1)} & 83.3 \tiny{(0.1)} & 60.1 \tiny{(0.3)} & 49.9 \tiny{(0.3)} & 86.0 \tiny{(0.7)} & \textcolor{blue}{\textbf{\textit{77.2}}} \tiny{(0.9)} & 27.7 \tiny{(1.5)} & 16.9 \tiny{(1.0)} & 85.4 \tiny{(1.6)} & 78.2 \tiny{(1.4)} & 65.9 \tiny{(1.6)} & 59.2 \tiny{(0.1)} & 57.0 \tiny{(3.4)} & 51.4 \tiny{(0.5)} & 4.5E-06 \\
     TransUNet \cite{chen2021transunet}        & 89.5 \tiny{(0.3)} & 82.1 \tiny{(0.4)} & 56.9 \tiny{(1.0)} & 48.0 \tiny{(0.7)} & 41.4 \tiny{(9.5)} & 32.1 \tiny{(4.2)} & 15.9 \tiny{(8.5)} & 9.6 \tiny{(5.5)} & 85.0 \tiny{(0.6)} & 77.3 \tiny{(0.3)} & 63.7 \tiny{(0.1)} & 58.4 \tiny{(0.3)} & 50.1 \tiny{(0.5)} & 44.0 \tiny{(2.3)} & 1.6E-06 \\
     FRCUNet \cite{azad2021deep}               & 90.6 \tiny{(0.1)} & 83.4 \tiny{(0.2)} & 62.9 \tiny{(1.1)} & 52.7 \tiny{(0.9)} & \textcolor{blue}{\textbf{\textit{86.5}}} \tiny{(2.3)} & \textcolor{blue}{\textbf{\textit{77.2}}} \tiny{(2.7)} & 26.1 \tiny{(5.6)} & 16.8 \tiny{(4.3)} & 86.7 \tiny{(0.7)} & 79.4 \tiny{(0.3)} & 69.1 \tiny{(1.0)} & 62.6 \tiny{(0.9)} & 65.1 \tiny{(1.0)} & 58.4 \tiny{(0.5)} & 2.3E-05\\
     MSRFNet \cite{srivastava2021msrf}         & 90.5 \tiny{(0.3)} & \textcolor{blue}{\textbf{\textit{83.5}}} \tiny{(0.3)} & 58.3 \tiny{(0.8)} & 48.4 \tiny{(0.6)} & 84.0 \tiny{(5.5)} & 75.2 \tiny{(8.2)} & 9.1 \tiny{(1.0)} & 5.3 \tiny{(0.7)} & 72.3 \tiny{(2.2)} & 65.4 \tiny{(2.2)} & 61.5 \tiny{(1.0)} & 54.8 \tiny{(0.8)} & 38.3 \tiny{(0.6)} & 33.7 \tiny{(0.7)} & 1.0E-07 \\
     HiFormer \cite{heidari2023hiformer}       & 86.9 \tiny{(1.6)} & 79.1 \tiny{(1.8)} & 54.1 \tiny{(1.0)} & 44.5 \tiny{(0.8)} & 80.7 \tiny{(2.9)} & 71.3 \tiny{(3.2)} & 21.9 \tiny{(8.9)} & 13.2 \tiny{(5.7)} & 84.7 \tiny{(1.1)} & 77.5 \tiny{(1.1)} & 67.6 \tiny{(1.4)} & 60.5 \tiny{(1.3)} & 56.7 \tiny{(3.2)} & 50.1 \tiny{(3.3)} & 2.5E-07 \\
     DCSAUNet \cite{xu2023dcsau}               & 89.0 \tiny{(0.4)} & 81.5 \tiny{(0.3)} & 52.4 \tiny{(1.2)} & 44.0 \tiny{(0.7)} & 86.1 \tiny{(0.5)} & 76.5 \tiny{(0.8)} & 4.3 \tiny{(0.3)} & 2.4 \tiny{(0.9)} & 68.9 \tiny{(4.0)} & 59.8 \tiny{(3.9)} & 57.8 \tiny{(0.4)} & 49.3 \tiny{(0.4)} & 42.9 \tiny{(3.0)} & 36.1 \tiny{(2.9)} & 1.3E-07	\\
     M2SNet \cite{zhao2023m}                   & \textcolor{blue}{\textbf{\textit{90.7}}} \tiny{(0.3)} & \textcolor{blue}{\textbf{\textit{83.5}}} \tiny{(0.5)} & \textcolor{blue}{\textbf{\textit{68.6}}} \tiny{(0.1)} & \textcolor{blue}{\textbf{\textit{58.9}}} \tiny{(0.2)} & 79.4 \tiny{(0.7)} & 69.3 \tiny{(0.6)} & 36.3 \tiny{(0.9)} & 23.1 \tiny{(0.8)} & \textcolor{red}{\textbf{\underline{89.9}}} \tiny{(0.2)} & \textcolor{red}{\textbf{\underline{83.2}}} \tiny{(0.3)} & \textcolor{blue}{\textbf{\textit{75.8}}} \tiny{(0.7)} & \textcolor{blue}{\textbf{\textit{68.5}}} \tiny{(0.5)} & \textcolor{blue}{\textbf{\textit{74.9}}} \tiny{(1.3)} & \textcolor{blue}{\textbf{\textit{67.8}}} \tiny{(1.4)} & 4.9E-02	\\
    \hline
    SFSSNet                                   & 89.8 \tiny{(0.2)} & 82.2 \tiny{(0.4)} & 65.1 \tiny{(1.6)} & 55.5 \tiny{(1.3)} & 59.1 \tiny{(0.3)} & 49.3 \tiny{(0.7)} & 21.5 \tiny{(7.2)} & 14.3 \tiny{(5.0)} & 81.7 \tiny{(0.3)} & 74.7 \tiny{(0.4)} & 65.6 \tiny{(0.4)} & 58.4 \tiny{(0.5)} & 56.4 \tiny{(0.7)} & 49.4 \tiny{(0.4)} & 2.0E-07 \\
    MFSSNet                                   & 90.2 \tiny{(0.8)} & 83.3 \tiny{(0.9)} & 67.6 \tiny{(0.5)} & 57.9 \tiny{(0.3)} & 66.1 \tiny{(0.8)} & 59.3 \tiny{(0.2)} & 30.1 \tiny{(7.5)} & 20.5 \tiny{(5.5)} & 83.3 \tiny{(1.4)} & 76.1 \tiny{(1.2)} & 66.0 \tiny{(0.7)} & 59.1 \tiny{(0.8)} & 59.3 \tiny{(0.2)} & 52.6 \tiny{(0.6)} & 3.9E-04 \\
    SFMSNet                                   & 90.8 \tiny{(0.3)} & 83.9 \tiny{(0.5)} & 67.7 \tiny{(1.1)} & 58.0 \tiny{(1.3)} & 84.5 \tiny{(0.2)} & 74.3 \tiny{(0.1)} & 28.1 \tiny{(9.9)} & 18.2 \tiny{(7.1)} & 84.2 \tiny{(1.2)} & 78.1 \tiny{(1.0)} & 75.9 \tiny{(0.8)} & 68.3 \tiny{(0.8)} & 68.9 \tiny{(0.3)} & 62.7 \tiny{(0.4)} & 7.9E-03 \\
    \hline
    \textbf{MADGNet}                          & \textcolor{red}{\textbf{\underline{91.3}}} \tiny{(0.1)} & \textcolor{red}{\textbf{\underline{84.6}}} \tiny{(0.1)} & \textcolor{red}{\textbf{\underline{72.2}}} \tiny{(0.3)} & \textcolor{red}{\textbf{\underline{62.6}}} \tiny{(0.3)} & \textcolor{red}{\textbf{\underline{88.4}}} \tiny{(1.0)} & \textcolor{red}{\textbf{\underline{79.9}}} \tiny{(1.5)} & \textcolor{red}{\textbf{\underline{46.7}}} \tiny{(4.3)} & \textcolor{red}{\textbf{\underline{32.0}}} \tiny{(2.9)} & \textcolor{blue}{\textbf{\textit{87.4}}} \tiny{(0.4)} & \textcolor{blue}{\textbf{\textit{79.9}}} \tiny{(0.4)} & \textcolor{red}{\textbf{\underline{77.5}}} \tiny{(1.1)} & \textcolor{red}{\textbf{\underline{69.7}}} \tiny{(1.2)} & \textcolor{red}{\textbf{\underline{77.0}}} \tiny{(0.3)} & \textcolor{red}{\textbf{\underline{69.7}}} \tiny{(0.5)} & - \\ \cline{1-7}
    \hline
    \end{tabular}
    \caption{Segmentation results on five different modalities with \textit{unseen} clinical settings. We also provide one tailed \textit{t}-Test results (\textit{P}-value) compared to our method and other methods. $( \cdot )$ denotes the standard deviations of multiple experiment results.}
    \label{tab:comparison_sota_out_domain}
\end{table*}

\noindent\textbf{Loss Function.} The loss function for multi-task learning with deep supervision in a ensemble manner is the same as that in a parallel manner, as follows: 

\begin{equation}
    \mathcal{L}_{total} = \sum_{i = 1}^{4}  \sum_{t \in \{ c, s_{1}, \dots, s_{L} \}}\lambda_{t} \mathcal{L}_{t} \left( \textbf{G}^{t}, \textbf{Up}_{5 - i} (\textbf{T}^{t}_{i}) \right)
\end{equation}

\noindent where $\textbf{G}^{t}$ and $\textbf{T}^{t}_{i}$ are the ground truth and predictions for task $t$ and $i$-th decoder, respectively. Additionally, $\mathcal{L}_{t}$ and $\lambda_{t}$ are the loss function and ratio for the task $t$, respectively.

For multi-task learning with deep supervision, we defined a core task as region $R$ and two sub-tasks as boundary $B$ and distance map $D$. The loss function for region prediction is defined as $\mathcal{L}_{R} = \mathcal{L}^{w}_{IoU} + \mathcal{L}^{w}_{bce}$, where $\mathcal{L}^{w}_{IoU}$ and $\mathcal{L}^{w}_{bce}$ are the weighted IoU and bce  loss functions, respectively. Additionally, we defined boundary and distance map loss function $\mathcal{L}_{B}$ and $\mathcal{L}_{D}$ as bce and mse loss, respectively. 
\section{Experiment Results}

\subsection{Experiment Settings}

In this study, we evaluated the models on seven medical image segmentation datasets with different modalities, including Dermoscopy \cite{gutman2016skin}, Radiology \cite{ma_jun_2020_3757476}, Ultrasound \cite{al2020dataset}, Microscopy \cite{caicedo2019nucleus}, Colonoscopy \cite{bernal2015wm, jha2020kvasir}, and Fundus Imaging \cite{orlando2020refuge}, to validate the modality-agnostic ability. Moreover, we evaluated the domain generalizability on eight external datasets with different modalities, including Dermoscopy \cite{mendoncca2013ph}, Radiology \cite{COVID19_2}, Ultrasound \cite{zhuang2019rdau}, Microscopy \cite{dinh2021breast}, Colonoscopy \cite{vazquez2017benchmark, tajbakhsh2015automated, silva2014toward}, and Fundus Imaging \cite{sivaswamy2015comprehensive}. To evaluate the performance of each model, we selected two metrics, DSC and mIoU, which are widely used in medical image segmentation. We compared the \textbf{MADGNet (Ours)} with ten medical image segmentation models, including UNet \cite{ronneberger2015u}, Attention UNet (AttUNet \cite{oktay1804attention}), UNet++ \cite{zhou2018unet++}, CENet \cite{gu2019net}, TransUNet \cite{chen2021transunet}, FRCUNet \cite{azad2021deep}, MSRFNet \cite{srivastava2021msrf}, HiFormer \cite{heidari2023hiformer}, DCSAUNet \cite{xu2023dcsau}, and M2SNet \cite{zhao2023m}. We report the \textit{mean} performance of three trials for all results. \textcolor{red}{\textbf{\underline{Red}}} and \textcolor{blue}{\textbf{\textit{Blue}}} are the first and second best performance results, respectively. Due to the page limit, we present the detailed dataset description and results using other metrics \cite{margolin2014evaluate, fan2017structure, fan2018enhanced} in Appendix \ref{appendix_dataset_descriptions} and \ref{appendix_additional_experiment_results_with_various_metrics}.

\subsection{Implementation Details}
\label{sec42_implementation_details}

\noindent\textbf{Training Settings.} We implemented MADGNet on a single NVIDIA RTX 3090 Ti in Pytorch  1.11 \cite{NEURIPS2019_9015}. We started with an initial learning rate of $10^{-4}$ using the Adam optimizer and reduced the parameters of each model to $10^{-6}$ using a cosine annealing learning rate scheduler \cite{loshchilov2016sgdr}. We optimized each model with a batch size of 16 and epochs of 50, 100, 100, 100, and 200 for Colonoscopy, Dermoscopy, Microscopy, Ultrasound, and Radiology modalities, respectively. During training, we used horizontal/vertical flipping, with a probability of 50\%, and rotation between $-5^{\circ}$ and $5^{\circ}$ based on the multi-scale training strategy, which is widely used in medical image segmentation \cite{fan2020pranet, zhao2021automatic, zhao2023m}. At this stage, because images in each dataset have different resolutions, all images were resized to $352 \times 352$.

\noindent\textbf{Hyperparameters of MADGNet.} Key hyperparameters for MADGNet on all datasets were set to $C_{e} = 64, C_{\text{min}} = 32, H_{\text{min}} = W_{\text{min}} = 8, \gamma = \frac{1}{2}, r = 16$ for efficiency and $S = 3$ and $K = 16$ for combining multi-scale and multi-frequency features. The aggregation function $\mathbf{A} ( \cdot )$ averaged refined feature maps from each scale branch. To reduce the sensitivity of hyperparameters, the loss ratio between tasks was fixed as $\lambda_{t} = 1$ for all tasks $t \in \{ R, D, B \}$. In this paper, the core task indicates the region prediction, while sub-tasks indicate the distance map and boundary predictions. In Appendix \ref{appendix_hyperparameter_on_MADGNet}, we provide the experiment results on various hyperparameter settings. 

\subsection{Comparison with State-of-the-art models}
\label{sec43_compare_with_sota}

\begin{figure}[t]
    \centering
    \includegraphics[width=0.5\textwidth]{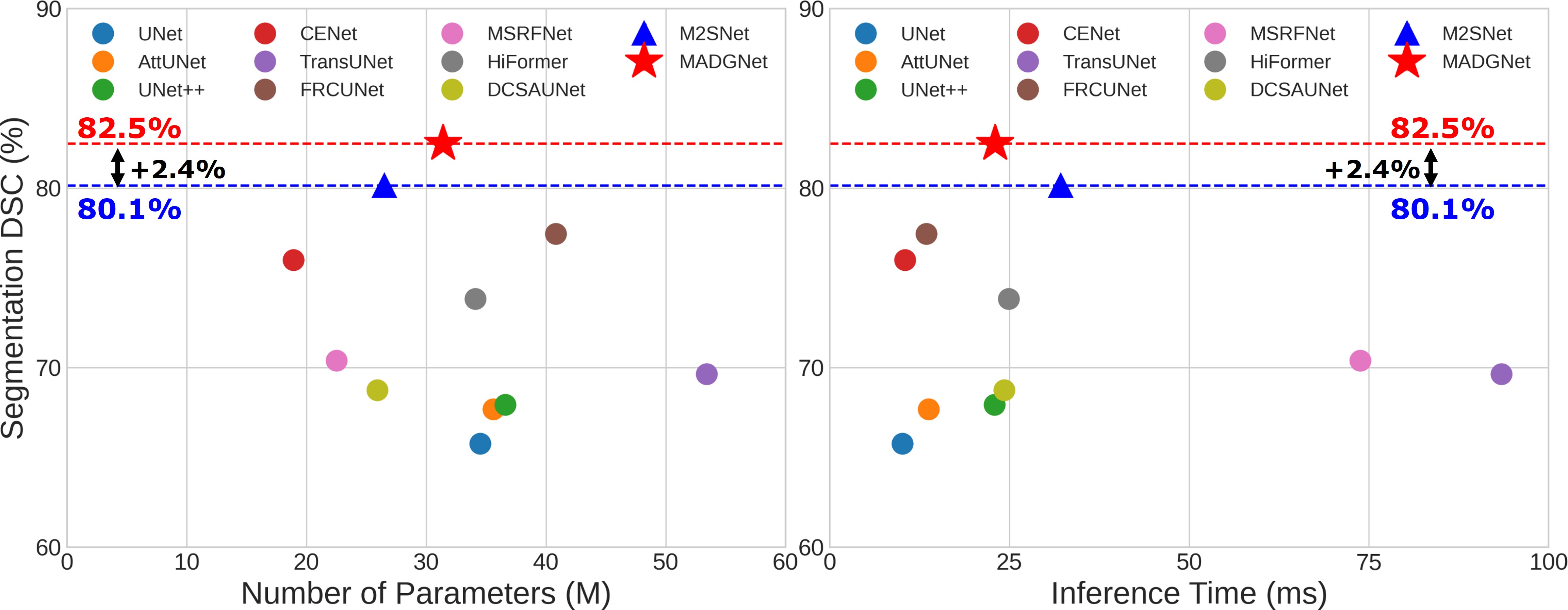}
    \caption{Comparison of parameters (M) and inference speed (ms) vs segmentation performance (DSC) on average for all datasets.}
    \label{fig:EfficiencyAnalysis}
\end{figure}

\begin{figure*}[t]
    \centering
    \includegraphics[width=\textwidth]{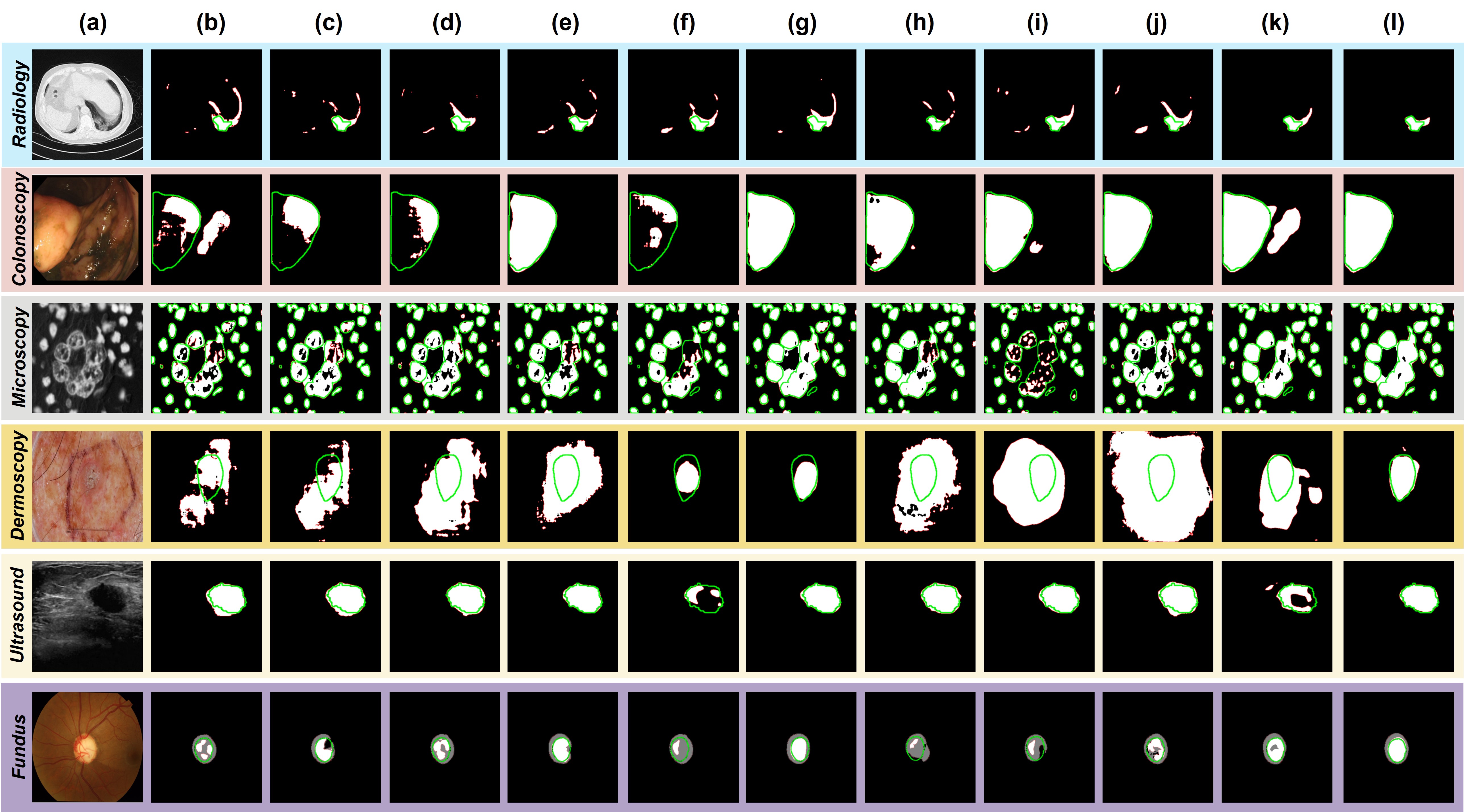}
    \caption{Qualitative comparison of other methods and MADGNet.  (a) Input images. (b) UNet \cite{ronneberger2015u}. (c) AttUNet \cite{oktay1804attention}. (d) UNet++ \cite{zhou2018unet++}. (e) CENet \cite{gu2019net}. (f) TransUNet \cite{chen2021transunet}. (g) FRCUNet \cite{azad2021deep}, (h) MSRFNet \cite{srivastava2021msrf}. (i) HiFormer \cite{heidari2023hiformer}. (j) DCSAUNet \cite{xu2023dcsau}. (k) M2SNet \cite{zhao2023m}. (l) \textbf{MADGNet (Ours)}. \textcolor{green}{\textbf{Green}} and \textcolor{red}{\textbf{Red}} lines denote the boundaries of the ground truth and prediction, respectively.}
    \label{fig:QualitativeResults}
\end{figure*}

\textbf{Binary Segmentation.} As listed in Tab. \ref{tab:comparison_sota_in_domain} and \ref{tab:comparison_sota_out_domain}, MADGNet achieved the highest segmentation performance on various modalities and clinical settings. In particular, compared to M2SNet, which acheived the second highest segmentation performance in most modalities in Tab. \ref{tab:comparison_sota_in_domain}, MADGNet improved the DSC and mIoU by 1.1\% and 1.0\% on average, respectively. Additionally, when compared to FRCUNet, which uses multi-frequency atttention, MADGNet improved the both DSC and mIoU by 2.0\% on average. To evaluate the domain generalizability of the model trained in each modality, we tested each model on external dataset from unseen clinical settings. As a result, MADGNet improved its performance by a higher margin than M2SNet, except for CVC-300. In addition, when compared FRCUNet and MADGNet on BUSI, the performance gap was tight. Nevertheless, MADGNet exhibits significant improvement of 1.9\% and 2.7\% on DSC and mIoU on STU, respectively. These results indicate that other models, which do not consider scale and frequency dimensions simultaneously, cannot comprehend intricate anatomical knowledge. Fig. \ref{fig:EfficiencyAnalysis} indicates that MADGNet contains almost 31M parameters with reasonable inference speed (0.024 sec/image), which has obvious advantages in terms of computational efficiency.

Fig. \ref{fig:QualitativeResults} illustrates the qualitative results on various modalities. The UNet relies solely on skip connection and progressive decoding; therefore, it produces noisy predictions. Such unreliable predictions are also observed in other models, such as AttUNet and UNet++, which do not consider multi-scale and multi-frequency features. And the other two transformer-based models, TransUNet and HiFormer, cannot segment in low-contrast Dermoscopy due to lack of inductive bias. As CENet, MSRFNet, DCSAUNet, and M2SNet utilize multi-scale information, they depict detailed boundary prediction; however, they struggle with severe noise in images from Radiology, Colonoscopy, and Ultrasound. Additionally, FRCUNet depicts detailed boundary on Ultrasound, which contains high-frequency images; but, has difficulties in predicting in Radiology and Dermoscopy. Despite this severe noise and lesions of various sizes, MADGNet successfully depicts detailed boundaries for all modalities due to the dual utilization of multi-scale and multi-frequency information.

\begin{table}
    \centering
    \scriptsize
    \begin{tabular}{c|cc|cc}
    \hline
    \multicolumn{1}{c|}{\multirow{2}{*}{Method}} & \multicolumn{2}{c|}{REFUGE \cite{orlando2020refuge} (\textit{seen})} & \multicolumn{2}{c}{Drishti-GS \cite{sivaswamy2015comprehensive} (\textit{unseen})} \\ \cline{2-5}
     & OD & OC & OD & OC \\
     \hline
     UNet \cite{ronneberger2015u}               & 79.9 \tiny{(0.9)} & 79.2 \tiny{(0.8)} & 62.2 \tiny{(1.1)} & 38.6 \tiny{(1.2)} \\
     AttUNet \cite{oktay1804attention}          & 80.8 \tiny{(0.1)} & \textcolor{blue}{\textbf{\textit{79.4}}} \tiny{(0.2)} & 72.6 \tiny{(2.6)} & 73.8 \tiny{(1.4)} \\
     UNet++ \cite{zhou2018unet++}               & 80.6 \tiny{(0.1)} & \textcolor{red}{\textbf{\underline{79.5}}} \tiny{(0.2)} & 76.1 \tiny{(1.4)} & 71.3 \tiny{(1.4)} \\
     CENet \cite{gu2019net}                     & 80.4 \tiny{(0.1)} & 74.1 \tiny{(0.4)} & \textcolor{blue}{\textbf{\textit{87.6}}} \tiny{(0.6)} & \textcolor{blue}{\textbf{\textit{78.8}}} \tiny{(1.2)} \\
     TransUNet \cite{chen2021transunet}         & 81.3 \tiny{(0.2)} & 40.7 \tiny{(0.2)} & 76.0 \tiny{(3.2)} & 38.0 \tiny{(3.1)} \\
     FRCUNet \cite{azad2021deep}                & \textcolor{blue}{\textbf{\textit{84.1}}} \tiny{(1.3)} & 45.2 \tiny{(0.1)} & 87.2 \tiny{(1.1)} & 44.4 \tiny{(0.1)} \\
     MSRFNet \cite{srivastava2021msrf}          & 83.6 \tiny{(0.6)} & 81.3 \tiny{(0.7)} & 71.3 \tiny{(0.8)} & 32.3 \tiny{(0.9)} \\
     HiFormer \cite{heidari2023hiformer}        & 80.5 \tiny{(0.1)} & 75.3 \tiny{(0.4)} & 79.9 \tiny{(1.3)} & 68.6 \tiny{(2.5)} \\
     DCSAUNet \cite{xu2023dcsau}                & 81.2 \tiny{(0.2)} & 59.6 \tiny{(0.5)} & 53.3 \tiny{(3.7)} & 28.3 \tiny{(3.6)} \\
     M2SNet \cite{zhao2023m}                    & 81.0 \tiny{(1.1)} & 60.1 \tiny{(1.2)} & 84.5 \tiny{(3.3)} & 69.4 \tiny{(1.2)} \\
     \hline
     \textbf{MADGNet}                           & \textcolor{red}{\textbf{\underline{84.9}}}  \tiny{(0.5)} & 78.5  \tiny{(0.4)} & \textcolor{red}{\textbf{\underline{88.8}}}  \tiny{(0.3)} & \textcolor{red}{\textbf{\underline{83.6}}}  \tiny{(0.4)} \\
     \hline
    \end{tabular}
    \caption{Segmentation results between methods on two Fundus Image segmentation datasets (REFUGE \cite{orlando2020refuge} and Drishti-GS \cite{sivaswamy2015comprehensive}).} \label{tab:comparison_sota_fundus}
\end{table}

\noindent\textbf{Multi-label Segmentation.} For medical image analysis, certain medical image segmentation datasets contain multi-label objects that need to be segmented. To satisfy this demand, we evaluated all models on two Fundus Imaging datasets, including REFUGE \cite{orlando2020refuge} (\textit{seen}) and Drishti-GS \cite{sivaswamy2015comprehensive} (\textit{unseen}), with different labels, Optic Disk (OD) and Optic Cup (OC). Except for data augmentation, we trained all models for 200 epochs with the same setting presented in Section \ref{sec42_implementation_details}. We cropped $512 \times 512$ ROIs centering OD and applied data augmentations, which are the same settings as those reported in the literature \cite{wang2019boundary}. 

As listed in Tab. \ref{tab:comparison_sota_fundus}, compared to CENet, MADGNet improved the average DSC of OD and OC by 2.9\% and 1.9\%, respectively. This result indicates that our model can be generalized to a multi-label segmentation task compared to other methods. The last row in Fig. \ref{fig:QualitativeResults} presents the qualitative results for each method. 

\subsection{Ablation Study on MADGNet}
To demonstrate the effectiveness of MADGNet, we conducted ablation studies on the MSMFA block and E-SDM.

\begin{figure}[t]
    \centering
    \includegraphics[width=0.5\textwidth]{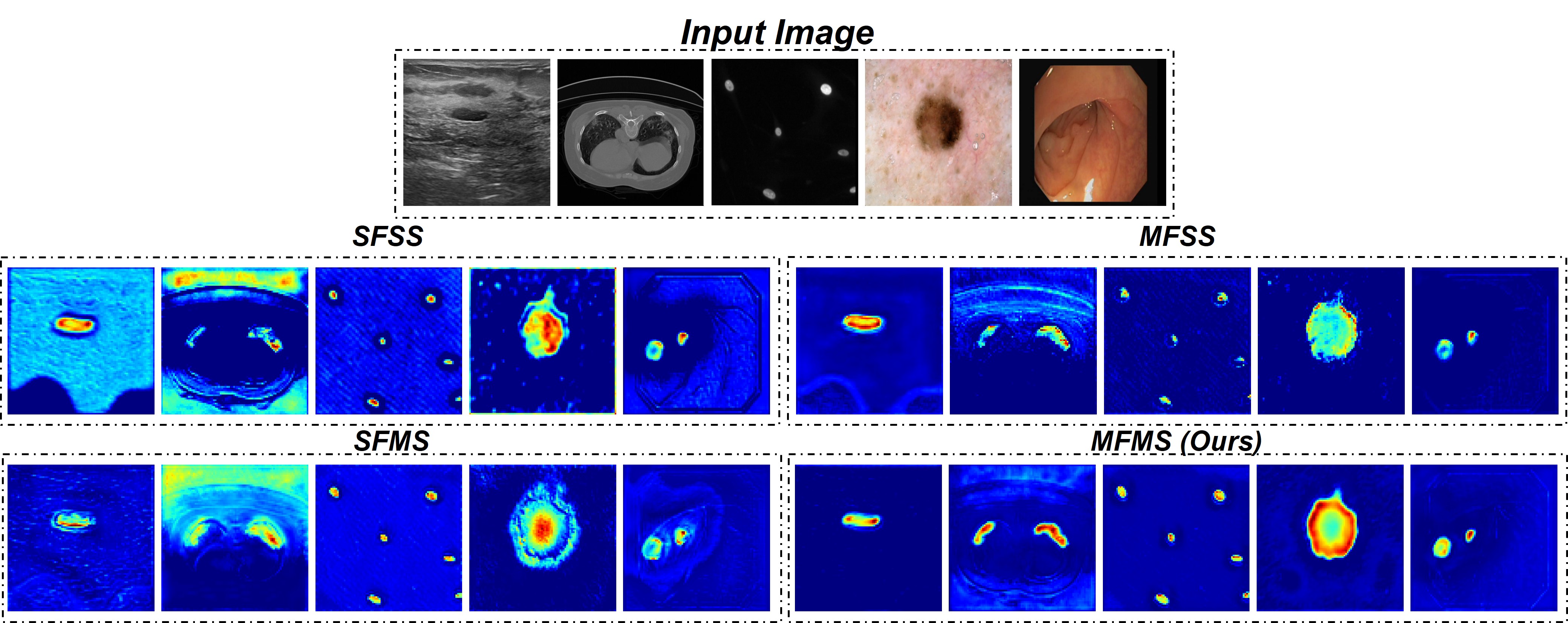} 
    \caption{Feature visualization of SFSS, MFSS, SFMS, \textbf{MFMS}.}
    \label{fig:SFSS_vs_MFSS_vs_SFMS_vs_MFMS}
\end{figure}

\noindent\textbf{Ablation Study on MFMSA Block.} To demonstrate that multi-scale and multi-frequency features are crucial in building the medical image segmentation model, we compared three variant models of MADGNet; Single-Frequency in Single-Scale Network (SFSSNet), Multi-Frequency in Single-Scale Network (MFSSNet), Single-Frequency in Multi-Scale Network (SFMSNet). These variants differs in $(S, K)$; SFSSNet, MFSSNet, and SFMSNet were set to $(1, 1)$, $(1, 16)$, and $(3, 1)$, respectively. As listed in Tab. \ref{tab:comparison_sota_in_domain} and \ref{tab:comparison_sota_out_domain}, as SFSSNet suffers from capturing the various sizes and detailed features in medical images, this approach has demonstrated unsatisfactory results in multiple clinical settings and modalities. The performance gain is achieved by utilizing multi-scale and multi-frequency features when these features are applied separately. This experimental evidence demonstrates the critical significance of both features in medical image segmentation, as illustrated in Fig. \ref{fig:scale_frequency_distribution}. Moreover, we observed that using both multi-scale and multi-frequency significantly improves the feature representation power and extracts enhanced boundary features, as illustrated in Fig. \ref{fig:SFSS_vs_MFSS_vs_SFMS_vs_MFMS}. Consequently, MADGNet simultaneously employs both features, resulting in significantly better performance across various modalities and clinical settings due to the enriched feature representation robust to noise and scale. 

\begin{table}
    \centering
    \scriptsize
    \begin{tabular}{c|c|c|cc|cc}
    \hline
    \multicolumn{1}{c|}{\multirow{2}{*}{DS}} & \multicolumn{1}{c|}{\multirow{2}{*}{Flow}} & \multicolumn{1}{c|}{\multirow{2}{*}{Task}} & \multicolumn{2}{c|}{\textit{Seen}} & \multicolumn{2}{c}{\textit{Unseen}} \\ \cline{4-7}
     & & & DSC & mIoU & DSC & mIoU \\
     \hline
     \multicolumn{1}{c|}{\multirow{2}{*}{\textcolor{red}{\xmark}}}  & \multicolumn{1}{c|}{-}        & \scriptsize{$R$}                                     & 90.8 & 85.7 & 75.2 & 68.2 \\
                                                                    & \multicolumn{1}{c|}{Parallel} & \scriptsize{$R \& D \& B$}                           & 91.5 & 86.6 & 76.2 & 69.9 \\
     \hline
     \multicolumn{1}{c|}{\multirow{3}{*}{\textcolor{blue}{\cmark}}} & \multicolumn{1}{c|}{Parallel} & \scriptsize{$R \& D \& B$}                           & 90.8 & 85.9 & 73.7 & 66.8 \\
                                                                    & \multicolumn{1}{c|}{Ensemble} & \scriptsize{$R \rightarrow D \rightarrow B$}         & 91.4 & 86.5 & 77.5 & 70.0 \\
                                                                    & \multicolumn{1}{c|}{Ensemble} & \scriptsize{$R \leftrightarrow D \leftrightarrow B$} & 92.0 & 87.3 & 80.9 & 73.3 \\
     \hline
    \end{tabular}
    \caption{Ablation study of E-SDM on the \textit{seen} (\cite{bernal2015wm, jha2020kvasir}) and \textit{unseen} (\cite{tajbakhsh2015automated, silva2014toward, vazquez2017benchmark}) datasets on Colonoscopy. DS denotes Deep Supervision. $R, D, B$ are region, distance map, and boundary task, respectively. $\rightarrow$ and $\leftrightarrow$ denote E-SDM without and with backward stream, respectively.}
    \label{tab:ablation_csd}
\end{table}

\begin{figure}[t]
    \centering
    \includegraphics[width=0.5\textwidth]{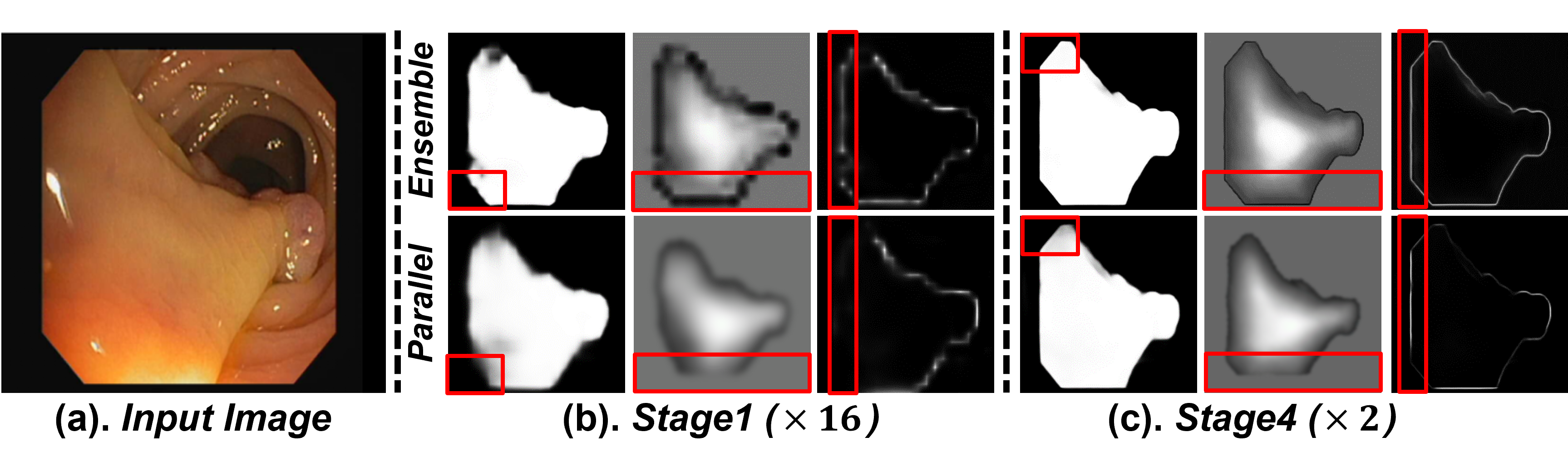}
    \caption{Qualitative results between ensemble and parallel manners. (a) Input Image, (b) and (c) Predictions from Stage1 ($\times 16 \textbf{ Up}$) and Stage4 ($\times 2 \textbf{ Up}$). First and second rows in (b) and (c) are predictions with \textbf{ensemble (Ours)} and parallel manners.}
    \label{fig:Qualitative_parallel_vs_cascade}
\end{figure}

\noindent\textbf{Ablation Study on E-SDM.} We performed ablation studies to demonstrate the effectiveness of E-SDM on Colonoscopy images. As listed in Tab. \ref{tab:ablation_csd}, our experimentation with multi-task learning and deep supervision in a parallel manner revealed a potential issue of performance degradation. This issue can be addressed by tackling the information loss due to drastic upsampling during deep supervision training. For this reason, when we trained a model in an ensemble manner, we observed that E-SDM had higher DSC of 1.2\% and 7.2\% on seen and unseen datasets, comparing two decoding flows (Parallel and Ensemble) with deep supervision, respectively. By using the ensemble method, we can overcome these challenges and obtain better results by leveraging an ensemble approach that combines the predictions from different tasks. This process enables us to identify complex boundaries $B$ and structures $D$ of the lesion $R$, resulting in higher performance compared to the parallel approach. Fig. \ref{fig:Qualitative_parallel_vs_cascade} illustrates the qualitative results of ensemble and parallel manners. E-SDM predict more reliably than in a parallel manner, even though it exhibits a high upsampling rate ($\times 16$). The ensemble method also depicts the detailed boundary and distance map predictions than in a parallel manner. Finally, we examined which forward  and backward streams had a larger impact on segmentation performance. Our study showed that the backward stream is crucial for maintaining the overall output quality as it compensates for the information loss caused by upsampling. While the forward stream only identifies areas requiring emphasis, the backward stream preserves critical information.
\section{Conclusion}

Based on the outcomes of extensive experiments on various modalities and clinical settings, we can summarize the effectiveness of MADGNet into three key aspects:  1) For the design of medical image segmentation, multi-scale and multi-frequency features prove pivotal. 2) MFCA’s adeptness at extracting discriminative features from noisy medical image feature maps translates into the acquisition of more sophisticated boundary cues through the MSSA. 3) E-SDM compensates for information loss caused by drastic upsampling when performing multi-task learning with deep supervision through an ensemble of tasks. Furthermore, our method demonstrated outstanding performance in segmenting OD and OC not only in binary classification but also in multi-label segmentation. In conclusion, we propose a novel medical image segmentation model called MADGNet, which can be utilized in various modalities and clinical settings. It includes two key components: the MFMSA block and E-SDM, which extract distinctive features and compensate for information loss during multi-task learning with deep supervision. Through rigorous experiment, we discovered that MADGNet is a highly effective model that surpasses other state-of-the-art options regarding segmentation performance. Furthermore, we will focus on increasing the memory-efficiency for real clinical use.
\section*{Acknowledgements}
This work was supported in part by the National Research Foundation of Korea (NRF) under Grant NRF-2021R1A2C2010893 and in part by Institute of Information and communications Technology Planning \& Evaluation (IITP) grant funded by the Korea government (MSIT) (No.RS-2022-00155915, Artificial Intelligence Convergence Innovation Human Resources Development (Inha University). 
{
    \small
    \bibliographystyle{ieeenat_fullname}
    \bibliography{main}
}

\clearpage
\setcounter{page}{1}
\maketitlesupplementary

\section{Dataset Descriptions}
\label{appendix_dataset_descriptions}

\begin{table}[h]
    \centering
    \scriptsize
    \setlength\tabcolsep{2pt} 
    \begin{tabular}{c|cccccc}
    \hline
    Dataset & Modality & Images & Resolutions & Train & Valid & Test \\
     \hline
     ISIC2018 \cite{gutman2016skin}                   & Dermoscopy   & 2594 & Variable           & 1868 & 465 & 261 \\
     COVID19-1 \cite{ma_jun_2020_3757476}             & Radiology    & 1277 & 512 $\times$ 512   & 643  & 251 & 383 \\
     BUSI \cite{al2020dataset}                        & Ultrasound   & 645  & Variable           & 324  & 160 & 161 \\
     2018 Data Science Bowl \cite{caicedo2019nucleus} & Microscopy   & 670  & Variable           & 483  & 120 & 67  \\
     CVC-ClinicDB \cite{bernal2015wm}                 & Colonoscopy  & 612  & 384 $\times$ 288   & 490  & 60  & 62  \\
     Kvasir-SEG \cite{jha2020kvasir}                  & Colonoscopy  & 1000 & Variable           & 800  & 100 & 100 \\
     REFUGE \cite{orlando2020refuge}                  & Fundus Image & 400  & 2124 $\times$ 2056 & 280  & 40  & 80  \\
    \hline
    \end{tabular}
    \caption{Details of the medical segmentation \textit{seen} clinical settings used in our experiments.}
    \label{tab:seen_clilical_dataset}
\end{table}

\begin{table}[h]
    \centering
    \scriptsize
    \begin{tabular}{c|cccc}
    \hline
    Dataset & Modality & Images & Resolutions & Test \\
     \hline
     PH2 \cite{mendoncca2013ph}                   & Dermoscopy   & 200  & 767 $\times$ 576  & 200  \\
     COVID19-2 \cite{COVID19_2}                   & Radiology    & 2535 & 512 $\times$ 512  & 2535 \\
     STU \cite{zhuang2019rdau}                    & Ultrasound   & 42   & Variable          & 42	  \\
     MonuSeg2018 \cite{dinh2021breast}            & Microscopy   & 82   & 256 $\times$ 256  & 82   \\
     CVC-300 \cite{vazquez2017benchmark}          & Colonoscopy  & 60   & 574 $\times$ 500  & 60   \\
     CVC-ColonDB \cite{tajbakhsh2015automated}    & Colonoscopy  & 380  & 574 $\times$ 500  & 380  \\
     ETIS \cite{silva2014toward}                  & Colonoscopy  & 196  & 1255 $\times$ 966 & 196  \\
     Drishti-GS \cite{sivaswamy2015comprehensive} & Fundus Image & 50   & Variable          & 50   \\
    \hline
    \end{tabular}
    \caption{Details of the medical segmentation \textit{unseen} clinical settings used in our experiments.}
    \label{tab:unseen_clilical_dataset}
\end{table}

\begin{itemize}
    \item \textit{Breast Ultrasound Segmentation:} The BUSI \cite{al2020dataset} comprises 780 images from 600 female patients, including 133 normal cases, 437 benign cases, and 210 malignant tumors. On the other hand, the STU \cite{zhuang2019rdau} includes only 42 breast ultrasound images collected by Shantou University. Due to the limited number of images in the STU, it is used only to evaluate the generalizability of each model across different datasets. \\

    \item \textit{Skin Lesion Segmentation:} The ISIC 2018 \cite{gutman2016skin} comprises 2,594 images with various sizes. We randomly selected train, validation, and test images with 1,868, 465, and 261, respectively. And, we used PH2 \cite{mendoncca2013ph} to evaluate the domain generalizability of each model. Note that ISIC2018 \cite{gutman2016skin} and PH2 \cite{mendoncca2013ph} are \textit{seen}, and \textit{unseen} clinical settings, respectively. \\
    
    \item \textit{COVID19 Lung Infection Segmentation:} COVID19-1 \cite{ma_jun_2020_3757476} comprises 1,277 high-quality CT images. We randomly selected train, validation, and test images with 643, 251, and 383, respectively. And, we used COVID19-2 \cite{COVID19_2} to evaluate the domain generalizability of each model. Note that COVID19-2 \cite{COVID19_2} is used for only testing. \\

    \item \textit{Cell Segmentation:} The 2018 Data Science Bowl dataset \cite{caicedo2019nucleus} comprises 670 microscopy images. The dataset consisted of training, validate, and test images with 483, 120, and 67, respectively. We also used MonuSeg2018 \cite{dinh2021breast} for evaluating the domain generalizability of each model. Note that MonuSeg2018 \cite{dinh2021breast} is used for only testing. \\
     
    \item \textit{Polyp Segmentation:} Colorectal cancer is the third most prevalent cancer globally and the second most common cause of death. It typically originates as small, non-cancerous (benign) clusters of cells known as polyps, which develop inside the colon. To evaluate the proposed model, we have used five benchmark datasets, namely CVC-ColonDB \cite{tajbakhsh2015automated}, ETIS \cite{silva2014toward}, Kvasir \cite{jha2020kvasir}, CVC-300 \cite{vazquez2017benchmark}, and CVC-ClinicDB \cite{bernal2015wm}. The same training set as the latest image polyp segmentation method has been adopted, consisting of 900 samples from Kvasir and 550 samples from CVC-ClinicDB for training. The remaining images and the other three datasets are used for only testing.  \\

    \item \textit{Fundus Image Segmentation:} To evaluate our method on multi-label segmentation, we utilize the training part of the REFUGE challenge dataset \cite{orlando2020refuge} as the training (280) and testing (80) dataset, and the public Drishti-GS \cite{sivaswamy2015comprehensive} dataset as the testing (50) dataset. 
\end{itemize}

\section{Efficiency Analysis}
\label{appendix_efficiency_analysis}

\begin{table}[ht]
    \centering
    \footnotesize
    \begin{tabular}{c|cc}
    \hline
    Method & Parameters (M) & inference speed (ms) \\
     \hline
     UNet \cite{ronneberger2015u}              & 34.5 & 10.1   \\
     AttUNet \cite{oktay1804attention}         & 35.6 & 13.8   \\
     UNet++ \cite{zhou2018unet++}              & 36.6 & 22.9   \\
     CENet \cite{gu2019net}                    & 18.9 & 10.5   \\
     TransUNet \cite{chen2021transunet}        & 53.4 & 93.4   \\
     FRCUNet \cite{azad2021deep}               & 40.8 & 13.4   \\
     MSRFNet \cite{srivastava2021msrf}         & 22.5 & 73.8   \\
     HiFormer \cite{heidari2023hiformer}       & 34.1 & 24.9   \\
     DCSAUNet \cite{xu2023dcsau}               & 25.9 & 24.3   \\
     M2SNet \cite{zhao2023m}                   & 26.5 & 32.1   \\
     \hline
     \textbf{MADGNet}                          & 31.4 & 24.0   \\
    \hline
    \end{tabular}
    \caption{The number of parameters (M) and inference speed (ms) of different models.}
    \label{tab:efficiency_analysis}
\end{table}

\begin{figure*}[t]
    \centering
    \includegraphics[width=\textwidth]{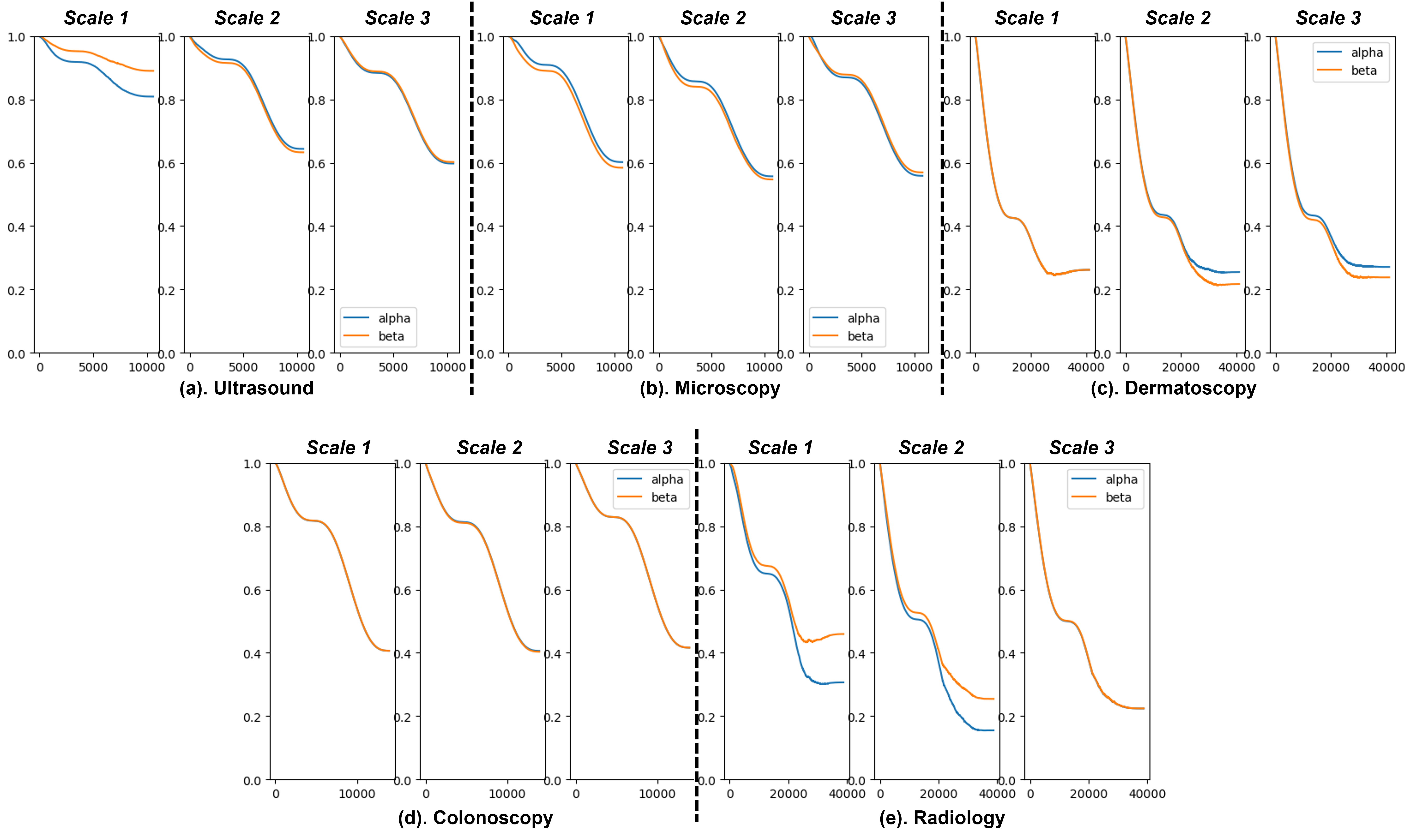}
    \caption{The training results of $\alpha^{s}_{3}$ and $\beta^{s}_{3}$ for each modalities ((a) Ultrasound, (b) Microscopy, (c) Dermoscopy, (d) Colonoscopy, and (e) Radiology) where $s \in \{ 1, 2, 3 \}$.}
    \label{fig:Sup_alpha_beta_history}
\end{figure*}

\begin{figure*}[t]
    \centering
    \includegraphics[width=\textwidth]{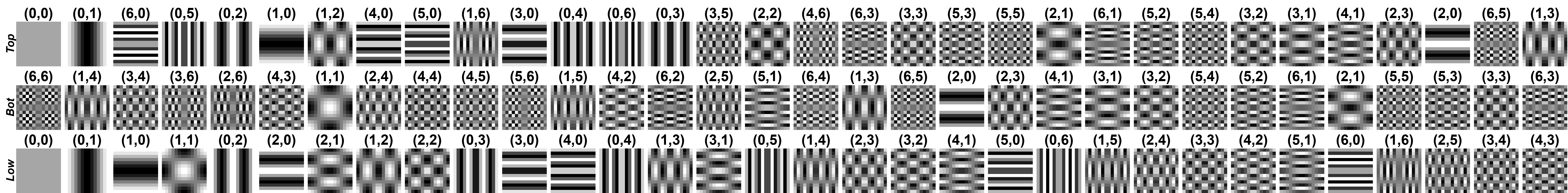}
    \caption{Frequency selection strategies (Top, Bot, Low) \cite{qin2021fcanet}. $(u_{k}, v_{k})$ denotes the frequency indices according to frequency selection strategy.}
    \label{fig:Sup_Frequency_Selection_Strategy}
\end{figure*}

\subsection{Parameter Efficiency Proof}
\label{appendix_parameter_efficiency_proof}
\noindent\textbf{Scale Decomposition.} For further efficiency, we replace the conventional convolution with kernel size $2s + 1$ by dilated convolution with a kernel size of 3 and dilation size of $s$. For instance, in Fig. \ref{fig:MFMSNet}. (b), the convolution with kernel size of 5 and 7 in second and third scale branch are replaced into the dilated convolution with a kernel size of 3 and dilation size of 2 and 3, respectively. This replacement can achieve a parameter reduction of $9 / (2s + 1)^{2}$ in each scale branch. Suppose that the number of channel at $i$-th MFMSA block is $C$. Then, by channel reduction ratio $\gamma \in (0, 1)$, the number of parameters for each scale branch is $9 \times C^{2}\gamma^{s - 1}$.  

\noindent\textbf{MFCA.} Note that MFCA contains two fully-connected layer $\mathbf{W}_{1} \in \mathbb{R}^{C\gamma^{s - 1} \times \frac{C\gamma^{s - 1}}{r}}$ and $\mathbf{W}_{2}  \in \mathbb{R}^{\frac{C\gamma^{s - 1}}{r} \times C\gamma^{s - 1}}$ with reduction ratio $r$. Then, the number of parameters in MFCA is $C\gamma^{s - 1} \times C\gamma^{s - 1} \times \frac{1}{r} \times 2$. 

\noindent\textbf{MSSA.} Firstly, to extract attention map from frequency-recalibrated feature map $\hat{\mathbf{X}}^{s}_{i}$, we apply a 2D convolution operation with kernel size 1. And then, to aggregate each refined features from different scale branches, we restore the number of channels into $C$. Hence, the number of parameters in MSSA is $C\gamma^{s - 1} + 9 \times C^{2}\gamma^{s - 1}$.

\noindent\textbf{MFMSA block.} Then, we can approximate the number of parameters at $s$-th scale branch as follows:

\begin{equation}
    \begin{split}
        p_{s} &= 9 \times C^{2}\gamma^{s - 1} + C\gamma^{s - 1} \times C\gamma^{s - 1} \times \frac{1}{r} \times 2 \\ &+ C\gamma^{s - 1} + 9 \times C^{2}\gamma^{s - 1} \\ 
              &= C\gamma^{s - 1} \left( 18C + C\gamma^{s - 1} \times \frac{2}{r} + 1 \right)
    \end{split}
\end{equation}

If we do not introduce the channel reduction ratio $\gamma$, then the number of parameters at each scale is $p = C (18C + 2C / r + 1)$. Then, we can calculate the parameter reduction ratio $\frac{p_{s}}{p}$ as follows:

\begin{equation}
    \begin{split}
        \frac{p_{s}}{p} &= \frac{C\gamma^{s - 1} \left( 18C + C\gamma^{s - 1} \times \frac{2}{r} + 1 \right)}{C (18C + \frac{2C}{r} + 1)} \\
                        &= \gamma^{s - 1} \left( \frac{18C + C\gamma^{s - 1} \times \frac{2}{r} + 1}{18C + \frac{2C}{r} + 1} \right) \\
                        &\approx \gamma^{s - 1} \left( \frac{18C + C\gamma^{s - 1} \times \frac{2}{r}}{18C + \frac{2C}{r}} \right) \\
                        &= \gamma^{s - 1} \left( \frac{18r + 2\gamma^{s - 1}}{18r + 2} \right) \\ 
                        &= \left( \gamma^{s - 1} \right)^{2} \left( \frac{\frac{18r}{\gamma^{s - 1}} + 2}{18r + 2} \right) \\
                        &\approx \left( \gamma^{s - 1} \right)^{2} \left( \frac{\frac{18r}{\gamma^{s - 1}}}{18r} \right) \\
                        &= \gamma^{s - 1} \\
    \end{split}
\end{equation}

\begin{algorithm}[t]
\caption{Ensemble Sub-Decoding Module for Multi-task Learning with Deep Supervision in Multi-label Segmentation}
\label{alg_CSD_multilabel}
\textbf{Input}: Refined feature map $\mathbf{Y}_{i}$ from $i$-th MFMSA block \\
\textbf{Output}: Core task prediction $\textbf{T}^{c, m}_{i}$ and sub-task predictions $\{ \mathbf{T}^{s_{1}, m}_{i}, \dots, \mathbf{T}^{s_{L}, m}_{i} \}$ at $i$-th decoder for each $m$-th label.
\begin{algorithmic}[1] 
\FOR{$m = 1, 2, \dots, M$}
    \STATE $\mathbf{P}_{i}^{c, m} = \textbf{Conv2D}_{1} (\mathbf{Y}_{i}) $ 
    \FOR{$l = 1, 2, \dots, L$ }
        \STATE $\mathbf{P}_{i}^{s_{l}, m} = \textbf{Conv2D}_{1} (\mathbf{Y}_{i} \times \sigma \left( \mathbf{P}_{i}^{s_{l - 1}, m}\right) ) $.
    \ENDFOR
    \STATE $\mathbf{T}_{i}^{s_{L}, m} = \textbf{Up}_{5 - i} \left( \mathbf{P}_{i}^{s_{L}, m} \right) $ 
    \FOR{$l = L - 1, \dots, 0$}
        \STATE $\mathbf{T}_{i}^{s_{l}, m} = \textbf{Up}_{5 - i} \left( \mathbf{P}_{i}^{s_{l}, m} \right) + \mathbf{T}_{i}^{s_{l + 1}, m}$
    \ENDFOR
    \STATE $\mathbf{O}_{i}^{m} = \{ \textbf{T}^{c, m}_{i}, \mathbf{T}^{s_{1}, m}_{i}, \dots, \mathbf{T}^{s_{L}, m}_{i} \}$
\ENDFOR
\STATE \textbf{return} $\{ \mathbf{O}_{i}^{1}, \dots, \mathbf{O}_{i}^{M} \}$
\end{algorithmic}
\end{algorithm}

\section{Ensemble Sub-Decoding Module for Multi-label Segmentation}

In this section, we show that E-SDM can be utilized to any segmentation dataset with $M$ multi-label. 

\noindent\textbf{Forward Stream.} During the forward stream, core and sub-task pseudo predictions $\{ \mathbf{P}^{c, m}_{i}, \mathbf{P}^{s_{1}, m}_{i}, \dots, \mathbf{P}^{s_{L}, m}_{i} \}$ for each label $m \in \{ 1, 2, \dots, M \}$ are produced at $i$-th decoder stage as follows:

\begin{equation}
    \begin{cases}
        &\mathbf{P}^{c, m}_{i} = \textbf{Conv2D}_{1} (\mathbf{Y}_{i}) \\ 
        &\mathbf{P}^{s_{l}, m}_{i} = \textbf{Conv2D}_{1} (\mathbf{Y}_{i} \times \sigma ( \mathbf{P}_{i}^{s_{l - 1}, m} )) \text{ for } l = 1, \dots, L
    \end{cases}
\end{equation}

where $\mathbf{P}_{i}^{s_{0}, m} = \mathbf{P}_{i}^{c, m}$. This stream enables the following sub-task prediction cascadingly focus on the region by spatial attention starting from core pseudo prediction $\mathbf{P}^{c, m}_{i}$. 

\noindent\textbf{Backward Stream.} After producing $L$-th sub-task pseudo prediction $\mathbf{P}^{s_{L}, m}_{i}$, to produce final core task prediction $\mathbf{T}^{c, m}_{i}$ for $m$-th label, we apply backward stream as follows:

\begin{equation}
\label{eq:backward_stream_multilabel}
    \begin{cases}
        &\mathbf{T}^{s_{L}, m}_{i} = \textbf{Up}_{5-i} \left( \mathbf{P}^{s_{L}, m}_{i} \right) \\ 
        &\mathbf{T}^{s_{l}, m}_{i} = \textbf{Up}_{5-i} \left( \mathbf{P}^{s_{l}, m}_{i} \right) + \mathbf{T}^{s_{l+1}, m}_{i}  \text{ for } l = 0, \dots, L - 1
    \end{cases}
\end{equation}

where $\mathbf{T}^{s_{0}, m}_{i} = \mathbf{T}^{c, m}_{i}$. To further analyze the Eq \ref{eq:backward_stream_multilabel}, we can recursively rewrite from core task $\mathbf{T}^{c, m}_{i}$ as follows:

\begin{equation}
\label{eq:backward_stream_recursive_multilabel}
    \begin{split}
        \mathbf{T}^{c, m}_{i} &= \mathbf{T}^{s_{0}, m}_{i} = \textbf{Up}_{5 - i} \left( \mathbf{P}^{s_{0}, m}_{i} \right) + \mathbf{T}^{s_{1}, m}_{i} \\ 
                           &= \left[ \textbf{Up}_{5 - i} \left( \mathbf{P}^{s_{0}, m}_{i} \right) + \textbf{Up}_{5 - i} \left( \mathbf{P}^{s_{1}, m}_{i} \right) \right] + \mathbf{T}^{s_{2}, m}_{i} = \cdots \\
                           &= \sum_{l = 0}^{L} \textbf{Up}_{5 - i} \left( \mathbf{P}^{s_{l}, m}_{i} \right)
    \end{split}
\end{equation}

\noindent Consequently, E-SDM can be interpreted as an ensemble of predictions between different tasks for describing the same legion for each $m$-th label. Algorithm \ref{alg_CSD_multilabel} describes the detailed training algorithm for E-SDM in multi-label segmentation.

\section{More Detailed Ablation Study on MADGNet}

\subsection{Ablation on Backbone Model}
\label{appendix_ablation_on_backbone_model}

\begin{table}[h]
    \centering
    \scriptsize
    \setlength\tabcolsep{3.5pt} 
    \begin{tabular}{c|cccccc}
    \hline
    \multicolumn{1}{c|}{\multirow{2}{*}{Backbone}} & \multicolumn{6}{c}{\textit{Seen} Datasets (\cite{al2020dataset, gutman2016skin, ma_jun_2020_3757476, caicedo2019nucleus, bernal2015wm, jha2020kvasir})} \\ \cline{2-7}
     & DSC \scriptsize{$\uparrow$} & mIoU \scriptsize{$\uparrow$} & $F_{\beta}^{w}$ \scriptsize{$\uparrow$}  & $S_{\alpha}$ \scriptsize{$\uparrow$} & $E_{\phi}^{max}$ \scriptsize{$\uparrow$} & MAE \scriptsize{$\downarrow$} \\
     \hline
     ResNet50 \cite{he2016deep}                        & 86.9 & 80.6 & 84.6 & 84.9 &	91.9 & 2.3 \\
     Res2Net50 \cite{gao2019res2net}                   & 87.1 & 80.9 & 83.1 & 85.0 &	92.0 & 2.4 \\
     ViT-B-16 \cite{dosovitskiy2020image}                   & 87.5 & 81.2 & 85.0 & 85.1 & 92.3 & 2.4 \\
     \textbf{ResNeSt50 \cite{zhang2022resnest} (Ours)} & 88.5 & 82.3 & 85.9 & 85.7 & 92.8 & 2.4 \\
    \hline
    \hline
    \multicolumn{1}{c|}{\multirow{2}{*}{Backbone}} & \multicolumn{6}{c}{\textit{Unseen} Datasets (\cite{mendoncca2013ph, COVID19_2, zhuang2019rdau, dinh2021breast, vazquez2017benchmark, tajbakhsh2015automated, silva2014toward})} \\ \cline{2-7}
     & DSC \scriptsize{$\uparrow$} & mIoU \scriptsize{$\uparrow$} & $F_{\beta}^{w}$ \scriptsize{$\uparrow$}  & $S_{\alpha}$ \scriptsize{$\uparrow$} & $E_{\phi}^{max}$ \scriptsize{$\uparrow$} & MAE \scriptsize{$\downarrow$} \\
     \hline
     ResNet50 \cite{he2016deep}                        & 69.1 & 61.0 & 67.6 & 73.8 &	80.7 & 6.2 \\
     Res2Net50 \cite{gao2019res2net}                   & 70.2 & 61.8 & 68.7 & 74.2 &	81.5 & 5.9 \\
     ViT-B-16 \cite{dosovitskiy2020image}                   & 69.0 & 61.6 & 67.8 & 75.4 & 81.1 & 4.9 \\
     \textbf{ResNeSt50 \cite{zhang2022resnest} (Ours)} & 77.1 & 68.1 & 75.0 & 77.2 &	87.0 & 6.2 \\
    \hline
    \end{tabular}
    \caption{Quantitative results for each \textit{Seen} (\cite{al2020dataset, gutman2016skin, ma_jun_2020_3757476, caicedo2019nucleus, bernal2015wm, jha2020kvasir}) and \textit{Unseen} (\cite{mendoncca2013ph, COVID19_2, zhuang2019rdau, dinh2021breast, vazquez2017benchmark, tajbakhsh2015automated, silva2014toward}) datasets according to backbone network. We presents the \textit{mean} performance for each dataset.}
    \label{tab:ablation_backbone_networks}
\end{table}

In this section, we present the performance of MADGNet according to various backbone network (ResNet50 \cite{he2016deep}, Res2Net50 \cite{gao2019res2net}, ViT-B-16 \cite{dosovitskiy2020image}, and \textbf{ResNeSt50 \cite{zhang2022resnest} (Ours)}) in Tab. \ref{tab:ablation_backbone_networks}. We report the \textit{mean} performance of \textit{seen} and \textit{unseen} datasets.

\subsection{Hyperparameter on MADGNet}
\label{appendix_hyperparameter_on_MADGNet}

\noindent\textbf{Number of Scale branch $S$.}

\begin{figure}[t]
    \centering
    \includegraphics[width=0.48\textwidth]{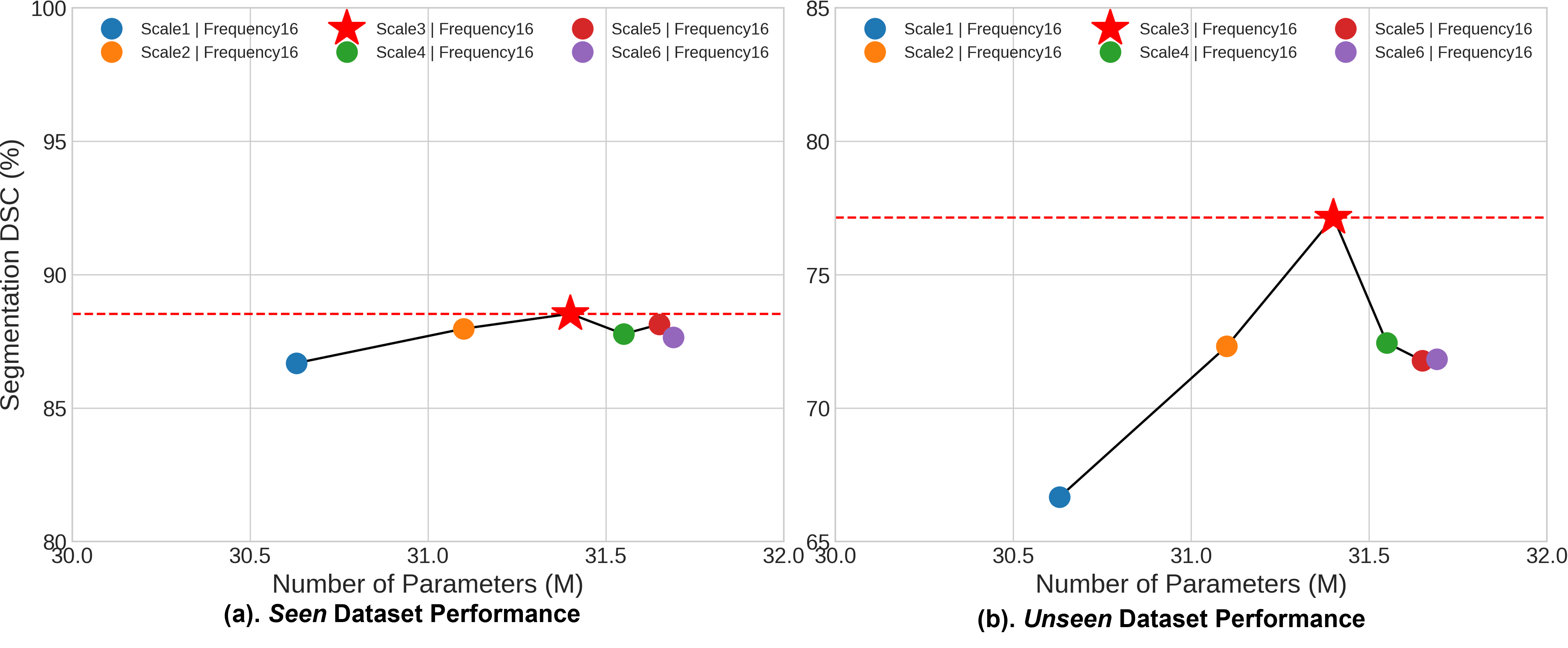}
    \caption{Comparison of parameters (M) vs segmentation performance (DSC) according to number of scale $S$ on average for (a) \textit{seen} and (b) \textit{unseen} datasets.}
    \label{fig:Sup_Scale_performance}
\end{figure}

\begin{table}[h]
    \centering
    \scriptsize
    \begin{tabular}{c|cccccc}
    \hline
    \multicolumn{1}{c|}{\multirow{2}{*}{Scale $S$}} & \multicolumn{6}{c}{\textit{Seen} Datasets (\cite{al2020dataset, gutman2016skin, ma_jun_2020_3757476, caicedo2019nucleus, bernal2015wm, jha2020kvasir})} \\ \cline{2-7}
     & DSC \scriptsize{$\uparrow$} & mIoU \scriptsize{$\uparrow$} & $F_{\beta}^{w}$ \scriptsize{$\uparrow$}  & $S_{\alpha}$ \scriptsize{$\uparrow$} & $E_{\phi}^{max}$ \scriptsize{$\uparrow$} & MAE \scriptsize{$\downarrow$} \\
     \hline
     1                 & 86.7 & 81.0 & 84.5 & 84.8 &	91.6 & 2.4 \\
     2                 & 88.0 & 81.7 & 85.5 & 85.4 &	92.6 & 2.3 \\
     \textbf{3 (Ours)} & 88.5 & 82.3 & 85.9 & 85.7 &	92.8 & 2.4 \\
     4                 & 87.8 & 81.5 & 85.3 & 85.3 &	92.6 & 2.3 \\
     5                 & 88.1 & 81.9 & 85.7 & 85.5 &	92.7 & 2.3 \\
     6                 & 87.6 & 81.9 & 85.3 & 85.3 &	92.7 & 2.4 \\
    \hline
    \hline
    \multicolumn{1}{c|}{\multirow{2}{*}{Scale $S$}} & \multicolumn{6}{c}{\textit{Unseen} Datasets (\cite{mendoncca2013ph, COVID19_2, zhuang2019rdau, dinh2021breast, vazquez2017benchmark, tajbakhsh2015automated, silva2014toward})} \\ \cline{2-7}
     & DSC \scriptsize{$\uparrow$} & mIoU \scriptsize{$\uparrow$} & $F_{\beta}^{w}$ \scriptsize{$\uparrow$}  & $S_{\alpha}$ \scriptsize{$\uparrow$} & $E_{\phi}^{max}$ \scriptsize{$\uparrow$} & MAE \scriptsize{$\downarrow$} \\
     \hline
     1                 & 66.7 & 59.1 & 66.0 & 73.7 &	79.4 & 9.6 \\
     2                 & 72.3 & 64.1 & 70.8 & 74.9 &	83.8 & 5.1 \\
     \textbf{3 (Ours)} & 77.1 & 68.1 & 75.0 & 77.2 &	87.0 & 6.2 \\
     4                 & 72.4 & 63.7 & 70.6 & 74.6 &	82.3 & 8.3 \\
     5                 & 71.8 & 63.6 & 70.4 & 75.8 & 83.5 & 5.4 \\
     6                 & 71.8 & 63.6 & 70.5 & 76.4 &	84.0 & 4.8 \\
    \hline
    \end{tabular}
    \caption{Quantitative results for each \textit{Seen} (\cite{al2020dataset, gutman2016skin, ma_jun_2020_3757476, caicedo2019nucleus, bernal2015wm, jha2020kvasir}) and \textit{Unseen} (\cite{mendoncca2013ph, COVID19_2, zhuang2019rdau, dinh2021breast, vazquez2017benchmark, tajbakhsh2015automated, silva2014toward}) datasets according to the number of scale $S$. We presents the \textit{mean} performance for each domain.}
    \label{tab:ablation_number_of_scale}
\end{table}

In this section, we present the performance of MADGNet according to the number of scales $S \in \{ 1, 2, 3, 4, 5, 6 \}$ with $F = 16$ in Tab. \ref{tab:ablation_number_of_scale} and Fig. \ref{fig:Sup_Scale_performance}. We report the \textit{mean} performance of \textit{seen} and \textit{unseen} datasets.

\noindent\textbf{Number of Frequency branch $K$.}

\begin{table}[h]
    \centering
    \scriptsize
    \begin{tabular}{c|cccccc}
    \hline
    \multicolumn{1}{c|}{\multirow{2}{*}{Frequency $K$}} & \multicolumn{6}{c}{\textit{Seen} Datasets (\cite{al2020dataset, gutman2016skin, ma_jun_2020_3757476, caicedo2019nucleus, bernal2015wm, jha2020kvasir})} \\ \cline{2-7}
     & DSC \scriptsize{$\uparrow$} & mIoU \scriptsize{$\uparrow$} & $F_{\beta}^{w}$ \scriptsize{$\uparrow$}  & $S_{\alpha}$ \scriptsize{$\uparrow$} & $E_{\phi}^{max}$ \scriptsize{$\uparrow$} & MAE \scriptsize{$\downarrow$} \\
     \hline
     1                  & 87.8 & 81.6 & 85.4 & 85.3 & 92.5 & 2.3 \\
     2                  & 87.7 & 81.5 & 85.3 & 85.2 & 92.6 & 2.3 \\
     4                  & 87.8 & 81.5 & 85.3 & 85.3 & 92.5 & 2.4 \\
     8                  & 87.8 & 81.5 & 85.3 & 85.3 & 92.4 & 2.3 \\
     \textbf{16 (Ours)} & 88.5 & 82.3 & 85.9 & 85.7 & 92.8 & 2.4 \\
     32                 & 87.9 & 81.7 & 85.5 & 85.3 & 92.7 & 2.3 \\
    \hline
    \hline
    \multicolumn{1}{c|}{\multirow{2}{*}{Frequency $K$}} & \multicolumn{6}{c}{\textit{Unseen} Datasets (\cite{mendoncca2013ph, COVID19_2, zhuang2019rdau, dinh2021breast, vazquez2017benchmark, tajbakhsh2015automated, silva2014toward})} \\ \cline{2-7}
     & DSC \scriptsize{$\uparrow$} & mIoU \scriptsize{$\uparrow$} & $F_{\beta}^{w}$ \scriptsize{$\uparrow$}  & $S_{\alpha}$ \scriptsize{$\uparrow$} & $E_{\phi}^{max}$ \scriptsize{$\uparrow$} & MAE \scriptsize{$\downarrow$} \\
     \hline
     1                  & 72.0 & 63.7 & 70.4 & 75.9 & 83.3 & 5.6 \\
     2                  & 71.8 & 63.7 & 70.4 & 76.3 & 83.8 & 5.3 \\
     4                  & 72.2 & 64.0 & 70.9 & 76.8 & 84.6 & 4.6 \\
     8                  & 71.3 & 63.0 & 69.7 & 75.3 & 82.9 & 6.1 \\
     \textbf{16 (Ours)} & 77.1 & 68.1 & 75.0 & 77.2 & 87.0 & 6.2 \\
     32                 & 73.1 & 64.1 & 70.9 & 74.6 & 82.5 & 7.8 \\
    \hline
    \end{tabular}
    \caption{Quantitative results for each \textit{Seen} (\cite{al2020dataset, gutman2016skin, ma_jun_2020_3757476, caicedo2019nucleus, bernal2015wm, jha2020kvasir}) and \textit{Unseen} (\cite{mendoncca2013ph, COVID19_2, zhuang2019rdau, dinh2021breast, vazquez2017benchmark, tajbakhsh2015automated, silva2014toward}) datasets according to the number of frequency $K$. We presents the \textit{mean} performance for each domain.}
    \label{tab:ablation_number_of_frequency}
\end{table}

\begin{figure}[t]
    \centering
    \includegraphics[width=0.48\textwidth]{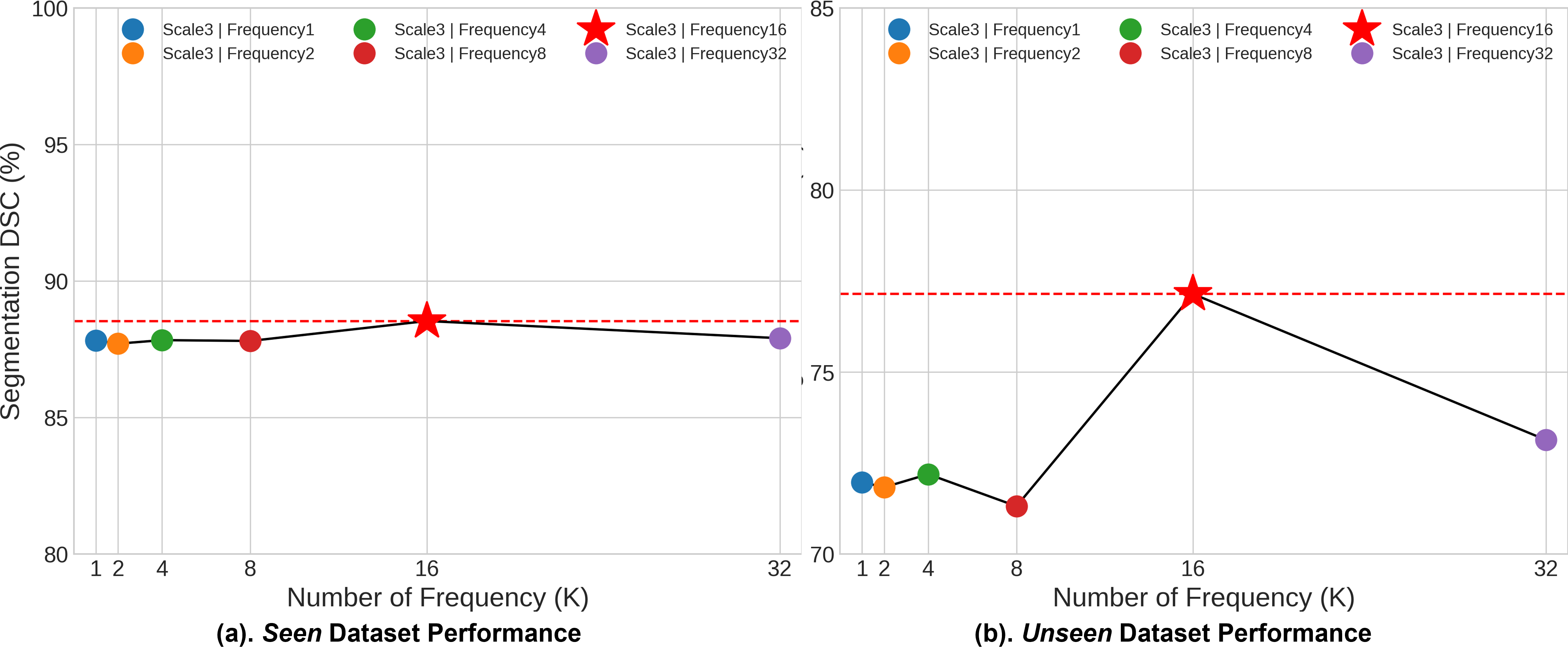}
    \caption{Number of frequency $F$ vs segmentation performance (DSC) on average for (a) \textit{seen} and (b) \textit{unseen} datasets.}
    \label{fig:Sup_Frequency_performance}
\end{figure}

In this section, we present the performance of MADGNet according to the number of frequencies $F \in \{ 1, 2, 4, 8, 16, 32 \}$ with $S = 3$ in Tab. \ref{tab:ablation_number_of_frequency} and Fig. \ref{fig:Sup_Frequency_performance}. We report the \textit{mean} performance of \textit{seen} and \textit{unseen} datasets.

\noindent\textbf{Frequency Selection Strategy (Top vs Bot vs Low).}

\begin{table}[h]
    \centering
    \scriptsize
    \begin{tabular}{c|cccccc}
    \hline
    \multicolumn{1}{c|}{\multirow{2}{*}{Strategy}} & \multicolumn{6}{c}{\textit{Seen} Datasets (\cite{al2020dataset, gutman2016skin, ma_jun_2020_3757476, caicedo2019nucleus, bernal2015wm, jha2020kvasir})} \\ \cline{2-7}
     & DSC \scriptsize{$\uparrow$} & mIoU \scriptsize{$\uparrow$} & $F_{\beta}^{w}$ \scriptsize{$\uparrow$}  & $S_{\alpha}$ \scriptsize{$\uparrow$} & $E_{\phi}^{max}$ \scriptsize{$\uparrow$} & MAE \scriptsize{$\downarrow$} \\
     \hline
     \textbf{Top (Ours)} & 88.5 & 82.3 &	85.9 & 85.7 & 92.8 & 2.4 \\
     Bot                 & 87.7 & 81.4 &	85.0 & 85.2 & 92.4 & 2.4 \\
     Low                 & 87.1 & 80.5 & 84.3 & 86.2 & 92.4 & 2.0 \\
    \hline
    \hline
    \multicolumn{1}{c|}{\multirow{2}{*}{Strategy}} & \multicolumn{6}{c}{\textit{Unseen} Datasets (\cite{mendoncca2013ph, COVID19_2, zhuang2019rdau, dinh2021breast, vazquez2017benchmark, tajbakhsh2015automated, silva2014toward})} \\ \cline{2-7}
     & DSC \scriptsize{$\uparrow$} & mIoU \scriptsize{$\uparrow$} & $F_{\beta}^{w}$ \scriptsize{$\uparrow$}  & $S_{\alpha}$ \scriptsize{$\uparrow$} & $E_{\phi}^{max}$ \scriptsize{$\uparrow$} & MAE \scriptsize{$\downarrow$} \\
     \hline
     \textbf{Top (Ours)} & 77.1 & 68.1 & 75.0 & 77.2 & 87.0 & 6.2 \\
     Bot                 & 72.1 & 63.5 &	70.3 & 74.3 & 82.9 & 7.3 \\
     Low                 & 70.1 & 62.5 &	69.2 & 75.8 & 82.8 & 5.0 \\
    \hline
    \end{tabular}
    \caption{Quantitative results for each \textit{Seen} (\cite{al2020dataset, gutman2016skin, ma_jun_2020_3757476, caicedo2019nucleus, bernal2015wm, jha2020kvasir}) and \textit{Unseen} (\cite{mendoncca2013ph, COVID19_2, zhuang2019rdau, dinh2021breast, vazquez2017benchmark, tajbakhsh2015automated, silva2014toward}) datasets according to frequency selection strategies. We presents the \textit{mean} performance for each domain.}
    \label{tab:ablation_frequency_strategy}
\end{table}

In this section, we present the performance of MADGNet according to frequency selections \textbf{Top (Ours)}, Bot, and Low with $S = 3$ and $F = 16$ in Tab. \ref{tab:ablation_frequency_strategy}. The set of DCT basis images according to each frequency selection strategy can be seen in the Fig. \ref{fig:Sup_Frequency_Selection_Strategy}. We report the \textit{mean} performance of \textit{seen} and \textit{unseen} datasets.

\section{Technical Innovation, Design Principle and Interpretability of MADGNet}
\label{technical_innovation}

Motivated by papers \cite{xu2020learning, li2015finding}, to extract discriminative features in both the frequency and spatial domains, we introduced dual attention modules with multiple statistic information of frequency (MFCA) and two learnable information flow parameters in multi-scale (MSSA). The causal effect of MFMSA block is interpreted as follows: 1) \textit{MFCA emphasizes the salient features while reducing the influence of noisy features by focusing on the frequency of interest}, characterized by high variance of frequency in medical domain (Fig. \ref{fig:scale_frequency_distribution}). 2) \textit{MSSA captures more reliable discriminative boundary cues (Fig. \ref{fig:SFSS_vs_MFSS_vs_SFMS_vs_MFMS}) for lesions of various sizes} by combining foreground and background attention with multi-scale attention and information flow parameters. Our approach distinguishes itself by successfully integrating both attentions with dilated convolution and downsampling, a pioneering endeavor in the medical domain.

\section{Metrics Descriptions}
\label{appendix_metric_descriptions}

\begin{itemize}
    \item \textit{Mean Dice Similarity Coefficient (DSC)} measures the similarity between two samples and is widely used in assessing the performance of segmentation tasks, such as image segmentation or object detection. Higher is better.\\

    \item \textit{Mean Intersection over Union (IoU)} measures the ratio of the intersection area to the union area of the predicted and ground truth masks in segmentation tasks. Higher is better.\\

    \item \textit{Mean Weighted F-Measure} $F_{\beta}^{w}$ is a metric that combines precision and recall into a single value by calculating the harmonic mean. "Weighted" often implies that it might be weighted by class frequency or other factors to provide a balanced measure across different classes. Higher is better. \\

    \item \textit{Mean S-Measure} $S_{\alpha}$ is used to evaluate the quality of image segmentation, specifically focusing on the structural similarity between the predicted and ground truth segmentation. Higher is better. \\

    \item \textit{Mean E-Measure} $E_{\phi}^{max}$ assesses the edge accuracy in edge detection or segmentation tasks. It evaluates how well the predicted edges align with the ground truth edges. Higher is better. \\

    \item \textit{Mean Mean Absolute Error (MAE)} calculates the average absolute differences between predicted and ground truth values. Lower is better.
\end{itemize}

\section{More Qualtative and Quantitative Results}
\label{appendix_additional_experiment_results_with_various_metrics}

\begin{figure*}[hbtp]
    \centering
    \includegraphics[width=\textwidth]{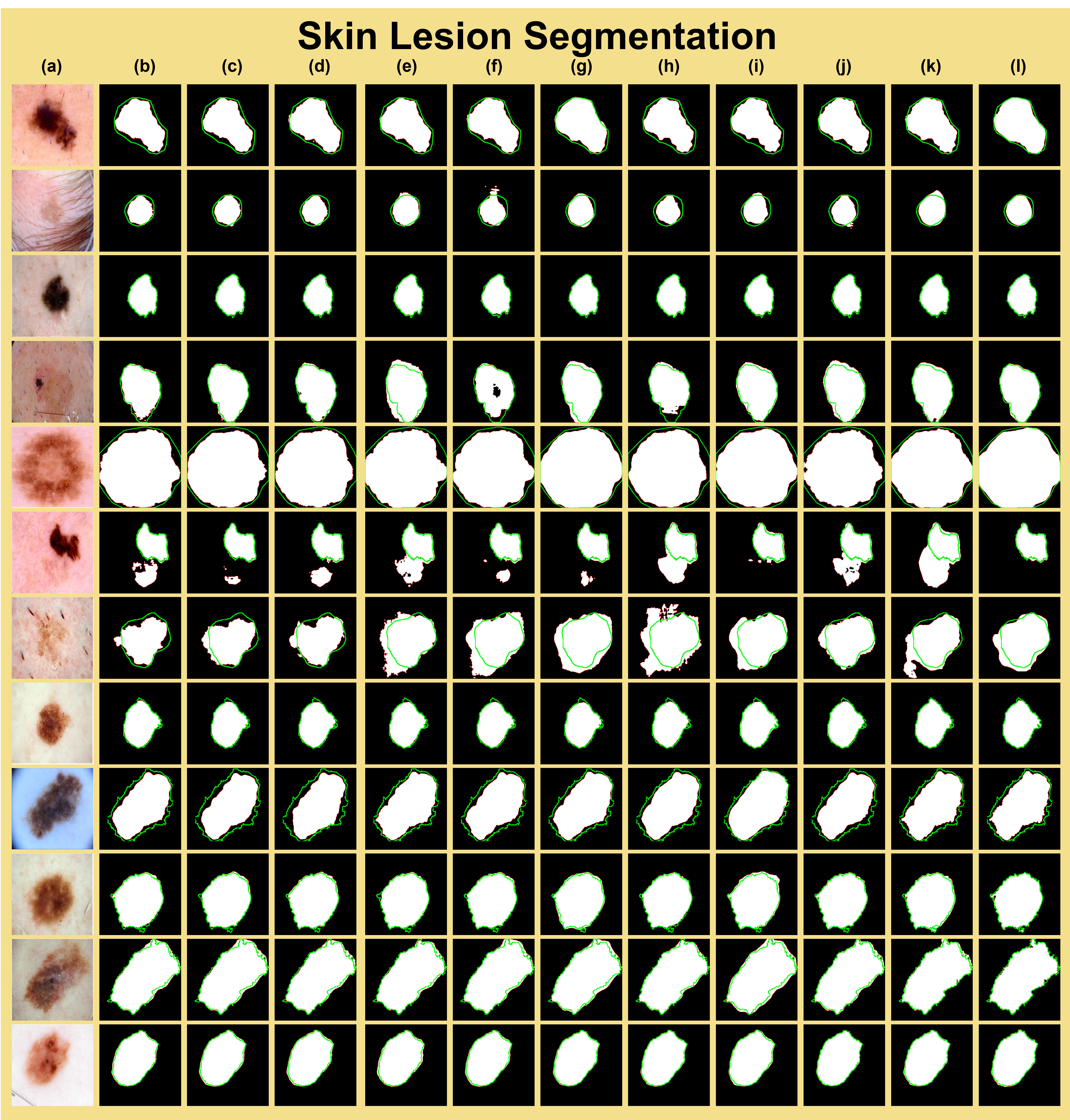}
    \caption{Qualitative comparison of other methods and MADGNet on \textbf{\underline{Skin Lesion Segmentation (Dermoscopy)}} \cite{gutman2016skin, mendoncca2013ph}. (a) Input images, (b) UNet \cite{ronneberger2015u}. (c) AttUNet \cite{oktay1804attention}, (d) UNet++ \cite{zhou2018unet++}, (e) CENet \cite{gu2019net}, (f) TransUNet \cite{chen2021transunet}, (g) FRCUNet \cite{azad2021deep}, (h) MSRFNet \cite{srivastava2021msrf}, (i) HiFormer \cite{heidari2023hiformer}, (j) DCSAUNet \cite{xu2023dcsau}, (k) M2SNet \cite{zhao2023m}, (l) \textbf{MADGNet (Ours)}. \textcolor{green}{\textbf{Green}} and \textcolor{red}{\textbf{Red}} lines denote the boundaries of the ground truth and prediction, respectively.}
    \label{fig:Sup_QualitativeResults_Dermatoscopy}
\end{figure*}

\begin{figure*}[hbtp]
    \centering
    \includegraphics[width=\textwidth]{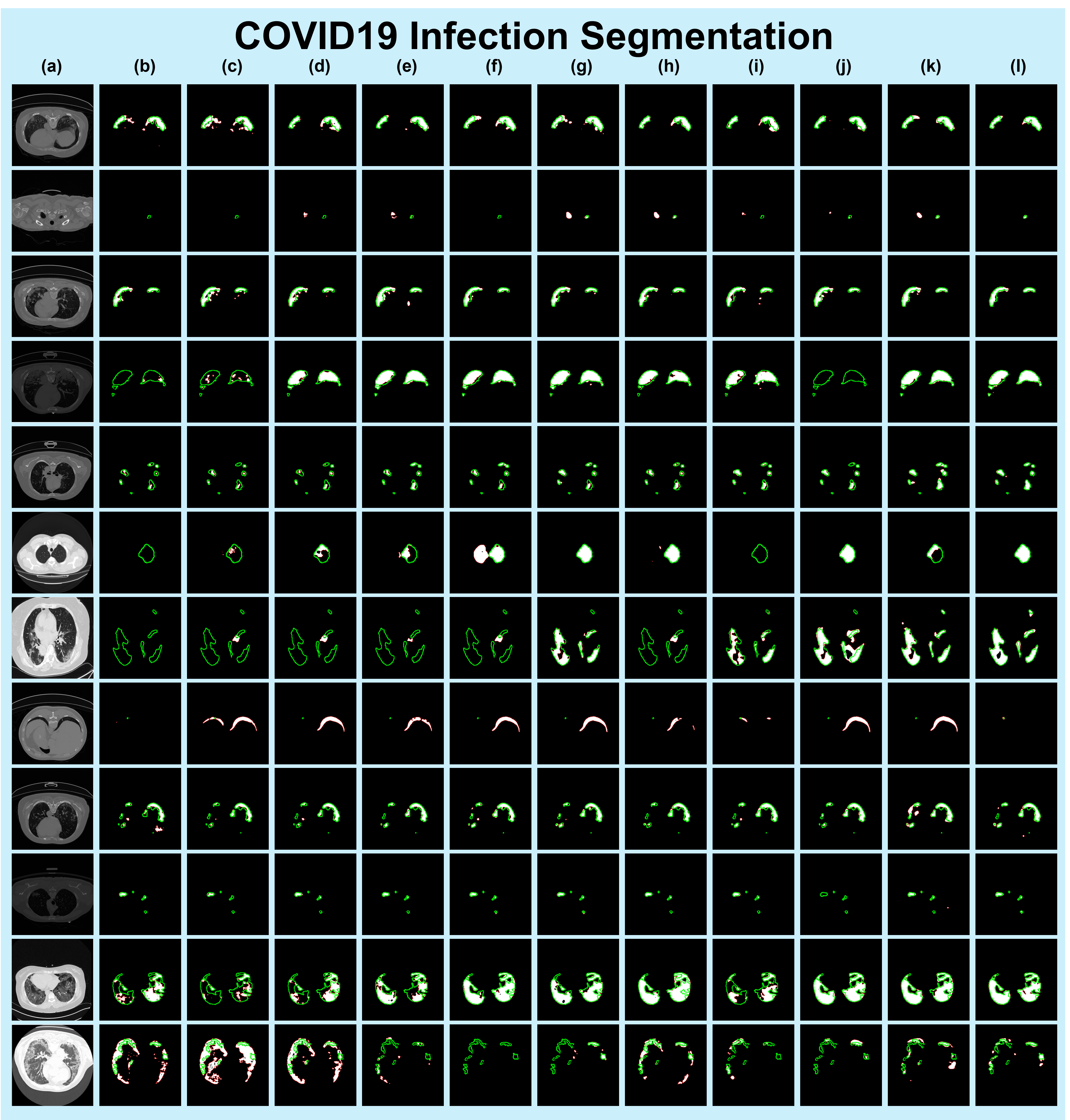}
    \caption{Qualitative comparison of other methods and MADGNet on \textbf{\underline{COVID19 Infection Segmentation (Radiology)}} \cite{ma_jun_2020_3757476, COVID19_2}. (a) Input images, (b) UNet \cite{ronneberger2015u}. (c) AttUNet \cite{oktay1804attention}, (d) UNet++ \cite{zhou2018unet++}, (e) CENet \cite{gu2019net}, (f) TransUNet \cite{chen2021transunet}, (g) FRCUNet \cite{azad2021deep}, (h) MSRFNet \cite{srivastava2021msrf}, (i) HiFormer \cite{heidari2023hiformer}, (j) DCSAUNet \cite{xu2023dcsau}, (k) M2SNet \cite{zhao2023m}, (l) \textbf{MADGNet (Ours)}.  \textcolor{green}{\textbf{Green}} and \textcolor{red}{\textbf{Red}} lines denote the boundaries of the ground truth and prediction, respectively.}
    \label{fig:Sup_QualitativeResults_Radiology}
\end{figure*}

\begin{figure*}[hbtp]
    \centering
    \includegraphics[width=\textwidth]{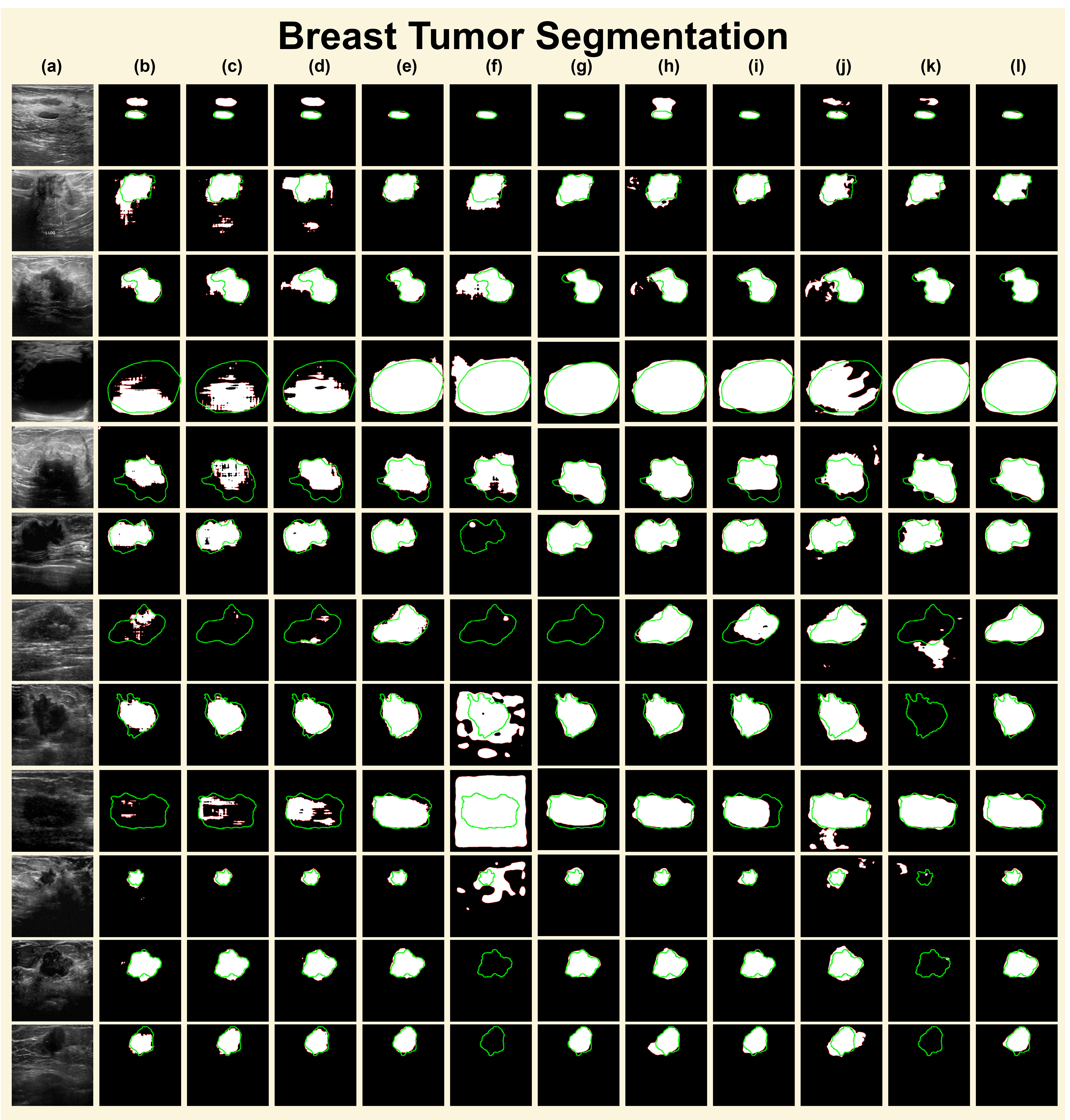}
    \caption{Qualitative comparison of other methods and MADGNet on \textbf{\underline{Breast Tumor Segmentation (Ultrasound)}} \cite{al2020dataset, zhuang2019rdau}. (a) Input images, (b) UNet \cite{ronneberger2015u}. (c) AttUNet \cite{oktay1804attention}, (d) UNet++ \cite{zhou2018unet++}, (e) CENet \cite{gu2019net}, (f) TransUNet \cite{chen2021transunet}, (g) FRCUNet \cite{azad2021deep}, (h) MSRFNet \cite{srivastava2021msrf}, (i) HiFormer \cite{heidari2023hiformer}, (j) DCSAUNet \cite{xu2023dcsau}, (k) M2SNet \cite{zhao2023m}, (l) \textbf{MADGNet (Ours)}. \textcolor{green}{\textbf{Green}} and \textcolor{red}{\textbf{Red}} lines denote the boundaries of the ground truth and prediction, respectively.}
    \label{fig:Sup_QualitativeResults_Ultrasound}
\end{figure*}

\begin{figure*}[hbtp]
    \centering
    \includegraphics[width=\textwidth]{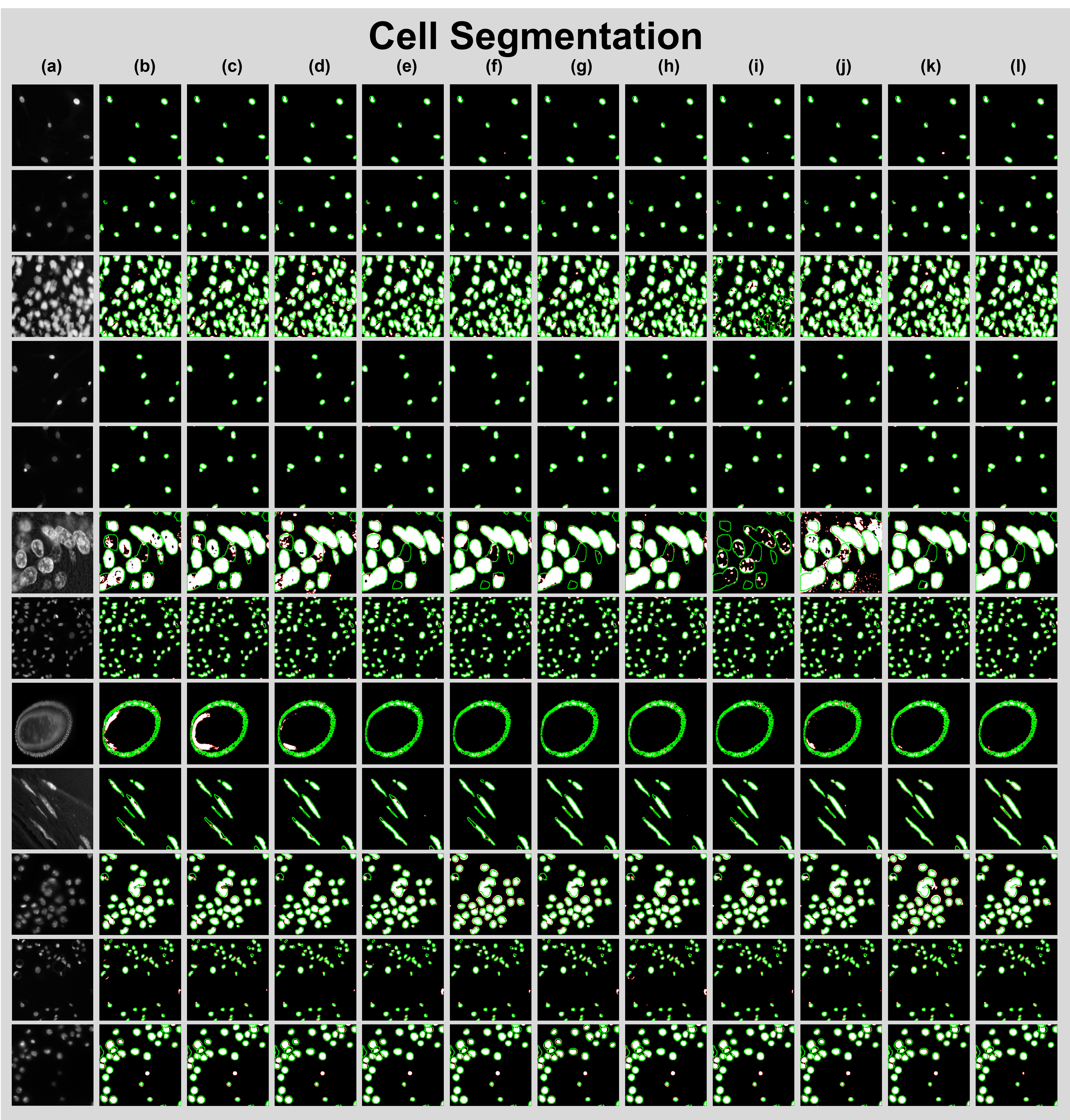}
    \caption{Qualitative comparison of other methods and MADGNet on \textbf{\underline{Cell Segmentation (Microscopy)}} \cite{caicedo2019nucleus, dinh2021breast}. (a) Input images, (b) UNet \cite{ronneberger2015u}. (c) AttUNet \cite{oktay1804attention}, (d) UNet++ \cite{zhou2018unet++}, (e) CENet \cite{gu2019net}, (f) TransUNet \cite{chen2021transunet}, (g) FRCUNet \cite{azad2021deep}, (h) MSRFNet \cite{srivastava2021msrf}, (i) HiFormer \cite{heidari2023hiformer}, (j) DCSAUNet \cite{xu2023dcsau}, (k) M2SNet \cite{zhao2023m}, (l) \textbf{MADGNet (Ours)}. \textcolor{green}{\textbf{Green}} and \textcolor{red}{\textbf{Red}} lines denote the boundaries of the ground truth and prediction, respectively.}
    \label{fig:Sup_QualitativeResults_Microscopy}
\end{figure*}

\begin{figure*}[hbtp]
    \centering
    \includegraphics[width=\textwidth]{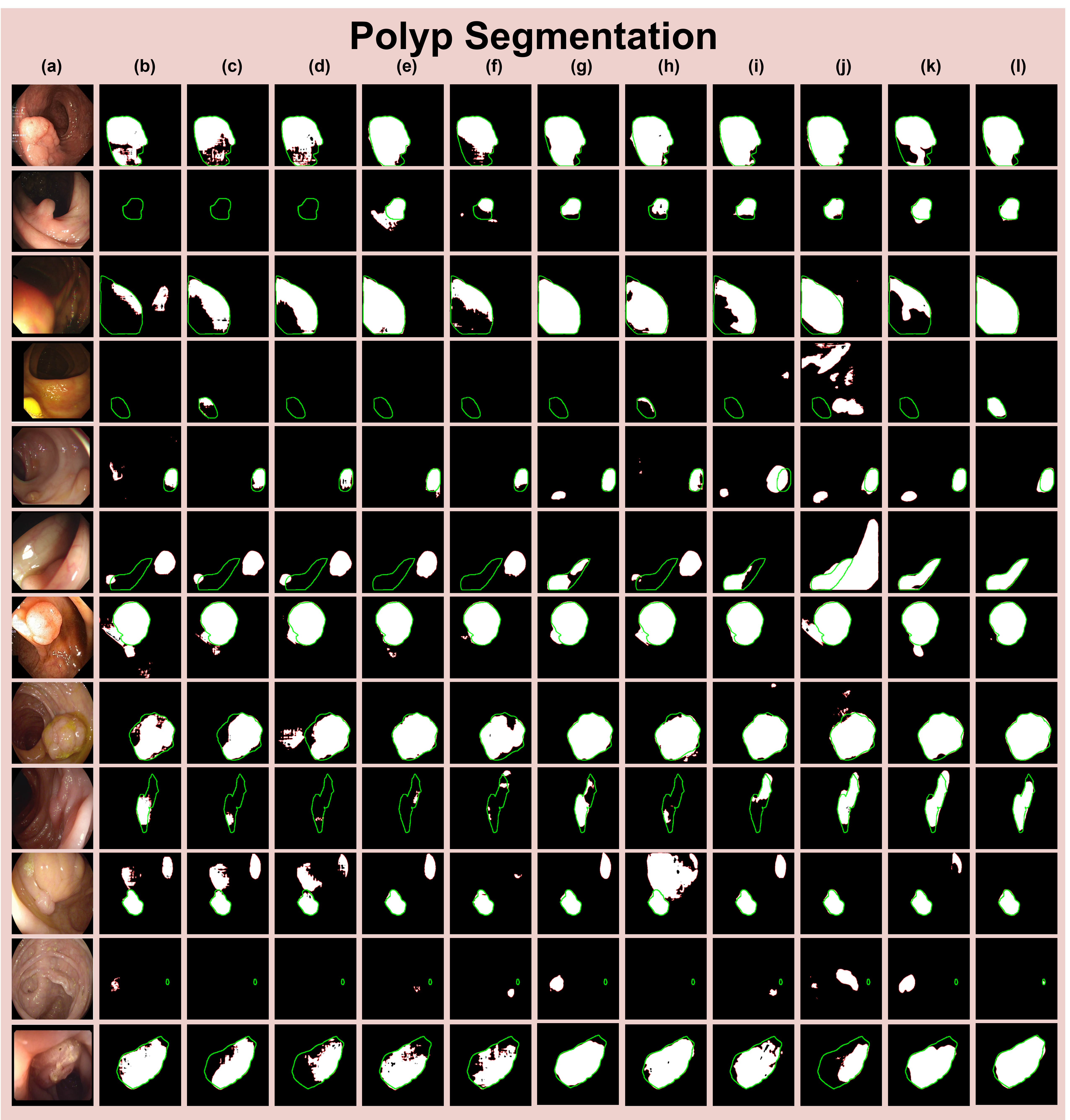}
    \caption{Qualitative comparison of other methods and MADGNet on \textbf{\underline{Polyp Segmentation (Colonoscopy)}} \cite{bernal2015wm, jha2020kvasir, vazquez2017benchmark, tajbakhsh2015automated, silva2014toward}. (a) Input images, (b) UNet \cite{ronneberger2015u}. (c) AttUNet \cite{oktay1804attention}, (d) UNet++ \cite{zhou2018unet++}, (e) CENet \cite{gu2019net}, (f) TransUNet \cite{chen2021transunet}, (g) FRCUNet \cite{azad2021deep}, (h) MSRFNet \cite{srivastava2021msrf}, (i) HiFormer \cite{heidari2023hiformer}, (j) DCSAUNet \cite{xu2023dcsau}, (k) M2SNet \cite{zhao2023m}, (l) \textbf{MADGNet (Ours)}. \textcolor{green}{\textbf{Green}} and \textcolor{red}{\textbf{Red}} lines denote the boundaries of the ground truth and prediction, respectively.}
    \label{fig:Sup_QualitativeResults_Colonoscopy}
\end{figure*}

In this section, we provide the quantitative results with various metrics in Tab. \ref{tab:comparison_sota_dermatoscopy_other_metrics}, \ref{tab:comparison_sota_radiology_other_metrics}, \ref{tab:comparison_sota_ultrasound_other_metrics}, \ref{tab:comparison_sota_microscopy_other_metrics}, and \ref{tab:comparison_sota_colonoscopy_other_metrics}. We report the \textit{mean} performance of three trials for all results. $(\cdot)$ denotes a standard deviation of three trials. \textcolor{red}{\textbf{\underline{Red}}} and \textcolor{blue}{\textbf{\textit{Blue}}} are the first and second best performance results, respectively. We also present more various qualitative results on datasets in Fig. \ref{fig:Sup_QualitativeResults_Dermatoscopy}, \ref{fig:Sup_QualitativeResults_Radiology}, \ref{fig:Sup_QualitativeResults_Ultrasound}, \ref{fig:Sup_QualitativeResults_Microscopy}, and \ref{fig:Sup_QualitativeResults_Colonoscopy}.

\begin{table}[h]
    \centering
    \scriptsize
    \setlength\tabcolsep{3pt} 
    \begin{tabular}{c|cccccc}
    \hline
    \multicolumn{1}{c|}{\multirow{2}{*}{Method}} & \multicolumn{6}{c}{ISIC2018 \cite{gutman2016skin} $\Rightarrow$ ISIC2018 \cite{gutman2016skin}} \\ \cline{2-7}
     & DSC \scriptsize{$\uparrow$} & mIoU \scriptsize{$\uparrow$} & $F_{\beta}^{w}$ \scriptsize{$\uparrow$}  & $S_{\alpha}$ \scriptsize{$\uparrow$} & $E_{\phi}^{max}$ \scriptsize{$\uparrow$} & MAE \scriptsize{$\downarrow$} \\
     \hline
     UNet \cite{ronneberger2015u}              & 87.3 \tiny{(0.8)} & 80.2 \tiny{(0.7)} & 87.9 \tiny{(0.0)} & 80.4 \tiny{(0.1)} & 91.3 \tiny{(0.0)} & 4.7 \tiny{(0.0)} \\
     AttUNet \cite{oktay1804attention}         & 87.8 \tiny{(0.1)} & 80.5 \tiny{(0.1)} & 86.5 \tiny{(0.2)} & 80.5 \tiny{(0.1)} & 92.0 \tiny{(0.1)} & 4.5 \tiny{(0.0)} \\
     UNet++ \cite{zhou2018unet++}              & 87.3 \tiny{(0.2)} & 80.2 \tiny{(0.1)} & 86.3 \tiny{(0.2)} & 80.1 \tiny{(0.1)} & 91.6 \tiny{(0.2)} & 4.7 \tiny{(0.0)} \\
     CENet \cite{gu2019net}                    & 89.1 \tiny{(0.2)} & 82.1 \tiny{(0.1)} & 88.1 \tiny{(0.2)} & 81.3 \tiny{(0.1)} & 93.0 \tiny{(0.2)} & 4.3 \tiny{(0.1)} \\
     TransUNet \cite{chen2021transunet}        & 87.3 \tiny{(0.2)} & 81.2 \tiny{(0.2)} & 88.6 \tiny{(0.2)} & 80.8 \tiny{(0.2)} & 91.9 \tiny{(0.2)} & 4.2 \tiny{(0.1)} \\
     FRCUNet \cite{azad2021deep}               & 88.9 \tiny{(0.1)} & 83.1 \tiny{(0.2)} & 89.3 \tiny{(0.0)} & 82.0 \tiny{(0.1)} & 93.9 \tiny{(0.2)} & 3.7 \tiny{(0.1)} \\
     MSRFNet \cite{srivastava2021msrf}         & 88.2 \tiny{(0.2)} & 81.3 \tiny{(0.2)} & 86.9 \tiny{(0.2)} & 80.7 \tiny{(0.1)} & 92.0 \tiny{(0.2)} & 4.7 \tiny{(0.1)} \\
     HiFormer \cite{heidari2023hiformer}       & 88.7 \tiny{(0.5)} & 81.9 \tiny{(0.5)} & 87.6 \tiny{(0.6)} & 80.8 \tiny{(0.5)} & 92.6 \tiny{(0.5)} & 4.4 \tiny{(0.3)} \\
     DCSAUNet \cite{xu2023dcsau}               & 89.0 \tiny{(0.3)} & 82.0 \tiny{(0.3)} & 87.8 \tiny{(0.3)} & 81.4 \tiny{(0.1)} & 92.9 \tiny{(0.3)} & 4.4 \tiny{(0.1)} \\
     M2SNet \cite{zhao2023m}                   & \textcolor{blue}{\textbf{\textit{89.2}}} \tiny{(0.2)} & \textcolor{blue}{\textbf{\textit{83.4}}} \tiny{(0.2)} & \textcolor{blue}{\textbf{\textit{88.9}}} \tiny{(0.1)} & \textcolor{blue}{\textbf{\textit{81.8}}} \tiny{(0.1)} & \textcolor{blue}{\textbf{\textit{93.8}}} \tiny{(0.1)} & \textcolor{blue}{\textbf{\textit{3.7}}} \tiny{(0.0)} \\
     \hline
     \textbf{MADGNet}                         & \textcolor{red}{\textbf{\underline{90.2}}} \tiny{(0.1)} & \textcolor{red}{\textbf{\underline{83.7}}} \tiny{(0.2)} & \textcolor{red}{\textbf{\underline{89.2}}} \tiny{(0.2)} & \textcolor{red}{\textbf{\underline{82.0}}} \tiny{(0.1)} & \textcolor{red}{\textbf{\underline{94.1}}} \tiny{(0.3)} & \textcolor{red}{\textbf{\underline{3.6}}} \tiny{(0.2)} \\
    \hline
    \multicolumn{1}{c|}{\multirow{2}{*}{Method}} & \multicolumn{6}{c}{ISIC2018 \cite{gutman2016skin} $\Rightarrow$ PH2 \cite{mendoncca2013ph}} \\ \cline{2-7}
     & DSC \scriptsize{$\uparrow$} & mIoU \scriptsize{$\uparrow$} & $F_{\beta}^{w}$ \scriptsize{$\uparrow$}  & $S_{\alpha}$ \scriptsize{$\uparrow$} & $E_{\phi}^{max}$ \scriptsize{$\uparrow$} & MAE \scriptsize{$\downarrow$} \\
     \hline
     UNet \cite{ronneberger2015u}              & 90.3 \tiny{(0.1)} & \textcolor{blue}{\textbf{\textit{83.5}}} \tiny{(0.1)} & \textcolor{red}{\textbf{\underline{88.4}}} \tiny{(0.1)} & 74.8 \tiny{(0.1)} & 90.8 \tiny{(0.1)} & 6.9 \tiny{(0.0)} \\
     AttUNet \cite{oktay1804attention}         & 89.9 \tiny{(0.2)} & 82.6 \tiny{(0.3)} & 87.3 \tiny{(0.3)} & 74.8 \tiny{(0.2)} & 90.8 \tiny{(0.2)} & 6.7 \tiny{(0.2)} \\
     UNet++ \cite{zhou2018unet++}              & 88.0 \tiny{(0.3)} & 80.1 \tiny{(0.3)} & 85.7 \tiny{(0.2)} & 73.2 \tiny{(0.1)} & 89.2 \tiny{(0.2)} & 7.9 \tiny{(0.1)} \\
     CENet \cite{gu2019net}                    & 90.5 \tiny{(0.1)} & 83.3 \tiny{(0.1)} & 87.3 \tiny{(0.1)} & 75.1 \tiny{(0.0)} & 91.5 \tiny{(0.1)} & 6.0 \tiny{(0.1)} \\
     TransUNet \cite{chen2021transunet}        & 89.5 \tiny{(0.3)} & 82.1 \tiny{(0.4)} & 86.9 \tiny{(0.4)} & 74.3 \tiny{(0.2)} & 90.3 \tiny{(0.2)} & 6.7 \tiny{(0.2)} \\
     FRCUNet \cite{azad2021deep}               & 90.6 \tiny{(0.1)} & 83.4 \tiny{(0.2)} & 87.4 \tiny{(0.2)} & 75.4 \tiny{(0.2)} & 91.7 \tiny{(0.1)} & \textcolor{blue}{\textbf{\textit{5.9}}} \tiny{(0.1)} \\
     MSRFNet \cite{srivastava2021msrf}         & 90.5 \tiny{(0.3)} & \textcolor{blue}{\textbf{\textit{83.5}}} \tiny{(0.3)} & 87.5 \tiny{(0.3)} & 75.0 \tiny{(0.0)} & 91.4 \tiny{(0.2)} & 6.0 \tiny{(0.3)} \\
     HiFormer \cite{heidari2023hiformer}       & 86.9 \tiny{(1.6)} & 79.1 \tiny{(1.8)} & 83.2 \tiny{(1.9)} & 72.9 \tiny{(1.1)} & 88.6 \tiny{(1.4)} & 8.0 \tiny{(0.9)} \\
     DCSAUNet \cite{xu2023dcsau}               & 89.0 \tiny{(0.4)} & 81.5 \tiny{(0.3)} & 85.7 \tiny{(0.2)} & 74.0 \tiny{(0.3)} & 90.2 \tiny{(0.3)} & 6.9 \tiny{(0.4)} \\
     M2SNet \cite{zhao2023m}                   & \textcolor{blue}{\textbf{\textit{90.7}}} \tiny{(0.3)} & \textcolor{blue}{\textbf{\textit{83.5}}} \tiny{(0.5)} & \textcolor{blue}{\textbf{\textit{87.6}}} \tiny{(0.4)} & \textcolor{blue}{\textbf{\textit{75.5}}} \tiny{(0.3)} & \textcolor{blue}{\textbf{\textit{92.0}}} \tiny{(0.2)} & \textcolor{blue}{\textbf{\textit{5.9}}} \tiny{(0.2)} \\
     \hline
     \textbf{MADGNet}                         & \textcolor{red}{\textbf{\underline{91.3}}} \tiny{(0.1)} & \textcolor{red}{\textbf{\underline{84.6}}} \tiny{(0.1)} & \textcolor{red}{\textbf{\underline{88.4}}} \tiny{(0.1)} & \textcolor{red}{\textbf{\underline{76.2}}} \tiny{(0.1)} & \textcolor{red}{\textbf{\underline{92.8}}} \tiny{(0.1)} & \textcolor{red}{\textbf{\underline{5.1}}} \tiny{(0.1)} \\
    \hline
    \end{tabular}
    \caption{Segmentation results on \textbf{Skin Lesion Segmentation (Dermatoscopy)} \cite{gutman2016skin, mendoncca2013ph}. We train each model on ISIC2018 \cite{gutman2016skin} train dataset and evaluate on ISIC2018 \cite{gutman2016skin} and PH2 \cite{mendoncca2013ph} test datasets.}
    \label{tab:comparison_sota_dermatoscopy_other_metrics}
\end{table}

\begin{table}[h]
    \centering
    \scriptsize
    \setlength\tabcolsep{3pt} 
    \begin{tabular}{c|cccccc}
    \hline
    \multicolumn{1}{c|}{\multirow{2}{*}{Method}} & \multicolumn{6}{c}{COVID19-1 \cite{ma_jun_2020_3757476} $\Rightarrow$ COVID19-1 \cite{ma_jun_2020_3757476}} \\ \cline{2-7}
     & DSC \scriptsize{$\uparrow$} & mIoU \scriptsize{$\uparrow$} & $F_{\beta}^{w}$ \scriptsize{$\uparrow$}  & $S_{\alpha}$ \scriptsize{$\uparrow$} & $E_{\phi}^{max}$ \scriptsize{$\uparrow$} & MAE \scriptsize{$\downarrow$} \\
     \hline
     UNet \cite{ronneberger2015u}              & 47.7 \tiny{(0.6)} & 38.6 \tiny{(0.6)} & 36.1 \tiny{(0.2)} & 69.6 \tiny{(0.1)} & 62.7 \tiny{(0.7)} & 2.1 \tiny{(0.0)} \\
     AttUNet \cite{oktay1804attention}         & 57.5 \tiny{(0.2)} & 48.4 \tiny{(0.2)} & 45.3 \tiny{(1.8)} & 74.5 \tiny{(1.2)} & 66.0 \tiny{(2.3)} & 1.7 \tiny{(0.0)} \\
     UNet++ \cite{zhou2018unet++}              & 65.6 \tiny{(0.7)} & 57.1 \tiny{(0.8)} & 54.4 \tiny{(8.9)} & 78.8 \tiny{(3.3)} & 73.2 \tiny{(5.2)} & 1.3 \tiny{(0.3)} \\
     CENet \cite{gu2019net}                    & 76.3 \tiny{(0.4)} & 69.2 \tiny{(0.5)} & 64.4 \tiny{(0.2)} & 83.2 \tiny{(0.2)} & 76.6 \tiny{(0.3)} & \textcolor{blue}{\textbf{\textit{0.6}}} \tiny{(0.0)} \\
     TransUNet \cite{chen2021transunet}        & 75.6 \tiny{(0.4)} & 68.8 \tiny{(0.2)} & 63.4 \tiny{(0.2)} & 82.7 \tiny{(0.3)} & 75.5 \tiny{(0.1)} & 0.7 \tiny{(0.0)} \\
     FRCUNet \cite{azad2021deep}               & 77.3 \tiny{(0.3)} & 70.4 \tiny{(0.2)} & 66.0 \tiny{(0.4)} & 84.0 \tiny{(0.2)} & 78.4 \tiny{(0.6)} & 0.7 \tiny{(0.0)} \\
     MSRFNet \cite{srivastava2021msrf}         & 75.2 \tiny{(0.4)} & 68.0 \tiny{(0.4)} & 63.4 \tiny{(0.4)} & 82.7 \tiny{(0.2)} & 76.3 \tiny{(0.6)} & 0.8 \tiny{(0.0)} \\
     HiFormer \cite{heidari2023hiformer}       & 72.9 \tiny{(1.4)} & 63.3 \tiny{(1.5)} & 60.2 \tiny{(1.0)} & 80.8 \tiny{(0.8)} & 76.0 \tiny{(1.0)} & 0.8 \tiny{(0.1)} \\
     DCSAUNet \cite{xu2023dcsau}               & 75.3 \tiny{(0.4)} & 68.2 \tiny{(0.4)} & 63.1 \tiny{(0.6)} & 83.0 \tiny{(0.3)} & 77.3 \tiny{(0.5)} & 0.7 \tiny{(0.0)} \\
     M2SNet \cite{zhao2023m}                   & \textcolor{blue}{\textbf{\textit{81.7}}} \tiny{(0.4)} & \textcolor{blue}{\textbf{\textit{74.7}}} \tiny{(0.5)} & \textcolor{blue}{\textbf{\textit{68.3}}} \tiny{(0.7)} & \textcolor{blue}{\textbf{\textit{85.7}}} \tiny{(0.2)} & \textcolor{blue}{\textbf{\textit{80.1}}} \tiny{(0.4)} & \textcolor{blue}{\textbf{\textit{0.6}}} \tiny{(0.0)} \\
     \hline
     \textbf{MADGNet}                         & \textcolor{red}{\textbf{\underline{83.7}}} \tiny{(0.2)} & \textcolor{red}{\textbf{\underline{76.8}}} \tiny{(0.2)} & \textcolor{red}{\textbf{\underline{70.2}}} \tiny{(0.2)} & \textcolor{red}{\textbf{\underline{86.3}}} \tiny{(0.2)} & \textcolor{red}{\textbf{\underline{81.5}}} \tiny{(0.1)} & \textcolor{red}{\textbf{\underline{0.5}}} \tiny{(0.0)} \\
    \hline
     \hline
    \multicolumn{1}{c|}{\multirow{2}{*}{Method}} & \multicolumn{6}{c}{COVID19-1 \cite{ma_jun_2020_3757476} $\Rightarrow$ COVID19-2 \cite{COVID19_2}} \\ \cline{2-7}
     & DSC \scriptsize{$\uparrow$} & mIoU \scriptsize{$\uparrow$} & $F_{\beta}^{w}$ \scriptsize{$\uparrow$}  & $S_{\alpha}$ \scriptsize{$\uparrow$} & $E_{\phi}^{max}$ \scriptsize{$\uparrow$} & MAE \scriptsize{$\downarrow$} \\
     \hline
     UNet \cite{ronneberger2015u}              & 47.1 \tiny{(0.7)} & 37.7 \tiny{(0.6)} & 46.7 \tiny{(0.8)} & 68.7 \tiny{(0.2)} & 68.6 \tiny{(1.0)} & \textcolor{blue}{\textbf{\textit{1.0}}} \tiny{(0.0)} \\
     AttUNet \cite{oktay1804attention}         & 43.7 \tiny{(0.8)} & 35.2 \tiny{(0.8)} & 44.5 \tiny{(0.7)} & 67.9 \tiny{(0.5)} & 64.0 \tiny{(0.6)} & \textcolor{blue}{\textbf{\textit{1.0}}} \tiny{(0.0)} \\
     UNet++ \cite{zhou2018unet++}              & 50.5 \tiny{(3.8)} & 40.9 \tiny{(3.7)} & 50.6 \tiny{(4.6)} & 69.8 \tiny{(1.3)} & 75.7 \tiny{(2.6)} & \textcolor{blue}{\textbf{\textit{1.0}}} \tiny{(0.2)} \\
     CENet \cite{gu2019net}                    & 60.1 \tiny{(0.3)} & 49.9 \tiny{(0.3)} & 61.1 \tiny{(0.3)} & 73.4 \tiny{(0.1)} & 80.1 \tiny{(0.3)} & 1.1 \tiny{(0.0)} \\
     TransUNet \cite{chen2021transunet}        & 56.9 \tiny{(1.0)} & 48.0 \tiny{(0.7)} & 58.0 \tiny{(0.1)} & 72.5 \tiny{(0.2)} & 80.0 \tiny{(1.3)} & \textcolor{red}{\textbf{\underline{0.8}}} \tiny{(0.0)} \\
     FRCUNet \cite{azad2021deep}               & 62.9 \tiny{(1.1)} & 52.7 \tiny{(0.9)} & 63.7 \tiny{(1.2)} & 74.5 \tiny{(0.3)} & 82.1 \tiny{(0.7)} & 1.4 \tiny{(0.2)} \\
     MSRFNet \cite{srivastava2021msrf}         & 58.3 \tiny{(0.8)} & 48.4 \tiny{(0.6)} & 59.1 \tiny{(0.9)} & 72.7 \tiny{(0.2)} & 79.8 \tiny{(0.8)} & \textcolor{blue}{\textbf{\textit{1.0}}} \tiny{(0.1)} \\
     HiFormer \cite{heidari2023hiformer}       & 54.1 \tiny{(1.0)} & 44.5 \tiny{(0.8)} & 55.2 \tiny{(1.0)} & 70.9 \tiny{(0.5)} & 78.0 \tiny{(0.7)} & \textcolor{blue}{\textbf{\textit{1.0}}} \tiny{(0.1)} \\
     DCSAUNet \cite{xu2023dcsau}               & 52.4 \tiny{(1.2)} & 44.0 \tiny{(0.7)} & 52.0 \tiny{(1.2)} & 71.3 \tiny{(0.1)} & 76.3 \tiny{(3.1)} & \textcolor{blue}{\textbf{\textit{1.0}}} \tiny{(0.1)} \\
     M2SNet \cite{zhao2023m}                   & \textcolor{blue}{\textbf{\textit{68.6}}} \tiny{(0.1)} & \textcolor{blue}{\textbf{\textit{58.9}}} \tiny{(0.2)} & \textcolor{blue}{\textbf{\textit{68.5}}} \tiny{(0.2)} & \textcolor{blue}{\textbf{\textit{76.9}}} \tiny{(0.1)} & \textcolor{blue}{\textbf{\textit{86.1}}} \tiny{(0.4)} & 1.1 \tiny{(0.1)} \\
     \hline
     \textbf{MADGNet}                         & \textcolor{red}{\textbf{\underline{72.2}}} \tiny{(0.3)} & \textcolor{red}{\textbf{\underline{62.6}}} \tiny{(0.3)} & \textcolor{red}{\textbf{\underline{72.3}}} \tiny{(0.5)} & \textcolor{red}{\textbf{\underline{78.2}}} \tiny{(0.1)} & \textcolor{red}{\textbf{\underline{88.1}}} \tiny{(0.2)} & \textcolor{blue}{\textbf{\textit{1.0}}} \tiny{(0.1)} \\
    \hline
    \end{tabular}
    \caption{Segmentation results on \textbf{COVID19 Infection Segmentation (Radiology)} \cite{ma_jun_2020_3757476, COVID19_2}. We train each model on COVID19-1 \cite{ma_jun_2020_3757476} train dataset and evaluate on COVID19-1 \cite{ma_jun_2020_3757476} and COVID19-2 \cite{COVID19_2} test datasets.}
    \label{tab:comparison_sota_radiology_other_metrics}
\end{table}

\begin{table}[h]
    \centering
    \scriptsize
    \setlength\tabcolsep{3pt} 
    \begin{tabular}{c|cccccc}
    \hline
    \multicolumn{1}{c|}{\multirow{2}{*}{Method}} & \multicolumn{6}{c}{BUSI \cite{al2020dataset} $\Rightarrow$ BUSI \cite{al2020dataset}} \\ \cline{2-7}
     & DSC \scriptsize{$\uparrow$} & mIoU \scriptsize{$\uparrow$} & $F_{\beta}^{w}$ \scriptsize{$\uparrow$}  & $S_{\alpha}$ \scriptsize{$\uparrow$} & $E_{\phi}^{max}$ \scriptsize{$\uparrow$} & MAE \scriptsize{$\downarrow$} \\
     \hline
     UNet \cite{ronneberger2015u}              & 69.5 \tiny{(0.3)} & 60.2 \tiny{(0.2)} & 67.2 \tiny{(0.3)} & 76.9 \tiny{(0.1)} & 83.2 \tiny{(0.2)} & 4.8 \tiny{(0.0)} \\
     AttUNet \cite{oktay1804attention}         & 71.3 \tiny{(0.4)} & 62.3 \tiny{(0.6)} & 68.9 \tiny{(0.5)} & 78.1 \tiny{(0.3)} & 84.4 \tiny{(0.1)} & 4.8 \tiny{(0.0)} \\
     UNet++ \cite{zhou2018unet++}              & 72.4 \tiny{(0.1)} & 62.5 \tiny{(0.2)} & 68.7 \tiny{(0.3)} & 78.4 \tiny{(0.2)} & 85.0 \tiny{(0.2)} & 5.0 \tiny{(0.1)} \\
     CENet \cite{gu2019net}                    & 79.7 \tiny{(0.6)} & 71.5 \tiny{(0.5)} & 78.1 \tiny{(0.6)} & 82.8 \tiny{(0.3)} & 91.1 \tiny{(0.2)} & 3.9 \tiny{(0.0)} \\
     TransUNet \cite{chen2021transunet}        & 75.5 \tiny{(0.5)} & 68.4 \tiny{(0.1)} & 73.8 \tiny{(0.2)} & 79.8 \tiny{(0.1)} & 88.6 \tiny{(0.6)} & 4.2 \tiny{(0.2)} \\
     FRCUNet \cite{azad2021deep}               & \textcolor{blue}{\textbf{\textit{81.2}}} \tiny{(0.2)} & \textcolor{blue}{\textbf{\textit{73.3}}} \tiny{(0.3)} & \textcolor{red}{\textbf{\underline{79.9}}} \tiny{(0.3)} & \textcolor{blue}{\textbf{\textit{83.5}}} \tiny{(0.2)} & \textcolor{red}{\textbf{\underline{91.9}}} \tiny{(0.1)} & \textcolor{blue}{\textbf{\textit{3.7}}} \tiny{(0.1)} \\
     MSRFNet \cite{srivastava2021msrf}         & 76.6 \tiny{(0.7)} & 68.1 \tiny{(0.7)} & 75.1 \tiny{(0.9)} & 80.9 \tiny{(0.3)} & 88.5 \tiny{(0.4)} & 4.2 \tiny{(0.1)} \\
     HiFormer \cite{heidari2023hiformer}       & 79.3 \tiny{(0.2)} & 70.8 \tiny{(0.1)} & 77.7 \tiny{(0.0)} & 82.3 \tiny{(0.1)} & 90.8 \tiny{(0.3)} & 4.1 \tiny{(0.1)} \\
     DCSAUNet \cite{xu2023dcsau}               & 73.7 \tiny{(0.5)} & 65.0 \tiny{(0.5)} & 71.5 \tiny{(0.4)} & 79.6 \tiny{(0.3)} & 86.0 \tiny{(0.3)} & 4.6 \tiny{(0.1)} \\
     M2SNet \cite{zhao2023m}                   & 80.4 \tiny{(0.8)} & 72.5 \tiny{(0.7)} & 78.7 \tiny{(0.6)} & 83.0 \tiny{(0.5)} & 91.2 \tiny{(0.4)} & 4.1 \tiny{(0.2)} \\
     \hline
     \textbf{MADGNet}                         & \textcolor{red}{\textbf{\underline{81.3}}} \tiny{(0.4)} & \textcolor{red}{\textbf{\underline{73.4}}} \tiny{(0.4)} & \textcolor{blue}{\textbf{\textit{79.5}}} \tiny{(0.4)} & \textcolor{red}{\textbf{\underline{83.8}}} \tiny{(0.2)} & \textcolor{blue}{\textbf{\textit{91.7}}} \tiny{(0.3)} & \textcolor{red}{\textbf{\underline{3.6}}} \tiny{(0.1)} \\
    \hline
    \hline
    \multicolumn{1}{c|}{\multirow{2}{*}{Method}} & \multicolumn{6}{c}{BUSI \cite{al2020dataset} $\Rightarrow$ STU \cite{zhuang2019rdau}} \\ \cline{2-7}
    & DSC \scriptsize{$\uparrow$} & mIoU \scriptsize{$\uparrow$} & $F_{\beta}^{w}$ \scriptsize{$\uparrow$}  & $S_{\alpha}$ \scriptsize{$\uparrow$} & $E_{\phi}^{max}$ \scriptsize{$\uparrow$} & MAE \scriptsize{$\downarrow$} \\
    \hline
     UNet \cite{ronneberger2015u}              & 71.6 \tiny{(1.0)} & 61.6 \tiny{(0.7)} & 71.6 \tiny{(0.8)} & 76.1 \tiny{(0.4)} & 82.4 \tiny{(0.9)} & 5.2 \tiny{(0.2)} \\
     AttUNet \cite{oktay1804attention}         & 77.0 \tiny{(1.6)} & 68.0 \tiny{(1.7)} & 76.4 \tiny{(1.2)} & 79.8 \tiny{(1.0)} & 86.7 \tiny{(1.4)} & 4.4 \tiny{(0.3)} \\
     UNet++ \cite{zhou2018unet++}              & 77.3 \tiny{(0.4)} & 67.8 \tiny{(0.3)} & 76.1 \tiny{(0.3)} & 79.4 \tiny{(0.5)} & 87.6 \tiny{(0.3)} & 4.4 \tiny{(0.1)} \\
     CENet \cite{gu2019net}                    & 86.0 \tiny{(0.7)} & \textcolor{blue}{\textbf{\textit{77.2}}} \tiny{(0.9)} & 84.2 \tiny{(0.6)} & 84.6 \tiny{(0.4)} & 93.7 \tiny{(0.4)} & \textcolor{blue}{\textbf{\textit{2.8}}} \tiny{(0.2)} \\
     TransUNet \cite{chen2021transunet}        & 41.4 \tiny{(9.5)} & 32.1 \tiny{(4.2)} & 40.8 \tiny{(8.7)} & 60.2 \tiny{(4.3)} & 58.1 \tiny{(8.4)} & 9.7 \tiny{(0.7)} \\
     FRCUNet \cite{azad2021deep}               & \textcolor{blue}{\textbf{\textit{86.5}}} \tiny{(2.3)} & \textcolor{blue}{\textbf{\textit{77.2}}} \tiny{(2.7)} & \textcolor{blue}{\textbf{\textit{84.9}}} \tiny{(2.1)} & \textcolor{blue}{\textbf{\textit{85.2}}} \tiny{(1.6)} & 94.1 \tiny{(2.0)} & \textcolor{blue}{\textbf{\textit{2.8}}} \tiny{(0.5)} \\
     MSRFNet \cite{srivastava2021msrf}         & 84.0 \tiny{(5.5)} & 75.2 \tiny{(8.2)} & 82.5 \tiny{(5.0)} & 83.5 \tiny{(5.7)} & 92.2 \tiny{(3.3)} & 3.1 \tiny{(0.2)} \\
     HiFormer \cite{heidari2023hiformer}       & 80.7 \tiny{(2.9)} & 71.3 \tiny{(3.2)} & 78.9 \tiny{(3.0)} & 81.2 \tiny{(1.6)} & 90.1 \tiny{(2.2)} & 3.7 \tiny{(0.5)} \\
     DCSAUNet \cite{xu2023dcsau}               & 86.1 \tiny{(0.5)} & 76.5 \tiny{(0.8)} & 82.7 \tiny{(0.8)} & 84.9 \tiny{(0.4)} & \textcolor{blue}{\textbf{\textit{94.7}}}\tiny{(0.5)} & 3.2 \tiny{(0.1)} \\
     M2SNet \cite{zhao2023m}                   & 79.4 \tiny{(0.7)} & 69.3 \tiny{(0.6)} & 76.4 \tiny{(0.8)} & 81.3 \tiny{(0.4)} & 90.7 \tiny{(0.9)} & 4.3 \tiny{(0.2)} \\
     \hline
     \textbf{MADGNet}                         & \textcolor{red}{\textbf{\underline{88.4}}} \tiny{(1.0)} & \textcolor{red}{\textbf{\underline{79.9}}} \tiny{(1.5)} & \textcolor{red}{\textbf{\underline{86.4}}} \tiny{(1.5)} & \textcolor{red}{\textbf{\underline{86.2}}} \tiny{(0.9)} & \textcolor{red}{\textbf{\underline{95.9}}} \tiny{(0.5)} & \textcolor{red}{\textbf{\underline{2.6}}} \tiny{(0.4)} \\
    \hline
    \end{tabular}
    \caption{Segmentation results on \textbf{Breast Tumor Segmentation (Ultrasound)} \cite{al2020dataset, zhuang2019rdau}. We train each model on BUSI \cite{al2020dataset} train dataset and evaluate on BUSI \cite{al2020dataset} and STU \cite{zhuang2019rdau} test datasets.}
    \label{tab:comparison_sota_ultrasound_other_metrics}
\end{table}

\begin{table}[h]
    \centering
    \scriptsize
    \setlength\tabcolsep{3pt} 
    \begin{tabular}{c|cccccc}
    \hline
    \multicolumn{1}{c|}{\multirow{2}{*}{Method}} & \multicolumn{6}{c}{DSB2018 \cite{caicedo2019nucleus} $\Rightarrow$ DSB2018 \cite{caicedo2019nucleus}} \\ \cline{2-7}
     & DSC \scriptsize{$\uparrow$} & mIoU \scriptsize{$\uparrow$} & $F_{\beta}^{w}$ \scriptsize{$\uparrow$}  & $S_{\alpha}$ \scriptsize{$\uparrow$} & $E_{\phi}^{max}$ \scriptsize{$\uparrow$} & MAE \scriptsize{$\downarrow$} \\
     \hline
     UNet \cite{ronneberger2015u}              & 91.1 \tiny{(0.2)} & 84.3 \tiny{(0.3)} & 92.1 \tiny{(0.1)} & 83.3 \tiny{(0.0)} & 96.8 \tiny{(0.0)} & 2.5 \tiny{(0.0)} \\
     AttUNet \cite{oktay1804attention}         & 91.6 \tiny{(0.1)} & 85.0 \tiny{(0.1)} & 92.5 \tiny{(0.0)} & \textcolor{blue}{\textbf{\textit{83.7}}} \tiny{(0.0)} & 97.2 \tiny{(0.0)} & 2.4 \tiny{(0.0)} \\
     UNet++ \cite{zhou2018unet++}              & 91.6 \tiny{(0.1)} & 85.0 \tiny{(0.1)} & 92.8 \tiny{(0.1)} & 83.6 \tiny{(0.0)} & 97.1 \tiny{(0.0)} & 2.4 \tiny{(0.0)} \\
     CENet \cite{gu2019net}                    & 91.3 \tiny{(0.1)} & 84.6 \tiny{(0.1)} & 92.5 \tiny{(0.1)} & 83.6 \tiny{(0.1)} & 97.2 \tiny{(0.1)} & \textcolor{blue}{\textbf{\textit{2.3}}} \tiny{(0.0)} \\
     TransUNet \cite{chen2021transunet}        & 91.8 \tiny{(0.3)} & 85.2 \tiny{(0.2)} & \textcolor{red}{\textbf{\underline{92.8}}} \tiny{(0.1)} & \textcolor{red}{\textbf{\underline{83.8}}} \tiny{(0.2)} & 97.3 \tiny{(0.2)} & \textcolor{blue}{\textbf{\textit{2.3}}} \tiny{(0.1)} \\
     FRCUNet \cite{azad2021deep}               & 90.8 \tiny{(0.3)} & 83.8 \tiny{(0.4)} & 92.1 \tiny{(0.2)} & 83.2 \tiny{(0.2)} & 97.0 \tiny{(0.2)} & 2.5 \tiny{(0.1)} \\
     MSRFNet \cite{srivastava2021msrf}         & \textcolor{blue}{\textbf{\textit{91.9}}} \tiny{(0.1)} & \textcolor{blue}{\textbf{\textit{85.3}}} \tiny{(0.1)} & \textcolor{blue}{\textbf{\textit{92.7}}} \tiny{(0.1)} & \textcolor{blue}{\textbf{\textit{83.7}}} \tiny{(0.1)} & \textcolor{red}{\textbf{\underline{97.5}}} \tiny{(0.0)} & \textcolor{blue}{\textbf{\textit{2.3}}} \tiny{(0.0)} \\
     HiFormer \cite{heidari2023hiformer}       & 90.7 \tiny{(0.2)} & 83.8 \tiny{(0.4)} & 91.1 \tiny{(1.1)} & 82.5 \tiny{(0.8)} & 96.0 \tiny{(1.1)} & 2.7 \tiny{(0.4)} \\
     DCSAUNet \cite{xu2023dcsau}               & 91.1 \tiny{(0.2)} & 84.4 \tiny{(0.2)} & 91.8 \tiny{(0.2)} & 82.9 \tiny{(0.3)} & 96.6 \tiny{(0.2)} & 2.8 \tiny{(0.2)} \\
     M2SNet \cite{zhao2023m}                   & 91.6 \tiny{(0.2)} & 85.1 \tiny{(0.3)} & 92.0 \tiny{(0.2)} & 83.5 \tiny{(0.1)} & \textcolor{red}{\textbf{\underline{97.5}}} \tiny{(0.1)} & \textcolor{red}{\textbf{\underline{2.2}}} \tiny{(0.1)} \\
     \hline
     \textbf{MADGNet}                         & \textcolor{red}{\textbf{\underline{92.0}}} \tiny{(0.0)} & \textcolor{red}{\textbf{\underline{85.5}}} \tiny{(0.1)} & 92.3 \tiny{(0.4)} & \textcolor{red}{\textbf{\underline{83.8}}} \tiny{(0.2)} & \textcolor{blue}{\textbf{\textit{97.4}}} \tiny{(0.3)} & \textcolor{blue}{\textbf{\textit{2.3}}} \tiny{(0.1)}  \\
    \hline
    \hline
    \multicolumn{1}{c|}{\multirow{2}{*}{Method}} & \multicolumn{6}{c}{DSB2018 \cite{caicedo2019nucleus} $\Rightarrow$ MonuSeg2018 \cite{dinh2021breast}} \\ \cline{2-7}
     & DSC \scriptsize{$\uparrow$} & mIoU \scriptsize{$\uparrow$} & $F_{\beta}^{w}$ \scriptsize{$\uparrow$}  & $S_{\alpha}$ \scriptsize{$\uparrow$} & $E_{\phi}^{max}$ \scriptsize{$\uparrow$} & MAE \scriptsize{$\downarrow$} \\
     \hline
     UNet \cite{ronneberger2015u}              & 29.2 \tiny{(5.1)} & 18.9 \tiny{(3.5)} & 28.0 \tiny{(5.2)} & 38.3 \tiny{(1.8)} & 49.9 \tiny{(6.0)} & 32.5 \tiny{(2.6)} \\
     AttUNet \cite{oktay1804attention}         & \textcolor{blue}{\textbf{\textit{39.0}}} \tiny{(3.1)} & \textcolor{blue}{\textbf{\textit{26.5}}} \tiny{(2.4)} & \textcolor{blue}{\textbf{\textit{40.9}}} \tiny{(2.2)} & \textcolor{blue}{\textbf{\textit{39.3}}} \tiny{(2.9)} & \textcolor{red}{\textbf{\underline{61.4}}} \tiny{(5.0)} & 24.5 \tiny{(4.6)} \\
     UNet++ \cite{zhou2018unet++}              & 25.4 \tiny{(0.8)} & 15.3 \tiny{(0.5)} & 21.1 \tiny{(0.4)} & 5.5 \tiny{(1.7)} & 21.1 \tiny{(3.1)} & 66.6 \tiny{(4.5)} \\
     CENet \cite{gu2019net}                    & 27.7 \tiny{(1.5)} & 16.9 \tiny{(1.0)} & 26.7 \tiny{(1.1)} & 37.7 \tiny{(1.9)} & 59.6 \tiny{(4.9)} & 27.5 \tiny{(5.1)} \\
     TransUNet \cite{chen2021transunet}        & 15.9 \tiny{(8.5)} & 9.6 \tiny{(5.5)} & 17.0 \tiny{(6.2)} & 32.6 \tiny{(4.1)} & 39.0 \tiny{(5.5)} & 23.4 \tiny{(7.6)} \\
     FRCUNet \cite{azad2021deep}               & 26.1 \tiny{(5.6)} & 16.8 \tiny{(4.3)} & 26.9 \tiny{(7.7)} & 40.6 \tiny{(6.7)} & 49.3 \tiny{(9.7)} & \textcolor{blue}{\textbf{\textit{22.9}}} \tiny{(10.5)} \\
     MSRFNet \cite{srivastava2021msrf}         & 9.1 \tiny{(1.0)} & 5.3 \tiny{(0.7)} & 12.0 \tiny{(0.9)} & 35.6 \tiny{(1.6)} & 46.4 \tiny{(0.4)} & \textcolor{red}{\textbf{\underline{18.7}}} \tiny{(0.4)} \\
     HiFormer \cite{heidari2023hiformer}       & 21.9 \tiny{(8.9)} & 13.2 \tiny{(5.7)} & 22.5 \tiny{(6.7)} & 39.9 \tiny{(1.4)} & 48.2 \tiny{(0.3)} & 23.9 \tiny{(9.1)} \\
     DCSAUNet \cite{xu2023dcsau}               & 4.3 \tiny{(0.3)} & 2.4 \tiny{(0.9)} & 5.2 \tiny{(3.2)} & 18.0 \tiny{(7.3)} & 37.0 \tiny{(1.2)} & 25.9 \tiny{(5.9)} \\
     M2SNet \cite{zhao2023m}                   & 36.3 \tiny{(0.9)} & 23.1 \tiny{(0.8)} & 29.1 \tiny{(1.6)} & 20.4 \tiny{(1.5)} & 34.5 \tiny{(6.8)} & 45.1 \tiny{(5.3)} \\
     \hline
     \textbf{MADGNet}                         & \textcolor{red}{\textbf{\underline{46.7}}} \tiny{(4.3)} & \textcolor{red}{\textbf{\underline{32.0}}} \tiny{(2.9)} & \textcolor{red}{\textbf{\underline{43.0}}} \tiny{(4.6)} & \textcolor{red}{\textbf{\underline{40.8}}} \tiny{(2.6)} & \textcolor{blue}{\textbf{\textit{60.7}}} \tiny{(6.3)} & 29.2 \tiny{(5.3)}  \\
    \hline
    \end{tabular}
    \caption{Segmentation results on \textbf{Cell Segmentation (Microscopy)} \cite{caicedo2019nucleus, dinh2021breast}. We train each model on DSB2018 \cite{caicedo2019nucleus} train dataset and evaluate on DSB2018 \cite{caicedo2019nucleus} and MonuSeg2018 \cite{dinh2021breast} test datasets.}
    \label{tab:comparison_sota_microscopy_other_metrics}
\end{table}

\begin{table}[h]
    \centering
    \scriptsize
    \setlength\tabcolsep{3pt} 
    \begin{tabular}{c|cccccc}
    \hline
    \multicolumn{1}{c|}{\multirow{2}{*}{Method}} & \multicolumn{6}{c}{CVC-ClinicDB \cite{bernal2015wm} + Kvasir-SEG \cite{jha2020kvasir} $\rightarrow$ CVC-ClinicDB \cite{bernal2015wm}} \\ \cline{2-7}
     & DSC \scriptsize{$\uparrow$} & mIoU \scriptsize{$\uparrow$} & $F_{\beta}^{w}$ \scriptsize{$\uparrow$}  & $S_{\alpha}$ \scriptsize{$\uparrow$} & $E_{\phi}^{max}$ \scriptsize{$\uparrow$} & MAE \scriptsize{$\downarrow$} \\
     \hline
     UNet \cite{ronneberger2015u}              & 76.5 \tiny{(0.8)} & 69.1 \tiny{(0.9)} & 75.1 \tiny{(0.8)} & 83.0 \tiny{(0.4)} & 86.4 \tiny{(0.6)} & 2.7 \tiny{(0.0)} \\
     AttUNet \cite{oktay1804attention}         & 80.1 \tiny{(0.6)} & 74.2 \tiny{(0.5)} & 79.8 \tiny{(0.7)} & 85.1 \tiny{(0.4)} & 88.5 \tiny{(0.5)} & 2.1 \tiny{(0.1)} \\
     UNet++ \cite{zhou2018unet++}              & 79.7 \tiny{(0.2)} & 73.6 \tiny{(0.4)} & 79.4 \tiny{(0.1)} & 85.1 \tiny{(0.2)} & 88.3 \tiny{(0.5)} & 2.2 \tiny{(0.0)} \\
     CENet \cite{gu2019net}                    & 89.3 \tiny{(0.3)} & 84.0 \tiny{(0.2)} & 89.1 \tiny{(0.2)} & 89.8 \tiny{(0.2)} & 96.0 \tiny{(0.6)} & \textcolor{blue}{\textbf{\textit{1.1}}} \tiny{(0.0)} \\
     TransUNet \cite{chen2021transunet}        & 87.4 \tiny{(0.2)} & 82.9 \tiny{(0.1)} & 87.2 \tiny{(0.1)} & 88.5 \tiny{(0.2)} & 95.2 \tiny{(0.1)} & 1.3 \tiny{(0.0)} \\
     FRCUNet \cite{azad2021deep}               & 91.8 \tiny{(0.2)} & 87.0 \tiny{(0.2)} & 91.3 \tiny{(0.3)} & 91.1 \tiny{(0.1)} & 97.1 \tiny{(0.3)} & 0.7 \tiny{(0.0)} \\
     MSRFNet \cite{srivastava2021msrf}         & 83.2 \tiny{(0.9)} & 76.5 \tiny{(1.1)} & 81.9 \tiny{(1.2)} & 86.4 \tiny{(0.5)} & 91.3 \tiny{(1.0)} & 1.7 \tiny{(0.0)} \\
     HiFormer \cite{heidari2023hiformer}       & 89.1 \tiny{(0.6)} & 83.7 \tiny{(0.6)} & 88.8 \tiny{(0.5)} & 89.5 \tiny{(0.2)} & 96.1 \tiny{(0.8)} & \textcolor{blue}{\textbf{\textit{1.1}}} \tiny{(0.2)} \\
     DCSAUNet \cite{xu2023dcsau}               & 80.5 \tiny{(1.2)} & 73.7 \tiny{(1.1)} & 79.6 \tiny{(1.1)} & 84.9 \tiny{(0.6)} & 89.9 \tiny{(1.0)} & 2.4 \tiny{(0.2)} \\
     M2SNet \cite{zhao2023m}                   & \textcolor{blue}{\textbf{\textit{92.8}}} \tiny{(0.8)} & \textcolor{blue}{\textbf{\textit{88.2}}} \tiny{(0.8)} & \textcolor{blue}{\textbf{\textit{92.3}}} \tiny{(0.7)} & \textcolor{blue}{\textbf{\textit{91.4}}} \tiny{(0.4)} & \textcolor{blue}{\textbf{\textit{97.7}}} \tiny{(0.5)} & \textcolor{red}{\textbf{\underline{0.7}}} \tiny{(0.1)} \\
     \hline
     \textbf{MADGNet}                         & \textcolor{red}{\textbf{\underline{93.9}}} \tiny{(0.6)} & \textcolor{red}{\textbf{\underline{89.5}}} \tiny{(0.5)} & \textcolor{red}{\textbf{\underline{93.6}}} \tiny{(0.6)} & \textcolor{red}{\textbf{\underline{92.2}}} \tiny{(0.2)} & \textcolor{red}{\textbf{\underline{98.5}}} \tiny{(0.7)} & \textcolor{red}{\textbf{\underline{0.7}}} \tiny{(0.0)} \\
    \hline
    \hline
    \multicolumn{1}{c|}{\multirow{2}{*}{Method}} & \multicolumn{6}{c}{CVC-ClinicDB \cite{bernal2015wm} + Kvasir-SEG \cite{jha2020kvasir} $\rightarrow$ Kvasir-SEG \cite{jha2020kvasir}} \\ \cline{2-7}
    & DSC \scriptsize{$\uparrow$} & mIoU \scriptsize{$\uparrow$} & $F_{\beta}^{w}$ \scriptsize{$\uparrow$}  & $S_{\alpha}$ \scriptsize{$\uparrow$} & $E_{\phi}^{max}$ \scriptsize{$\uparrow$} & MAE \scriptsize{$\downarrow$} \\
    \hline
     UNet \cite{ronneberger2015u}              & 80.5 \tiny{(0.3)} & 72.6 \tiny{(0.4)} & 78.2 \tiny{(0.4)} & 79.9 \tiny{(0.2)} & 88.2 \tiny{(0.2)} & 5.2 \tiny{(0.2)} \\
     AttUNet \cite{oktay1804attention}         & 83.9 \tiny{(0.1)} & 77.1 \tiny{(0.1)} & 83.1 \tiny{(0.0)} & 81.9 \tiny{(0.0)} & 90.0 \tiny{(0.1)} & 4.4 \tiny{(0.1)} \\
     UNet++ \cite{zhou2018unet++}              & 84.3 \tiny{(0.3)} & 77.4 \tiny{(0.2)} & 83.1 \tiny{(0.3)} & 82.1 \tiny{(0.1)} & 90.5 \tiny{(0.2)} & 4.6 \tiny{(0.1)} \\
     CENet \cite{gu2019net}                    & 89.5 \tiny{(0.7)} & 83.9 \tiny{(0.7)} & 88.9 \tiny{(0.7)} & \textcolor{blue}{\textbf{\textit{85.3}}} \tiny{(0.3)} & 94.1 \tiny{(0.4)} & \textcolor{blue}{\textbf{\textit{3.0}}} \tiny{(0.2)} \\
     TransUNet \cite{chen2021transunet}        & 86.4 \tiny{(0.4)} & 81.3 \tiny{(0.4)} & 85.4 \tiny{(0.4)} & 83.0 \tiny{(0.4)} & 92.1 \tiny{(0.5)} & 4.0 \tiny{(0.3)} \\
     FRCUNet \cite{azad2021deep}               & 88.8 \tiny{(0.4)} & 83.5 \tiny{(0.6)} & 88.4 \tiny{(0.6)} & 85.1 \tiny{(0.2)} & 93.6 \tiny{(0.4)} & 3.3 \tiny{(0.1)} \\
     MSRFNet \cite{srivastava2021msrf}         & 86.1 \tiny{(0.5)} & 79.3 \tiny{(0.4)} & 84.9 \tiny{(0.7)} & 82.8 \tiny{(0.1)} & 92.0 \tiny{(0.4)} & 4.0 \tiny{(0.1)} \\
     HiFormer \cite{heidari2023hiformer}       & 88.1 \tiny{(1.0)} & 82.3 \tiny{(1.2)} & 87.3 \tiny{(1.1)} & 84.6 \tiny{(0.5)} & 93.9 \tiny{(0.6)} & 3.1 \tiny{(0.3)} \\
     DCSAUNet \cite{xu2023dcsau}               & 82.6 \tiny{(0.5)} & 75.2 \tiny{(0.5)} & 80.7 \tiny{(0.3)} & 81.3 \tiny{(0.7)} & 90.1 \tiny{(0.1)} & 4.9 \tiny{(0.2)} \\
     M2SNet \cite{zhao2023m}                   & \textcolor{blue}{\textbf{\textit{90.2}}} \tiny{(0.5)} & \textcolor{blue}{\textbf{\textit{85.1}}} \tiny{(0.6)} & \textcolor{blue}{\textbf{\textit{89.4}}} \tiny{(0.8)} & \textcolor{red}{\textbf{\underline{85.6}}} \tiny{(0.5)} & \textcolor{blue}{\textbf{\textit{94.6}}} \tiny{(0.7)} & \textcolor{red}{\textbf{\underline{2.8}}} \tiny{(0.1)} \\
     \hline
     \textbf{MADGNet}                         & \textcolor{red}{\textbf{\underline{90.7}}} \tiny{(0.8)} & \textcolor{red}{\textbf{\underline{85.3}}} \tiny{(0.8)} & \textcolor{red}{\textbf{\underline{89.9}}} \tiny{(0.8)} & \textcolor{red}{\textbf{\underline{85.6}}} \tiny{(0.5)} & \textcolor{red}{\textbf{\underline{94.7}}} \tiny{(1.0)} & 3.1 \tiny{(0.2)} \\
    \hline
    \hline
    \multicolumn{1}{c|}{\multirow{2}{*}{Method}} & \multicolumn{6}{c}{CVC-ClinicDB \cite{bernal2015wm} + Kvasir-SEG \cite{jha2020kvasir} $\rightarrow$ CVC-300 \cite{vazquez2017benchmark}} \\ \cline{2-7}
    & DSC \scriptsize{$\uparrow$} & mIoU \scriptsize{$\uparrow$} & $F_{\beta}^{w}$ \scriptsize{$\uparrow$}  & $S_{\alpha}$ \scriptsize{$\uparrow$} & $E_{\phi}^{max}$ \scriptsize{$\uparrow$} & MAE \scriptsize{$\downarrow$} \\
    \hline
     UNet \cite{ronneberger2015u}              & 66.1 \tiny{(2.3)} & 58.5 \tiny{(2.1)} & 65.0 \tiny{(2.2)} & 79.7 \tiny{(1.0)} & 80.0 \tiny{(2.4)} & 1.7 \tiny{(0.1)} \\
     AttUNet \cite{oktay1804attention}         & 63.0 \tiny{(0.3)} & 57.2 \tiny{(0.4)} & 62.4 \tiny{(0.4)} & 79.1 \tiny{(0.3)} & 76.6 \tiny{(0.9)} & 1.8 \tiny{(0.0)} \\
     UNet++ \cite{zhou2018unet++}              & 64.3 \tiny{(2.2)} & 58.4 \tiny{(2.0)} & 63.7 \tiny{(2.3)} & 79.5 \tiny{(1.1)} & 77.4 \tiny{(1.5)} & 1.8 \tiny{(0.1)} \\
     CENet \cite{gu2019net}                    & 85.4 \tiny{(1.6)} & 78.2 \tiny{(1.4)} & 84.2 \tiny{(1.8)} & 90.2 \tiny{(0.5)} & 94.0 \tiny{(1.5)} & \textcolor{blue}{\textbf{\textit{0.8}}} \tiny{(0.1)} \\
     TransUNet \cite{chen2021transunet}        & 85.0 \tiny{(0.6)} & 77.3 \tiny{(0.3)} & 83.1 \tiny{(0.7)} & 89.4 \tiny{(0.3)} & \textcolor{blue}{\textbf{\textit{95.2}}} \tiny{(0.7)} & 1.1 \tiny{(0.1)} \\
     FRCUNet \cite{azad2021deep}               & 86.7 \tiny{(0.7)} & 79.4 \tiny{(0.3)} & 85.1 \tiny{(0.2)} & 90.5 \tiny{(0.3)} & 95.0 \tiny{(0.2)} & 0.8 \tiny{(0.1)} \\
     MSRFNet \cite{srivastava2021msrf}         & 72.3 \tiny{(2.2)} & 65.4 \tiny{(2.2)} & 71.2 \tiny{(2.0)} & 83.5 \tiny{(1.6)} & 84.6 \tiny{(1.7)} & 1.4 \tiny{(0.1)} \\
     HiFormer \cite{heidari2023hiformer}       & 84.7 \tiny{(1.1)} & 77.5 \tiny{(1.1)} & 83.2 \tiny{(0.7)} & 89.7 \tiny{(0.6)} & 94.0 \tiny{(1.3)} & \textcolor{blue}{\textbf{\textit{0.8}}} \tiny{(0.3)} \\
     DCSAUNet \cite{xu2023dcsau}               & 68.9 \tiny{(4.0)} & 59.8 \tiny{(3.9)} & 66.3 \tiny{(3.8)} & 81.1 \tiny{(2.1)} & 83.8 \tiny{(2.9)} & 2.0 \tiny{(0.3)} \\
     M2SNet \cite{zhao2023m}                   & \textcolor{red}{\textbf{\underline{89.9}}} \tiny{(0.2)} & \textcolor{red}{\textbf{\underline{83.2}}} \tiny{(0.3)} & \textcolor{red}{\textbf{\underline{88.3}}} \tiny{(0.2)} & \textcolor{red}{\textbf{\underline{93.0}}} \tiny{(0.2)} & \textcolor{red}{\textbf{\underline{96.9}}} \tiny{(0.2)} & \textcolor{red}{\textbf{\underline{0.6}}} \tiny{(0.0)} \\
     \hline
     \textbf{MADGNet}                         & \textcolor{blue}{\textbf{\textit{87.4}}} \tiny{(0.4)} & \textcolor{blue}{\textbf{\textit{79.9}}} \tiny{(0.4)} & \textcolor{blue}{\textbf{\textit{84.5}}} \tiny{(0.5)} & \textcolor{blue}{\textbf{\textit{92.0}}} \tiny{(0.2)} & 94.7 \tiny{(0.5)} & 0.9 \tiny{(0.1)} \\
    \hline
    \hline
    \multicolumn{1}{c|}{\multirow{2}{*}{Method}} & \multicolumn{6}{c}{CVC-ClinicDB \cite{bernal2015wm} + Kvasir-SEG \cite{jha2020kvasir} $\rightarrow$ CVC-ColonDB \cite{tajbakhsh2015automated}} \\ \cline{2-7}
    & DSC \scriptsize{$\uparrow$} & mIoU \scriptsize{$\uparrow$} & $F_{\beta}^{w}$ \scriptsize{$\uparrow$}  & $S_{\alpha}$ \scriptsize{$\uparrow$} & $E_{\phi}^{max}$ \scriptsize{$\uparrow$} & MAE \scriptsize{$\downarrow$} \\
    \hline
     UNet \cite{ronneberger2015u}              & 56.8 \tiny{(1.3)} & 49.0 \tiny{(1.2)} & 55.9 \tiny{(1.2)} & 72.6 \tiny{(0.6)} & 73.9 \tiny{(1.6)} & 5.1 \tiny{(0.1)} \\
     AttUNet \cite{oktay1804attention}         & 56.8 \tiny{(1.6)} & 50.0 \tiny{(1.5)} & 56.2 \tiny{(1.7)} & 73.0 \tiny{(0.7)} & 72.3 \tiny{(1.3)} & 4.9 \tiny{(0.1)} \\
     UNet++ \cite{zhou2018unet++}              & 57.5 \tiny{(0.4)} & 50.2 \tiny{(0.4)} & 56.6 \tiny{(0.3)} & 73.3 \tiny{(0.3)} & 73.9 \tiny{(0.5)} & 5.0 \tiny{(0.1)} \\
     CENet \cite{gu2019net}                    & 65.9 \tiny{(1.6)} & 59.2 \tiny{(0.1)} & 65.8 \tiny{(0.1)} & 77.7 \tiny{(0.1)} & 79.5 \tiny{(0.4)} & 4.0 \tiny{(0.2)} \\
     TransUNet \cite{chen2021transunet}        & 63.7 \tiny{(0.1)} & 58.4 \tiny{(0.4)} & 62.8 \tiny{(0.9)} & 75.8 \tiny{(0.4)} & 79.3 \tiny{(1.6)} & 4.8 \tiny{(0.0)} \\
     FRCUNet \cite{azad2021deep}               & 69.1 \tiny{(1.0)} & 62.6 \tiny{(0.9)} & 68.5 \tiny{(1.0)} & 79.3 \tiny{(0.6)} & 81.4 \tiny{(0.6)} & 4.0 \tiny{(0.1)} \\
     MSRFNet \cite{srivastava2021msrf}         & 61.5 \tiny{(1.0)} & 54.8 \tiny{(0.8)} & 60.8 \tiny{(0.8)} & 75.4 \tiny{(0.5)} & 76.1 \tiny{(0.9)} & 4.5 \tiny{(0.1)} \\
     HiFormer \cite{heidari2023hiformer}       & 67.6 \tiny{(1.4)} & 60.5 \tiny{(1.3)} & 66.9 \tiny{(1.4)} & 78.6 \tiny{(0.7)} & 81.2 \tiny{(1.4)} & 4.2 \tiny{(0.0)} \\
     DCSAUNet \cite{xu2023dcsau}               & 57.8 \tiny{(0.4)} & 49.3 \tiny{(0.4)} & 54.9 \tiny{(0.6)} & 73.3 \tiny{(0.3)} & 76.0 \tiny{(1.3)} & 5.8 \tiny{(0.3)} \\
     M2SNet \cite{zhao2023m}                   & \textcolor{blue}{\textbf{\textit{75.8}}} \tiny{(0.7)} & \textcolor{blue}{\textbf{\textit{68.5}}} \tiny{(0.5)} & \textcolor{blue}{\textbf{\textit{73.7}}} \tiny{(0.7)} & \textcolor{red}{\textbf{\underline{84.2}}} \tiny{(0.3)} & \textcolor{blue}{\textbf{\textit{86.9}}} \tiny{(0.1)} & \textcolor{blue}{\textbf{\textit{3.8}}} \tiny{(0.1)} \\
     \hline
     \textbf{MADGNet}                         & \textcolor{red}{\textbf{\underline{77.5}}} \tiny{(1.1)} & \textcolor{red}{\textbf{\underline{69.7}}} \tiny{(1.2)} & \textcolor{red}{\textbf{\underline{76.2}}} \tiny{(1.2)} & \textcolor{blue}{\textbf{\textit{83.3}}} \tiny{(0.8)} & \textcolor{red}{\textbf{\underline{88.0}}} \tiny{(1.0)} & \textcolor{red}{\textbf{\underline{3.2}}} \tiny{(0.2)} \\
     \hline
    \hline
    \multicolumn{1}{c|}{\multirow{2}{*}{Method}} & \multicolumn{6}{c}{CVC-ClinicDB \cite{bernal2015wm} + Kvasir-SEG \cite{jha2020kvasir} $\rightarrow$ ETIS \cite{silva2014toward}} \\ \cline{2-7}
    & DSC \scriptsize{$\uparrow$} & mIoU \scriptsize{$\uparrow$} & $F_{\beta}^{w}$ \scriptsize{$\uparrow$}  & $S_{\alpha}$ \scriptsize{$\uparrow$} & $E_{\phi}^{max}$ \scriptsize{$\uparrow$} & MAE \scriptsize{$\downarrow$} \\
    \hline
     UNet \cite{ronneberger2015u}              & 41.6 \tiny{(1.1)} & 35.4 \tiny{(1.0)} & 39.5 \tiny{(1.0)} & 67.2 \tiny{(0.6)} & 61.7 \tiny{(0.2)} & 2.7 \tiny{(0.2)} \\
     AttUNet \cite{oktay1804attention}         & 38.4 \tiny{(0.3)} & 33.5 \tiny{(0.1)} & 37.6 \tiny{(0.4)} & 65.4 \tiny{(0.2)} & 59.7 \tiny{(1.2)} & 2.6 \tiny{(0.1)} \\
     UNet++ \cite{zhou2018unet++}              & 39.1 \tiny{(2.4)} & 34.0 \tiny{(2.1)} & 38.3 \tiny{(2.4)} & 65.8 \tiny{(1.0)} & 59.3 \tiny{(1.9)} & 2.7 \tiny{(0.1)} \\
     CENet \cite{gu2019net}                    & 57.0 \tiny{(3.4)} & 51.4 \tiny{(0.5)} & 56.0 \tiny{(0.4)} & 74.9 \tiny{(0.1)} & 73.7 \tiny{(0.4)} & 2.2 \tiny{(0.2)} \\
     TransUNet \cite{chen2021transunet}        & 50.1 \tiny{(0.5)} & 44.0 \tiny{(2.3)} & 48.8 \tiny{(1.8)} & 70.7 \tiny{(1.3)} & 68.7 \tiny{(2.0)} & 2.6 \tiny{(0.1)} \\
     FRCUNet \cite{azad2021deep}               & 65.1 \tiny{(1.0)} & 58.4 \tiny{(0.5)} & 62.9 \tiny{(0.6)} & 78.7 \tiny{(0.4)} & 81.0 \tiny{(1.3)} & 2.2 \tiny{(0.3)} \\
     MSRFNet \cite{srivastava2021msrf}         & 38.3 \tiny{(0.6)} & 33.7 \tiny{(0.7)} & 36.9 \tiny{(0.6)} & 66.0 \tiny{(0.2)} & 58.4 \tiny{(0.9)} & 3.6 \tiny{(0.5)} \\
     HiFormer \cite{heidari2023hiformer}       & 56.7 \tiny{(3.2)} & 50.1 \tiny{(3.3)} & 55.2 \tiny{(3.0)} & 74.1 \tiny{(1.7)} & 74.7 \tiny{(2.5)} & 1.8 \tiny{(0.1)} \\
     DCSAUNet \cite{xu2023dcsau}               & 42.9 \tiny{(3.0)} & 36.1 \tiny{(2.9)} & 40.5 \tiny{(3.5)} & 67.9 \tiny{(1.4)} & 69.3 \tiny{(2.0)} & 4.1 \tiny{(0.9)} \\
     M2SNet \cite{zhao2023m}                   & \textcolor{blue}{\textbf{\textit{74.9}}} \tiny{(1.3)} & \textcolor{blue}{\textbf{\textit{67.8}}} \tiny{(1.4)} & \textcolor{blue}{\textbf{\textit{71.2}}} \tiny{(1.6)} & \textcolor{red}{\textbf{\underline{84.6}}} \tiny{(0.1)} & \textcolor{blue}{\textbf{\textit{87.2}}} \tiny{(0.7)} & \textcolor{blue}{\textbf{\textit{1.7}}} \tiny{(0.3)} \\
     \hline
     \textbf{MADGNet}                         & \textcolor{red}{\textbf{\underline{77.0}}} \tiny{(0.3)} & \textcolor{red}{\textbf{\underline{69.7}}} \tiny{(0.5)} & \textcolor{red}{\textbf{\underline{75.3}}} \tiny{(0.2)} & \textcolor{blue}{\textbf{\textit{84.6}}} \tiny{(0.5)} & \textcolor{red}{\textbf{\underline{88.4}}} \tiny{(0.6)} & \textcolor{red}{\textbf{\underline{1.6}}} \tiny{(0.4)} \\
     \hline
    \end{tabular}
    \caption{Segmentation results on \textbf{Polyp Segmentation (Colonoscopy)} \cite{bernal2015wm, jha2020kvasir, vazquez2017benchmark, tajbakhsh2015automated, silva2014toward}. We train each model on CVC-ClinicDB \cite{bernal2015wm} + Kvasir-SEG \cite{jha2020kvasir} train dataset and evaluate on CVC-ClinicDB \cite{bernal2015wm}, Kvasir-SEG \cite{jha2020kvasir}, CVC-300 \cite{vazquez2017benchmark}, CVC-ColonDB \cite{tajbakhsh2015automated}, and ETIS \cite{silva2014toward} test datasets.}
    \label{tab:comparison_sota_colonoscopy_other_metrics}
\end{table}

\end{document}